\documentclass{arxiv}
\usepackage{tikz}
\usepackage[toc,page]{appendix}

\begin{document}

\title{Far-Field Aeroacoustic Shape Optimization Using Large Eddy Simulation}
\author{Mohsen Hamedi and Brian C. Vermeire\\\\
\textit{Department of Mechanical, Industrial, and Aerospace Engineering}\\
\textit{Concordia University} \\
\textit{Montr\'eal, QC, Canada}} 
\date{}

\maketitle

\begin{abstract}

This study presents a shape optimization framework that combines a Flux Reconstruction (FR) spatial discretization, Large Eddy Simulation (LES), the Ffowcs-Williams and Hawkings (FW-H) formulation, and the gradient-free Mesh Adaptive Direct Search (MADS) optimization algorithm. We emphasize the necessity of duplicating the data surface to achieve accurate far-field noise prediction in spanwise periodic problems using the FW-H formulation. The proposed parallel implementation of the optimization framework ensures consistent runtime per optimization iteration, regardless of the number of design parameters, thereby addressing a common limitation of many gradient-free algorithms. The framework is demonstrated through far-field aeroacoustic shape optimization of NACA 4-digit airfoils at a Reynolds number of $23,000$. The objective function minimizes the Overall Sound Pressure Level (OASPL) at a far-field observer positioned $10$ unit chords below the trailing edge, while preserving the mean lift coefficient and reducing the mean drag coefficient. The optimized airfoil achieves an OASPL reduction of $5.9~dB$ and over $14\%$ decrease in mean drag, while maintaining the mean lift coefficient. These results underscore the feasibility and effectiveness of the proposed approach for practical shape optimization applications.


\end{abstract}

\section{Introduction}

Aeroacoustic shape optimization has gained significant attention due to its diverse applications, including reducing wind turbine noise, minimizing aviation noise near airports, and designing quiet urban air taxis. This optimization is crucial for enhancing environmental sustainability and community comfort. The adverse impacts of noise on the environment and human health have been well established \cite{mahashabde2011assessing, basner2017aviation}. Environmental impacts include disruptions to wildlife behavior and habitat \cite{pepper2003review}, while human health impacts can range from hearing loss and sleep disturbance to increased stress levels and cardiovascular diseases \cite{basner2017aviation}. Addressing these issues necessitates reducing noise pollution, underscoring the need for advanced aeroacoustic optimization frameworks. Aeroacoustic shape optimization thus plays a critical role in mitigating these negative effects, emphasizing its significance for ecological sustainability and public health. In this study, a far-field aeroacoustic shape optimization framework is proposed, consisting of three components: a Large Eddy Simulation (LES) flow solver, an acoustic solver, and an optimization algorithm. To our knowledge, this is the first work to demonstrate far-field aeroacoustic optimization using LES.

Aeroacoustic shape optimization frameworks employ various computational methods to minimize noise while ensuring aerodynamic performance. XFOIL \cite{drela1989xfoil} simulations are commonly used in aeroacoustic shape optimization for aerodynamic analysis, employing panel methods for cost-effective exploration of design spaces \cite{kou2023aeroacoustic, volkmer2018aeroacoustic, jones2000aerodynamic}. However, these methods lack the precision required for reliable optimal designs \cite{volkmer2018aeroacoustic}. An alternative to panel methods is Reynolds-Averaged Navier-Stokes (RANS) simulations. However, due to the inherent unsteady nature of noise phenomena, RANS simulations cannot effectively capture unsteady flow characteristics \cite{slotnick2014cfd} and have limitations in representing the complete acoustic spectrum of noise generation \cite{zhang2022acoustic}. Consequently, scale-resolving techniques, i.e., LES and Direct Numerical Simulation (DNS), offer a more detailed representation of flow physics, albeit with added computational costs \cite{colonius2004computational, marsden2007trailing, marsden2001shape}. Common Computational Fluid Dynamics (CFD) codes, such as OpenFOAM \cite{openfoam}, SU2 \cite{su2, palacios2013stanford}, and CHARLES \cite{charles}, rely on Finite Volume (FV) methods with second-order spatial accuracy, which, despite handling complex geometries, are limited in harnessing the full computational power of modern hardware \cite{vincent2016towards}. These industry-standard FV methods achieve only $3\%$ of theoretical peak performance and Graphical Processing Units (GPUs) \cite{langguth2013gpu}, while the Flux Reconstruction (FR) approach \cite{huynh2007flux} has demonstrated over $55\%$ efficiency \cite{vincent2016towards}, making it computationally superior with additionally reduced numerical dispersion and dissipation errors through high-order accuracy \cite{abgrall2017high, hesthaven2017numerical, wang2013high}. In addition, the FR approach has been shown to be suitable for scale-resolving simulations, leveraging the behaviour of its numerical error for Implicit LES (ILES) \cite{vermeire2016implicit}, and via filtering approaches for highly under-resolved problems \cite{hamedi2022optimized}. In this study, an in-house High-ORder Unstructured Solver (HORUS) is used, which employs the FR approach for spatial discretization of the governing equations and ILES for turbulence modelling. 

In general, there are two approaches to sound prediction. The first, highly accurate but computationally demanding, is the direct approach. This approach involves computing the sound field along with unsteady turbulent flow, requiring the observer to be inside the computational domain, making it computationally expensive for far-field sound computation. Therefore, even if the current growth level in supercomputers' performance remains the same in the forthcoming years, this method remains prohibitively expensive for general aeroacoustic problems in the aviation industry. Alternatively, the hybrid approach is more computationally efficient for far-field aeroacoustics. In this approach, the sound waves are generated and resolved in the near-field within the flow solver, and then propagated to the far-field within the acoustic solver. This method proves computationally efficient and significantly less expensive compared to employing a flow solver for the whole domain. The Ffowcs Williams and Hawkings (FW-H) equations \cite{ffowcs1969sound} are widely used as an acoustic analogy in the aviation industry \cite{lockard2002comparison, magagnato2003far, spalart2009variants, mendez2013use, naqavi2016far, bodling2018implementation, ribeiro2023lessons}. 

Optimization techniques can be broadly classified into gradient-based and gradient-free methods. The choice of method depends on factors such as the cost of function evaluation, availability of gradient information, function noise level, and implementation complexity. Gradient-based methods require gradient information and are efficient for smooth, continuous functions. Gradient-free methods, while generally more robust to noisy functions and simpler to implement, may require more function evaluations. The gradient-free Mesh Adaptive Direct Search (MADS) \cite{audet2006mesh} and its extension, Orthogonal MADS (OrthoMADS) \cite{abramson2009orthomads}, are highly effective for optimization, particularly in non-smooth and chaotic flows. MADS has demonstrated significant performance improvements in aerodynamic \cite{karbasian2022gradient,aubry2022high} and aeroacoustic \cite{hamedi2024near, hamedi2025gradient} shape optimization when integrated with high-order LES techniques. OrthoMADS, an advancement of MADS, introduces deterministic and structured polling directions, improving design space exploration and computational efficiency without compromising robustness. Both algorithms are robust against complex flow behaviors and do not rely on gradient information. However, their scalability remains a challenge, as runtime and computational costs increase linearly with the number of design variables, making them prohibitive for large-scale problems. To address this, our proposed framework employs parallelization, enabling concurrent CFD simulations during each optimization iteration. This approach eliminates runtime dependency on the number of design parameters, provided sufficient computational resources are available.

Despite advancements, the challenge of accurately predicting and minimizing far-field aeroacoustic emissions persists. Addressing this issue is essential for advancing the design of quieter aerodynamic structures. In this study, we introduce an aeroacoustic shape optimization framework based on the FR approach, FW-H formulation, and the gradient-free OrthoMADS optimization algorithm. Building upon our prior works \cite{hamedi2024near, hamedi2025gradient}, which assessed MADS optimization algorithm for aeroacoustic shape optimization via high-order FR in two and three dimensions, we extend its application to far-field aeroacoustic shape optimization. To our knowledge, no previous work has combined gradient-free OrthoMADS algorithm with a high-order LES solver for far-field aeroacoustic shape optimization.

This paper is outlined as follows. Section \ref{sec:Methodology} presents the methodology, followed by NACA 4-digit airfoil shape optimization in Section \ref{sec:Optimization}. The conclusions and future work recommendations are given in Section \ref{sec:Conclusions}. Finally, acoustic solver formulation, implementation, verification, and validation are explained in Appendices \ref{sec:AcousticFormulation}, \ref{sec:AcousticSolver}, \ref{sec:AcousticVerification}, and \ref{sec:AcousticValidation}, respectively, followed by explaining the OrthoMADS optimization algorithm in Appendix \ref{sec:OrthoMads}.

\section{Methodology}
\label{sec:Methodology}

This section presents an overview of the methodology employed to solve the unsteady Navier-Stokes equations in HORUS, along with the aeroacoustic shape optimization framework.

\subsection{Governing Equations}

In this study, the instantaneous compressible Navier-Stokes equations are solved without explicit filtering or SubGrid-Scale (SGS) modeling, following the ILES methodology. The high-order FR discretization provides implicit filtering and dissipation, where the computational mesh effectively acts as the filter and the numerical dissipation characteristics of the FR scheme provide the necessary energy removal at the smallest resolved scales \cite{vermeire2016implicit, hamedi2022optimized}. This approach leverages the inherent dissipative and stabilizing properties of the high-order numerical method to model unresolved subgrid scales without requiring explicit SGS models or filtering operations. The compressible unsteady Navier–Stokes equations can be cast in the following general form
\begin{equation}
\frac{\partial \pmb{u}}{\partial t} + \pmb{\nabla} \cdot \pmb{F} = 0,
\label{equation_conservation_law}
\end{equation}
where $t$ is time and $\pmb{u}$ is a vector of conserved variables
\begin{equation}
\pmb{u} = 
\begin{bmatrix}
\rho \\
\rho u_i \\
\rho E
\end{bmatrix},
\end{equation}
where $\rho$ is density, $\rho u_i$ is a component of the momentum, $u_i$ are velocity components, and $\rho E$ is the total energy. The inviscid and viscous Navier-Stokes fluxes are 
\begin{equation}
\pmb{F}_{inv,j} (\pmb{u}) = 
\begin{bmatrix}
\rho u_j \\
\rho u_i u_j + \delta_{ij} p \\
u_j ( \rho E + p)
\end{bmatrix},
\end{equation}
and
\begin{equation}
\pmb{F}_{vis, j} (\pmb{u}, \nabla \pmb{u}) =
\begin{bmatrix}
0 \\
\tau_{ij} \\
-q_j - u_i \tau_{ij}
\end{bmatrix} ,
\end{equation}
respectively, where $\delta_{ij}$ is the Kronecker delta. The pressure is determined via the ideal gas law as
\begin{equation}
p = (\gamma - 1) \rho \left( E - \frac{1}{2} u_k u_k \right) ,
\end{equation}
where $\gamma=1.4$ is the ratio of the specific heat at constant pressure, $c_p$, to the specific heat at constant volume, $c_v$. The viscous stress tensor is
\begin{equation}
\tau_{ij} = \mu \left( \frac{\partial u_i}{\partial x_j} + \frac{\partial u_j}{\partial x_i} - \frac{2}{3} \frac{\partial u_k}{\partial x_k} \delta_{ij} \right) ,
\end{equation}
and, the heat flux is
\begin{equation}
q_j = - \frac{\mu}{Pr} \frac{\partial}{\partial x_j} \left( E + \frac{p}{\rho} - \frac{1}{2} u_k u_k \right) ,
\end{equation}
where $\mu$ is the dynamic viscosity and $Pr=0.71$ is the Prandtl number. 

\subsection{Aeroacoustic Shape Optimization Framework}

\begin{figure}
\centering
\includegraphics[width=\textwidth]{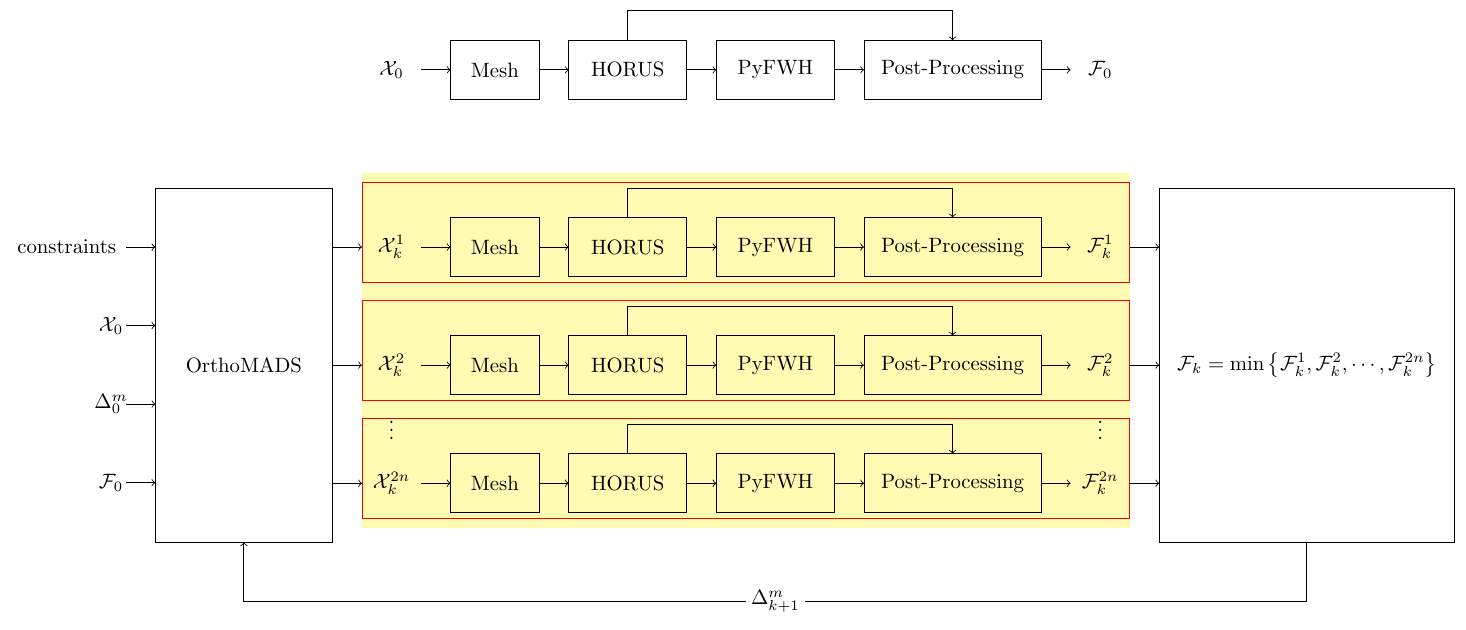}
\caption{Visualization of the proposed far-field aeroacoustic shape optimization framework. The two-layer parallel part of the framework is highlighted in yellow, in which, each red rectangle is run on multiple GPUs while all the red rectangles are also performed concurrently.}
\label{fig-alg}
\end{figure}

The proposed aeroacoustic shape optimization framework, depicted in Figure \ref{fig-alg}, integrates several computational tools to achieve optimal aerodynamic and aeroacoustic performance. This framework is designed to leverage high-performance computing and state-of-the-art optimization algorithms, ensuring both accuracy and efficiency.

The process begins with the generation of a computational mesh for the baseline design, denoted as $\pmb{\mathcal{X}}_0$. Using HORUS, the flow field is computed in parallel on GPUs, significantly reducing computation time. The computed flow fields serve as inputs to the acoustic solver, PyFWH. The objective function, $\mathcal{F}_0$, is evaluated by combining aerodynamic characteristics from HORUS and the overall sound pressure level (OASPL) from PyFWH. Next, the optimization algorithm is initialized with an initial mesh size parameter ($\Delta^m_0$), the baseline design ($\pmb{\mathcal{X}}_0$), and the computed objective function ($\mathcal{F}_0$). The algorithm identifies $2n$ candidate designs, where $n$ represents the total number of design parameters. For each candidate design, a new mesh is generated, and the flow fields are computed using HORUS. These flow fields are then used as inputs to the PyFWH solver to compute the OASPL at the observer location(s). Each CFD simulation with HORUS is executed in parallel across multiple GPUs, and the entire optimization iteration is also parallelized, creating  two-layers of parallelism. This approach effectively reduces the runtime of $2n$ CFD simulations per optimization iteration to that of a single CFD simulation, provided that sufficient computational resources are available. Upon evaluating the objective functions of the candidate designs, the optimal design is selected and compared to the incumbent design. Depending on whether a superior design is identified, the mesh size parameter is updated, and the optimization process continues. The optimization converges when the mesh size parameter drops below $10^{-6}$ and the changes in design parameter values between consecutive iterations are less than one percent. These convergence criteria indicate the algorithm has successfully identified an optimal design. 

The OrthoMADS optimization algorithm is explained in details in Appendix \ref{sec:OrthoMads}, and the formulation of the PyFWH solver and its implementation, validation, and verification are further discussed in Appendices \ref{sec:AcousticFormulation} to \ref{sec:AcousticValidation}. Furthermore, for a comprehensive understanding of the proposed far-field aeroacoustic shape optimization, the complete algorithm is presented in Algorithm \ref{algorithm_mads}. The proposed framework exemplifies the integration of high-order CFD solvers with optimization algorithms, demonstrating a robust and efficient methodology for aeroacoustic shape optimization. The parallel execution of CFD simulations and optimization iterations not only accelerates the process but also ensures scalability for complex aerodynamic and aeroacoustic problems. 

\begin{algorithm}[htbp]
\SetAlgoLined
\caption{The far-field aeroacoustic shape optimization framework.}
\label{algorithm_mads}
$k=0$; \\
OrthoMADS Iteration, $iter=0$; \\

Run Baseline Design; \\
Evaluate $\mathcal{F}_0$; \\
Define Incumbent $\mathcal{I}_0 = \mathcal{F}_0$; \\

Define $\Delta^m_0$; \\

\While{True}{

	\If{$\Delta^m_k > \Delta^m_0$}{
		$\Delta^m_k = \Delta^m_0$;\\}
		
	Generate Candidate Designs, $\pmb{p}^1_k, ..., \pmb{p}^{2n}_k$; \\
	
	\For{$i=1,...,2n$}{
		Run HORUS and PyFWH for $\pmb{p}^i_k$; \\
		Evaluate $\mathcal{F}^i_k$; \\}
		
	\eIf{$\min\left\lbrace \mathcal{F}^1_k, ..., \mathcal{F}^{2n}_k \right\rbrace < \mathcal{I}_{iter}$}{
		$\Delta^m_{k+1} = 4 \Delta^m_k$;\\
		$iter$\texttt{+=}$1$;\\
		$\mathcal{I}_{iter} = \min\left\lbrace \mathcal{F}^1_k, ..., \mathcal{F}^{2n}_k \right\rbrace$;\\}
		{$\Delta^m_{k+1} = \frac{1}{4} \Delta^m_k$;\\}

	$k$\texttt{+=}$1$; \\
	
	\If{$\Delta^m_k < 10^{-6}$ and $\left| \frac{  \pmb{\mathcal{X}}_k - \pmb{\mathcal{X}}_{k-1} }{\pmb{\mathcal{X}}_{k-1}} \right| < 0.01$}{
	\texttt{break;}}
}
\end{algorithm}

\section{Aeroacoustic Shape Optimization of a NACA 4-Digit Airfoil}
\label{sec:Optimization}

This study investigates NACA 4-digit airfoils at a Reynolds number of $23,000$, which is representative of several low-speed aerodynamic applications including small-scale Unmanned Aerial Vehicles (UAVs), Micro Air Vehicles (MAVs), and small wind turbines operating in low-wind environments. At this Reynolds number, characteristic of the transitional flow regime ($10,000 < Re < 100,000$), the flow exhibits complex phenomena including laminar separation bubbles, boundary layer transition, and unsteady vortex shedding that significantly influence both aerodynamic performance and noise generation mechanisms. Furthermore, this Reynolds number provides an ideal benchmark for validating high-fidelity computational methods and optimization frameworks in transitional flow regimes, where accurate prediction of flow separation, reattachment, and associated acoustic phenomena remains challenging.

This section validates the PyFWH solver against direct acoustic computation using HORUS. The NACA0012 airfoil at a $6^\circ$ angle of attack serves as the baseline for far-field aeroacoustic shape optimization.

\subsection{Computational Details}

\begin{figure}
\centering
\begin{subfigure}{0.45\textwidth}
\centering
\includegraphics[width=\textwidth]{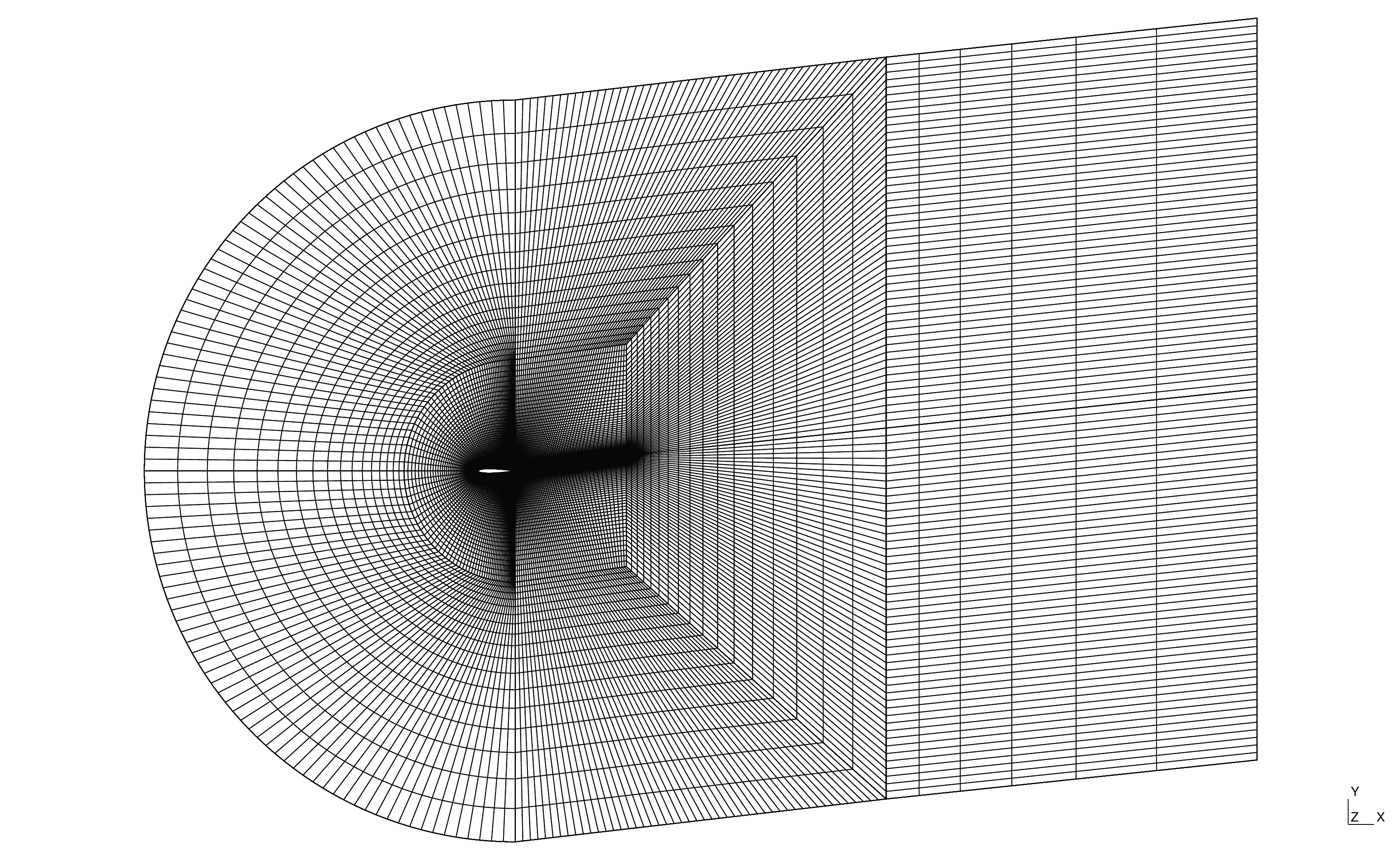}
\subcaption{The computational domain.}
\end{subfigure}
\begin{subfigure}{0.45\textwidth}
\centering
\includegraphics[width=\textwidth]{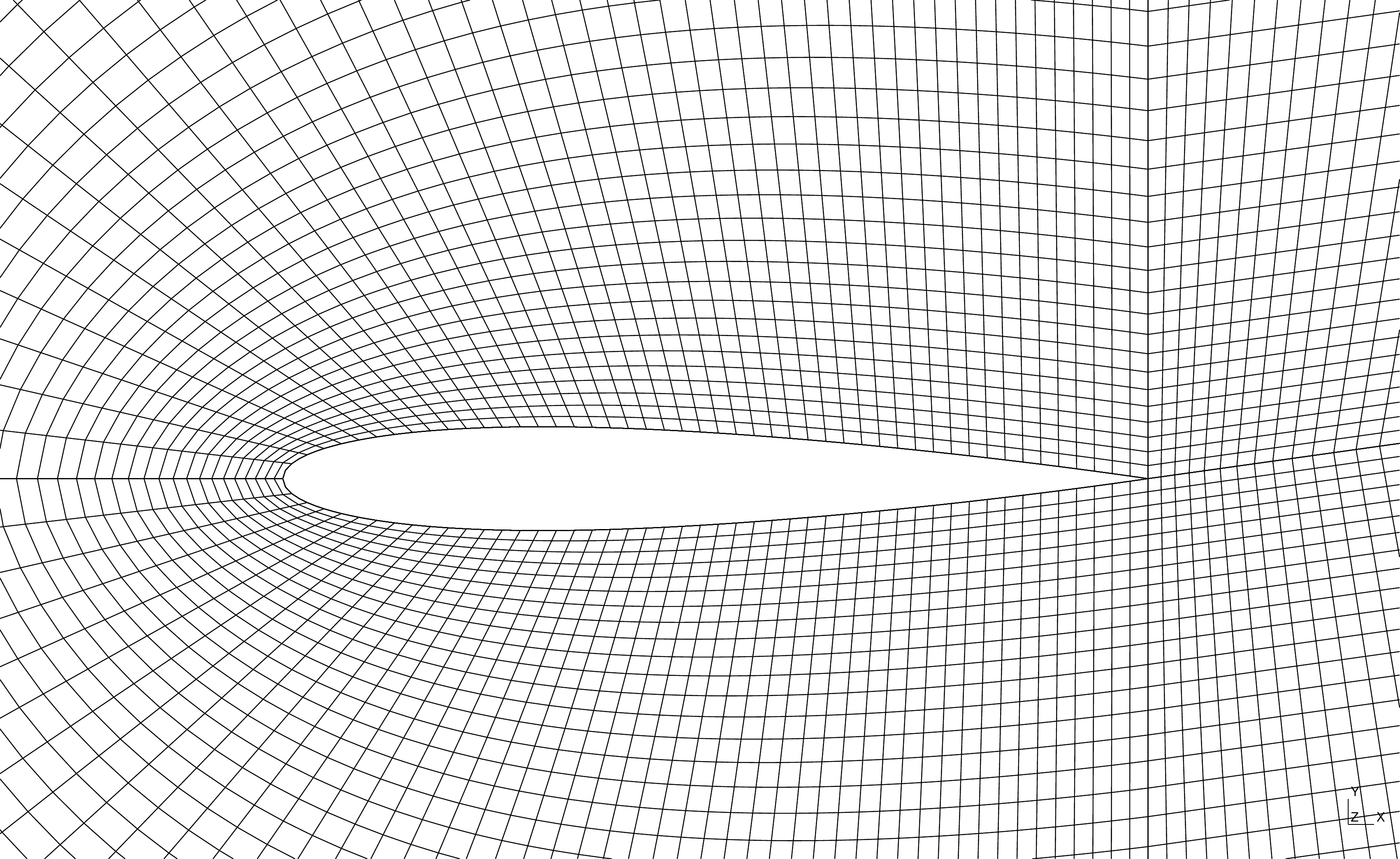}
\subcaption{The vicinity of the airfoil.}
\end{subfigure}
\caption{The computational grid for NACA0012 airfoil at $\alpha = 6^\circ$.}
\label{fig_naca_3d_domain}
\end{figure}

\begin{figure}
\centering
\begin{subfigure}{0.45\textwidth}
\centering
\includegraphics[width=\textwidth]{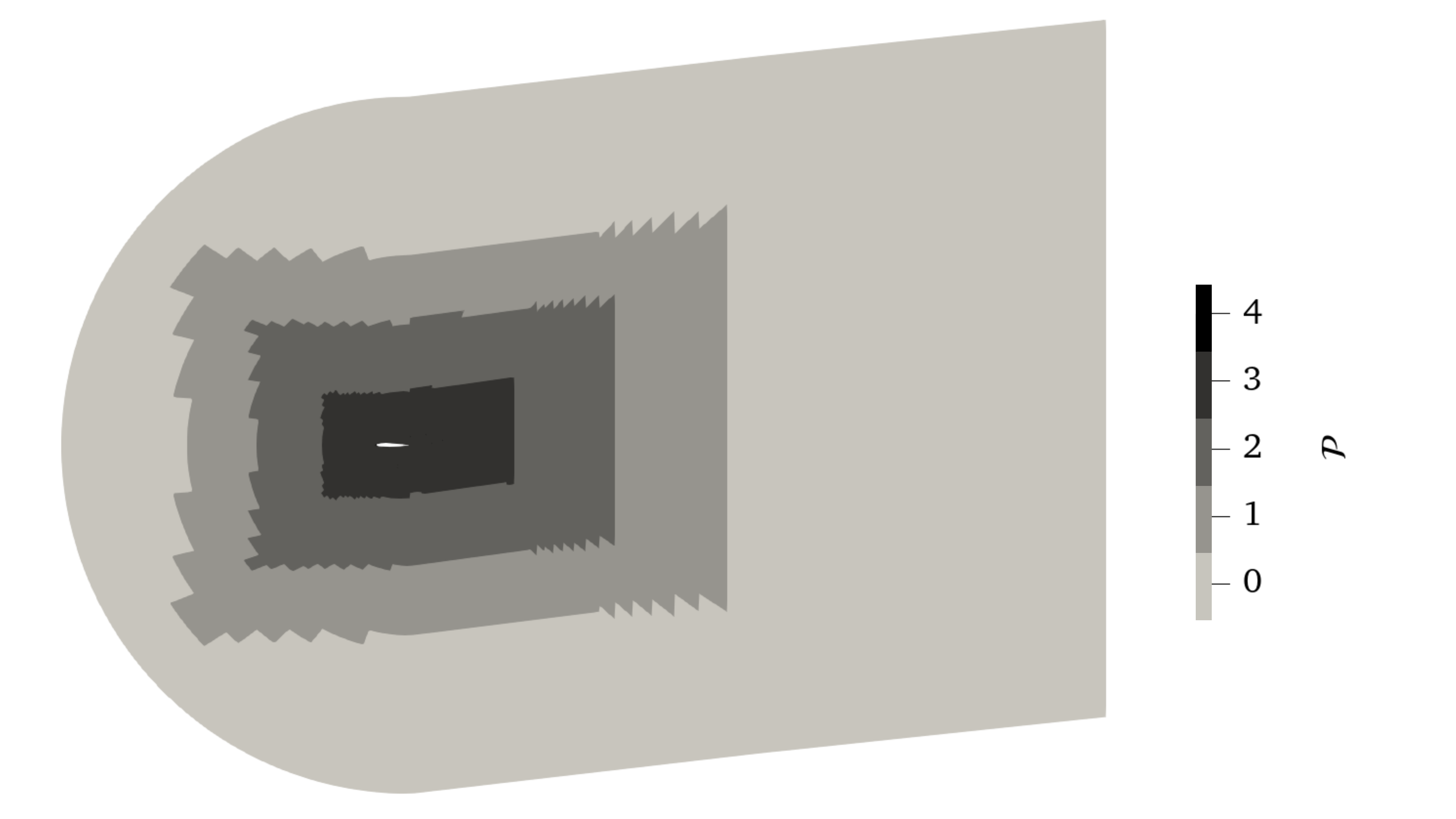}
\subcaption{Low resolution, $\mathcal{P}0 - \mathcal{P}3$.}
\label{fig_naca_3d_p_distribution_p3}
\end{subfigure}
\begin{subfigure}{0.45\textwidth}
\centering
\includegraphics[width=\textwidth]{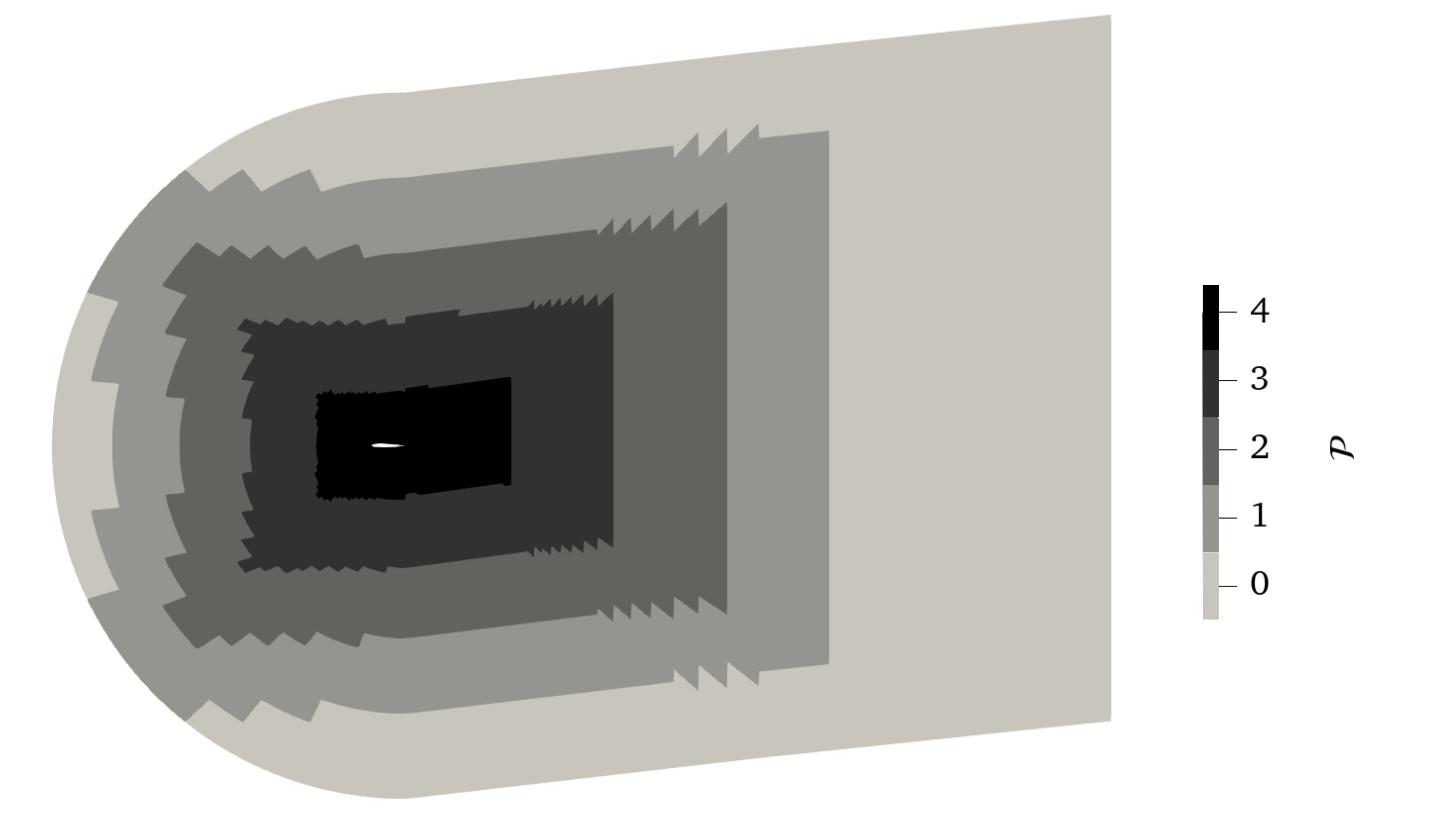}
\subcaption{High resolution, $\mathcal{P}0 - \mathcal{P}4$.}
\end{subfigure}
\caption{Different solution polynomial distributions for grid independence study of NACA0012 airfoil at $\alpha = 6^\circ$.}
\label{fig_naca_3d_p_distribution}
\end{figure}

The computational grid consists of $121,520$ hexahedral elements, illustrated in Figure \ref{fig_naca_3d_domain}. The domain extends to $20c$ in the $x$-direction, $10c$ in the $y$-direction, and $0.2c$ in the $z$-direction, with $c=1$ representing the airfoil chord. Notably, elements in the wake region are inclined at the angle of attack to accurately capture trailing-edge vortices. The flow conditions are characterized by a Reynolds number of $23,000$, a free-stream Mach number of $M=0.2$, and Prandtl number is $Pr=0.71$. The simulation is run for $10$ convective times to allow the initial transition disappears and then run for another $70$ convective times for flow statistics averaging. Additionally, a variable solution polynomial degree is implemented to eliminate acoustic wave reflections from boundaries, as demonstrated in Figure \ref{fig_naca_3d_p_distribution}.

\begin{figure}
\centering
\includegraphics[width=\textwidth]{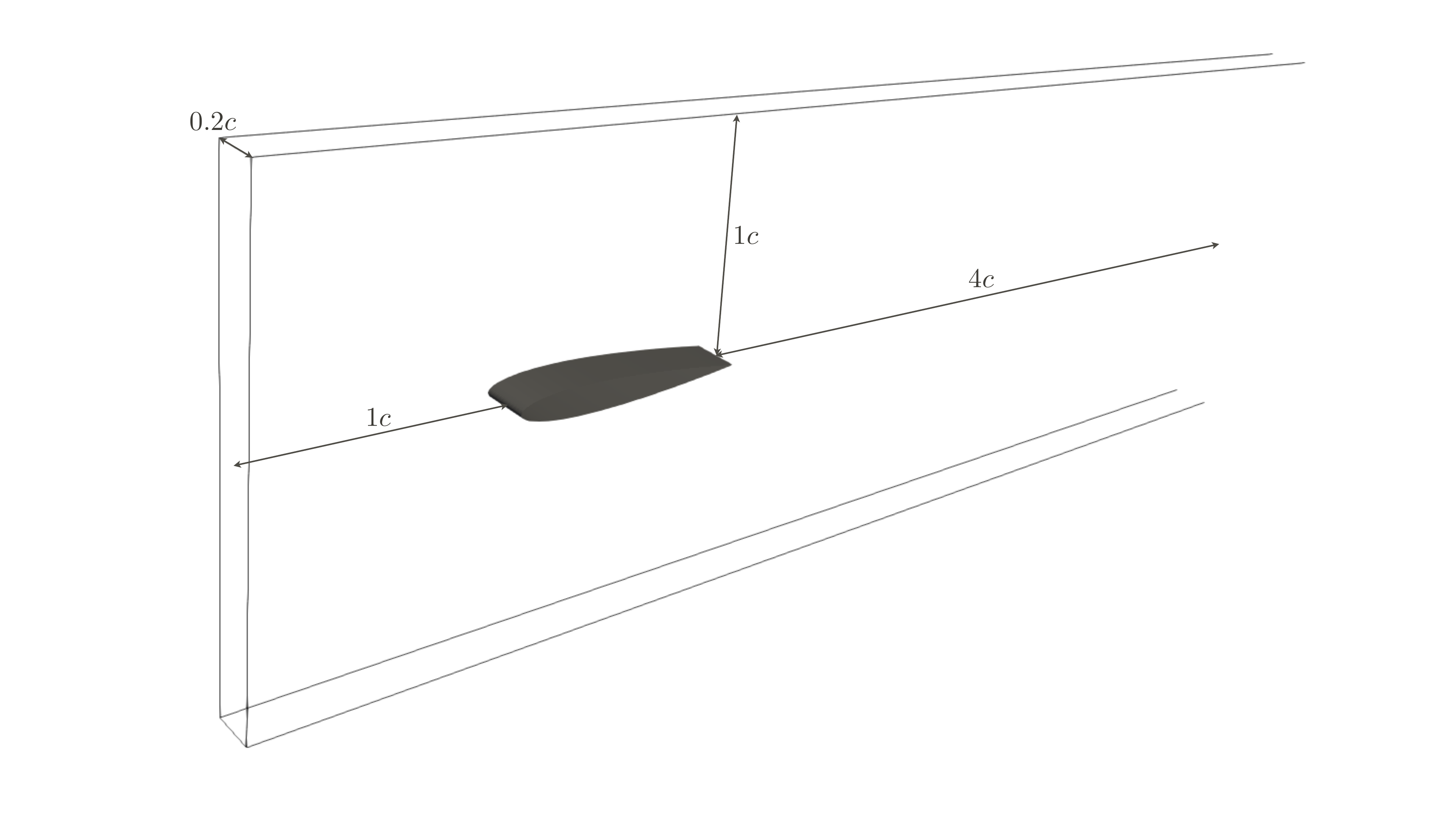}
\caption{Schematic diagram of the data surface with $L_z = 0.2c$.}
\label{fig_naca_3d_data_surface_0dup}
\end{figure}

An open permeable data surface gathers flow field data for sound computation in the PyFWH solver. This surface extends to one chord length in the $y$-direction and four chord lengths into the wake region, and covers the entire airfoil span, effectively capturing relevant turbulent structures in the near-field region, as illustrated in Figure \ref{fig_naca_3d_data_surface_0dup}. The surface remains open-ended to prevent erroneous acoustic wave generation associated with vortices crossing it. The spacing between sample points on the data surface is set at $0.01c$ to ensure a uniform distribution, with points positioned away from periodic planes to avoid spurious noise. Consequently, the first and last points in the spanwise direction are situated $0.005c$ away from these planes.

The second-order Nasab-Pereira-Vermeire scheme \cite{hedayati2021optimal} is employed with adaptive time-stepping \cite{vermeire2020optimal}, featuring an averaged time-step size of approximately $\Delta t_{avg}=0.001561 t_c$ non-dimensionalized by $t_c = c / U_\infty$, where $U_\infty$ is the free-stream velocity. Data collection occurs every $25$ time-steps, resulting in a sampling rate of $\Delta t = 0.009233 t_c$, providing $4332$ flow snapshots over a $40t_c$ averaging period. 

\subsection{Grid Independence Study}

\begin{figure}
\centering
\includegraphics[width=0.7\textwidth]{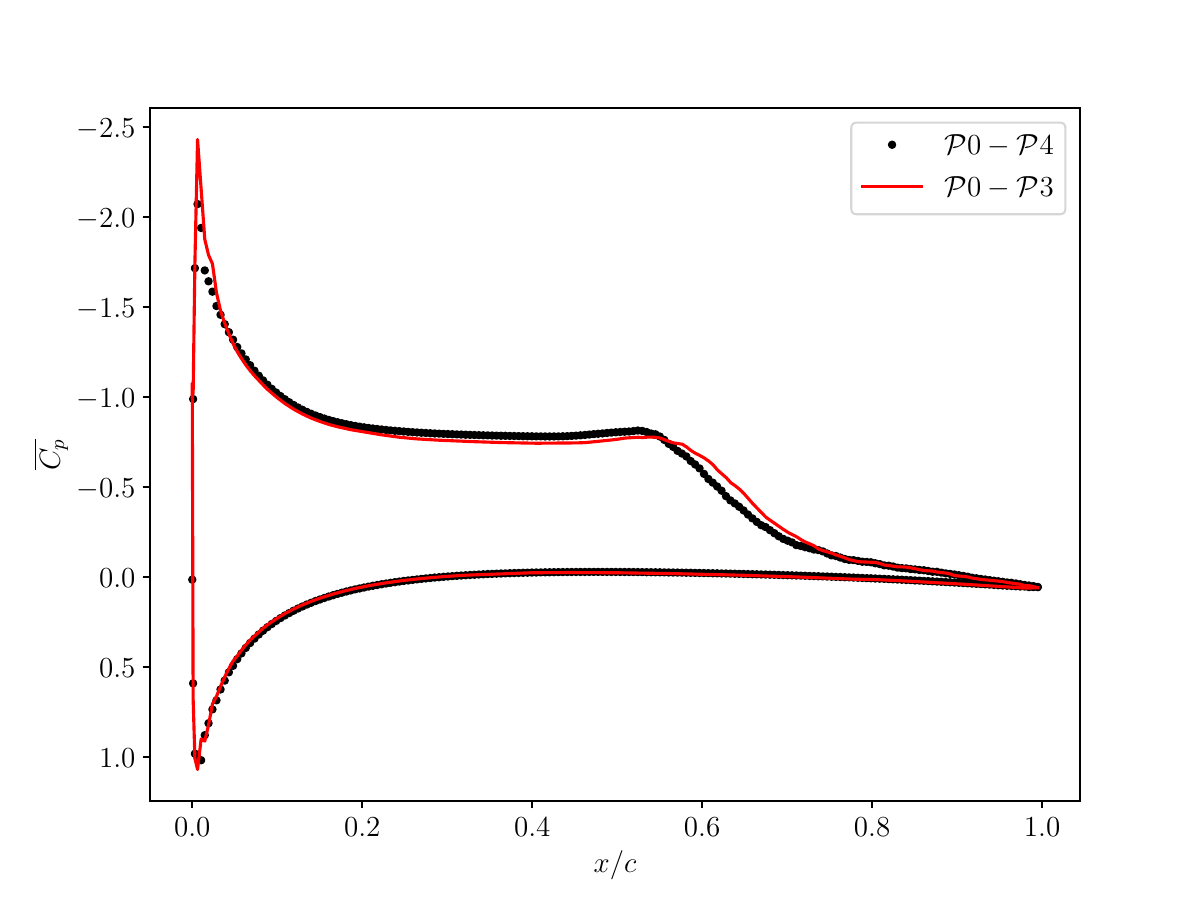}
\caption{The time-averaged pressure coefficient for both $\mathcal{P}3$ and $\mathcal{P}4$ simulations.}
\label{fig_naca_cp}
\end{figure}

\begin{figure}
\centering
\includegraphics[width=0.7\textwidth]{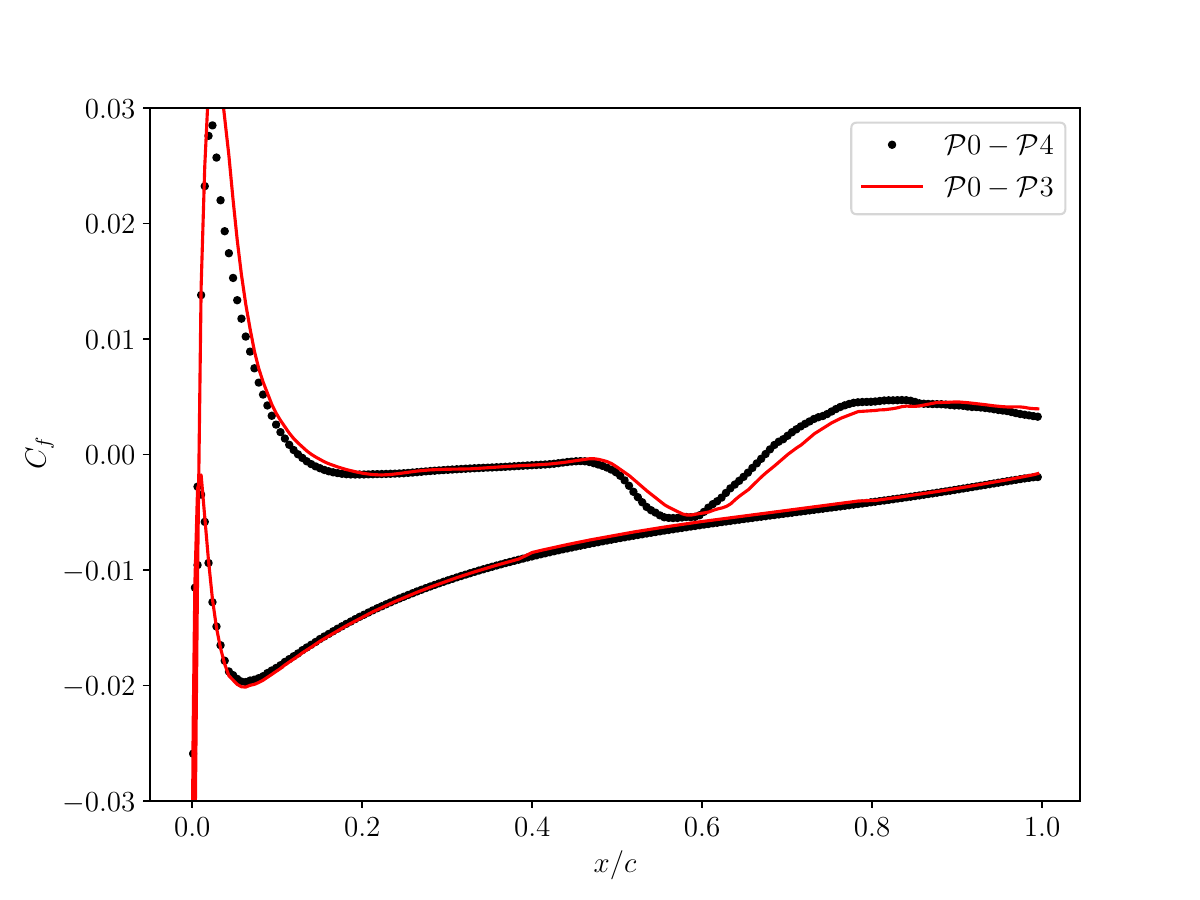}
\caption{The skin friction coefficient for both $\mathcal{P}3$ and $\mathcal{P}4$ simulations.}
\label{fig_naca_cf}
\end{figure}

Two distinct grid resolutions are employed with maximum solution polynomial degrees of $\mathcal{P}3$ and $\mathcal{P}4$, as depicted in Figure \ref{fig_naca_3d_p_distribution}. The time-averaged lift and drag coefficients are compared to the ILES reference data \cite{kojima2013large}, presented in Table \ref{table_naca_3d_timehistories}. The difference between the time-averaged lift and drag coefficients obtained from the $\mathcal{P}3$ simulation and the reference data is around $2\%$, affirming the adequacy of the $\mathcal{P}3$ simulation's grid resolution. 
The root-mean-square of the pressure fluctuations scaled with the free-stream pressure, $p^\prime_{rms}/p_\infty$, at an observer located two unit chord lengths below the trailing edge is computed for both $\mathcal{P}3$ and $\mathcal{P}4$ simulations. 
Various averaging window lengths are applied, and the results are summarized in Table \ref{table_naca_3d_spl_independence}. 
The time-averaged pressure coefficient, $\overline{C_p}$, and the skin friction coefficient, $C_f$, for both resolutions are shown in Figures \ref{fig_naca_cp} and \ref{fig_naca_cf}, respectively. These plots show that the separation point, identified with each simulation, are very close and differ by less than $2\%$. Considering the findings presented in Tables \ref{table_naca_3d_timehistories} and \ref{table_naca_3d_spl_independence}, and Figures \ref{fig_naca_cp} and \ref{fig_naca_cf}, we opt to conduct $\mathcal{P}3$ simulation for a total duration of $70$ convective times for the optimization study.

\begin{table}
\centering
\caption{The time-averaged lift and drag coefficients of NACA0012 airfoil at $\alpha = 6^\circ$.}
\begin{tabular}{cccc}
\hline
 & $\mathcal{P}0 - \mathcal{P}3$ & $\mathcal{P}0 - \mathcal{P}4$ & reference \cite{kojima2013large} \\
\hline
$\overline{C_L}$ & $0.6534$ & $0.6399$ & $0.6402$ \\
$\overline{C_D}$ & $0.0553$ & $0.0548$ & $0.0541$ \\
\hline
\end{tabular}
\label{table_naca_3d_timehistories}
\end{table}

\begin{table}
\centering
\caption{The grid independence study of $p^\prime_{rms}/p_\infty$ using different averaging window lengths for NACA0012 airfoil at $\alpha = 6^\circ$, for an observer located two unit-chord below the trailing edge.}
\begin{tabular}{ccc}
\hline
\multirow{2}{*}{Averaging Window Length} & \multicolumn{2}{c}{$p^\prime_{rms}/p_\infty$} \\
 & $\mathcal{P}0 - \mathcal{P}3$ & $\mathcal{P}0 - \mathcal{P}4$ \\
\hline
$20t_c$ & 1.10E-4 & 1.29E-4 \\
$40t_c$ & 1.22E-4 & 1.29E-4 \\
$60t_c$ & 1.22E-4 & 1.27E-4 \\
$80t_c$ & 1.22E-4 & 1.27E-4 \\
\hline
\end{tabular}
\label{table_naca_3d_spl_independence}
\end{table}

\subsection{PyFWH Validation}

The mathematical formulation, implementation, verification, and validation details of the PyFWH solver used in this work are presented in the Appendices \ref{sec:AcousticFormulation}, \ref{sec:AcousticSolver}, \ref{sec:AcousticVerification}, and \ref{sec:AcousticValidation}, respectively.
Here, the PyFWH solver is validated by comparing its results with those from HORUS for a NACA0012 airfoil at a $6^\circ$ angle of attack. The acoustic pressure is first directly computed via HORUS for a near-field observer located two unit chords below the trailing edge. The PyFWH solver then computes the acoustic pressure at the same location. The resulting pressure perturbations and its corresponding Power Spectral Density (PSD) from the PyFWH solver are compared with those obtained from HORUS. The PSD of pressure fluctuations is computed using Welch's method \cite{welch1967use} with three Hanning windows and $50\%$ overlap.

\subsubsection{Data Surface Duplication}

\begin{figure}
\centering
\includegraphics[width=\textwidth]{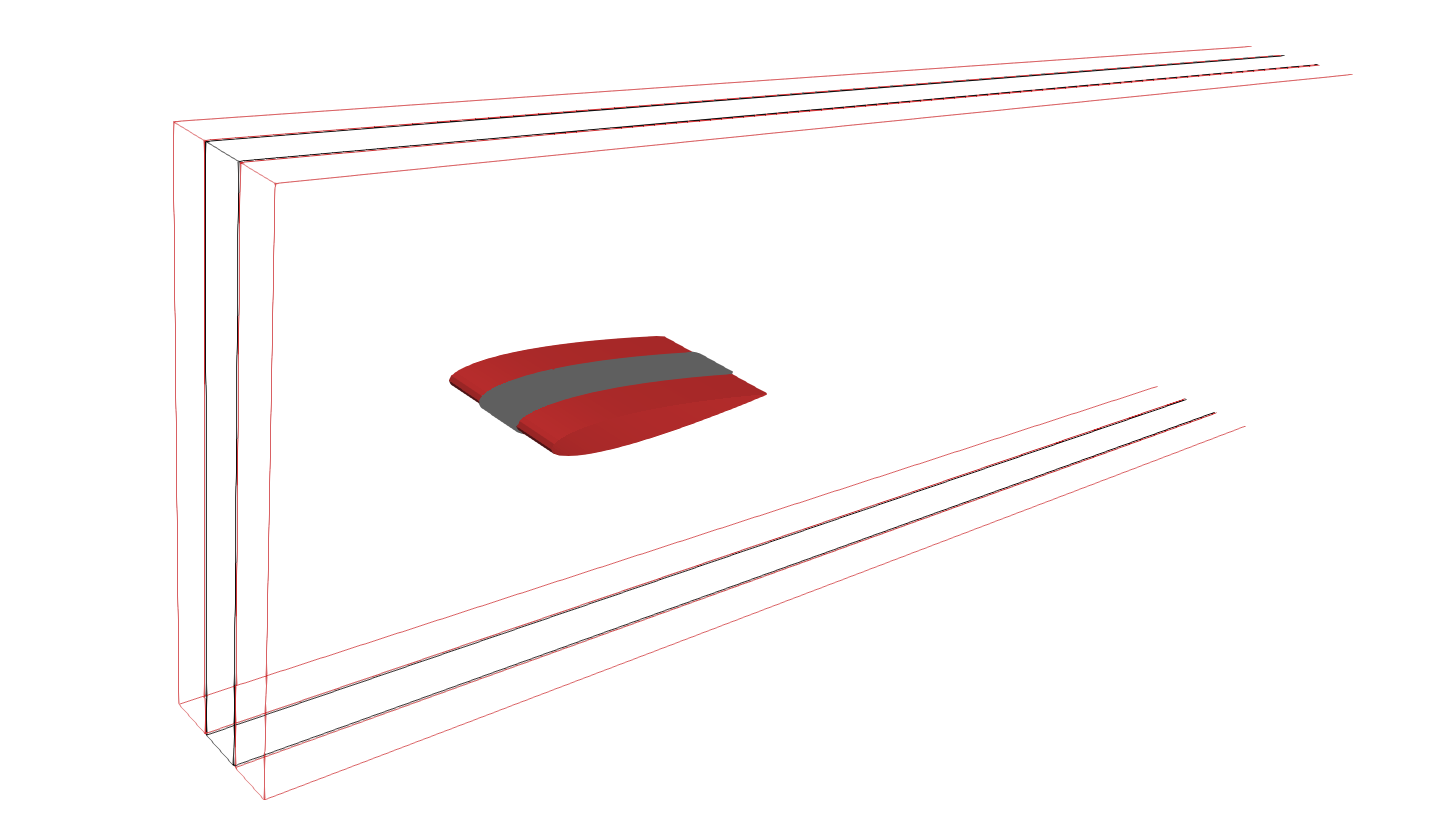}
\caption{Schematic diagram of the repeated data surface in the periodic spanwise direction with $L_z = 0.6c$.}
\label{fig_naca_3d_data_surface_1dup}
\end{figure}

\begin{figure}
\centering
\begin{subfigure}{0.7\textwidth}
\includegraphics[width=\textwidth]{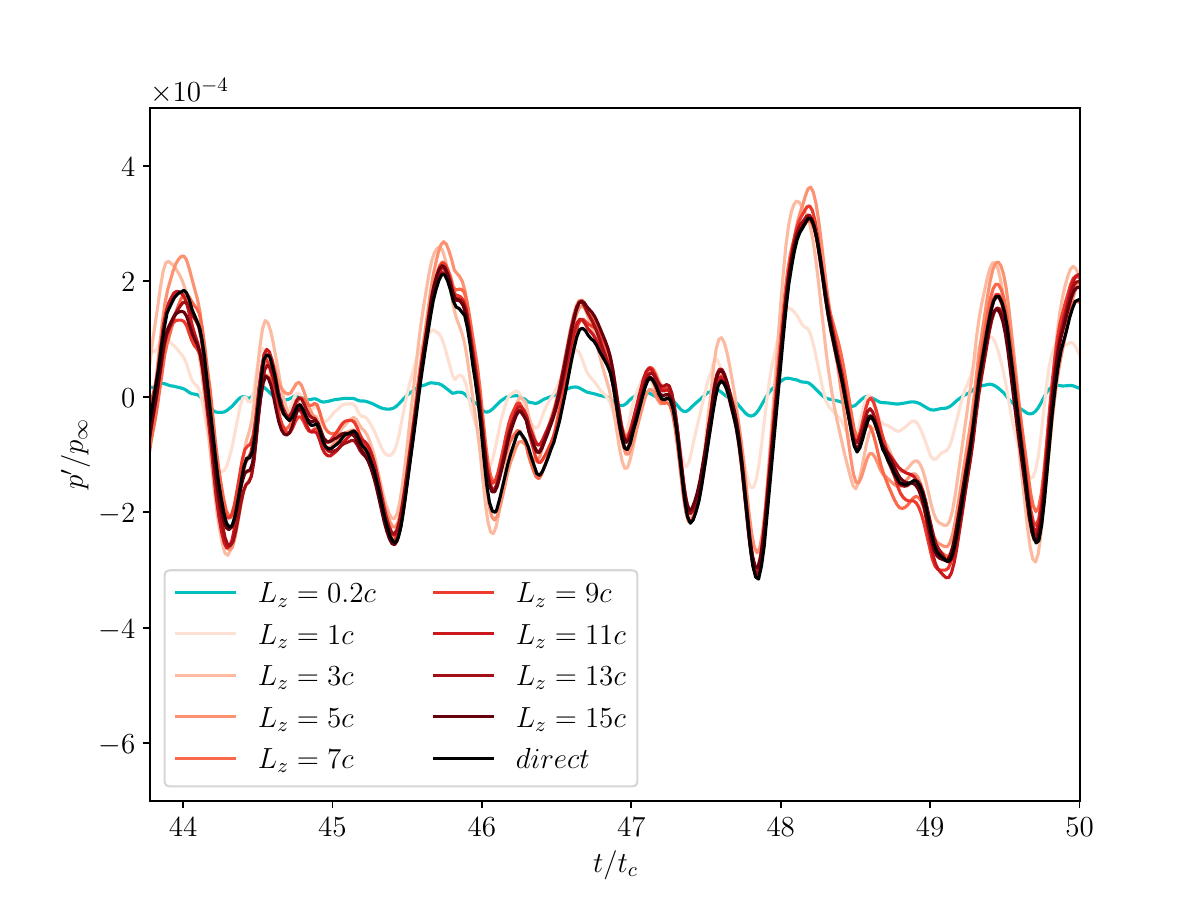}
\caption{The pressure perturbation time history.}
\label{fig_naca_fwh_validation_pacs}
\end{subfigure}
\begin{subfigure}{0.7\textwidth}
\includegraphics[width=\textwidth]{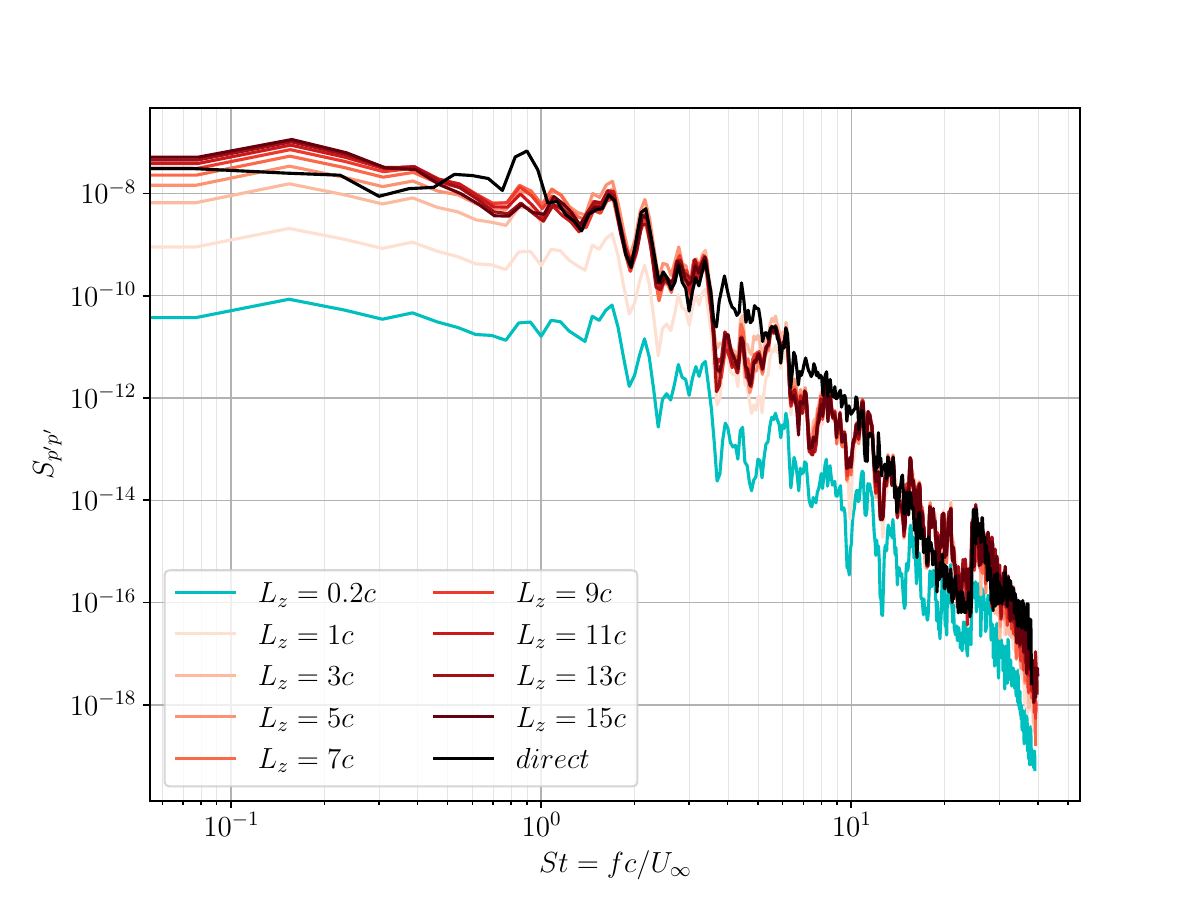}
\caption{The PSD of pressure perturbations.}
\label{fig_naca_fwh_validation_psdspl}
\end{subfigure}
\caption{The pressure perturbation time history and its corresponding PSD at the near-field observer using multiple sets of data surface duplications.}
\label{fig_naca_fwh_validation}
\end{figure}

The acoustic pressure time history along with the PSD of pressure fluctuations for the near-field observer using different spanwise data surface duplications are illustrated in Figure \ref{fig_naca_fwh_validation}. It is apparent that the acoustic solver fails to accurately predict pressure fluctuations and its PSD when the data surface is not duplicated in the spanwise direction. This observation highlights that relying solely on the computational domain is insufficient for capturing far-field noise. The primary issue stems from the spanwise periodic boundary conditions, which are implemented to mimic an infinite span in the CFD simulation. However, while this approach is appropriate for the flow solver, the FW-H formulation requires a different treatment to properly represent infinite span conditions. Specifically, the acoustic method necessitates duplication of the data surface in the spanwise direction to correctly account for acoustic wave propagation in this dimension within the hybrid approach. To address this discrepancy, an iterative integration of the data surface is necessary on domains shifted either side of the airfoil over a sufficient distance. The data surface is subsequently duplicated in the spanwise direction, extending to various sets of $L_z$ values. It is evident that extending the data surface to $L_z=15c$ proves sufficient for accurate noise prediction. 
Table \ref{table_naca_fwh_spl_convergence} summarizes the $p^\prime_{rms}/p_\infty$ for the near-field observer when using different data surface duplications. A comparison to the direct result confirms the effectiveness of data surface duplication up to $L_z=15c$.
Note that according to the inverse square law of acoustic wave dissipation, as the observer is placed further away from the data surface, more duplication of the data surface in the periodic spanwise direction is required (see Figure \ref{fig_naca_3d_data_surface_1dup}).

\begin{table}
\centering
\caption{The $p^\prime_{rms}/p_\infty$ of the near-field observer using different sets of data surface duplications.}
\begin{tabular}{cc}
\hline
Duplication Length $(L_z)$ & $p^\prime_{rms}/p_\infty$ \\
\hline
$0.2c$ & 1.33E-5 \\
$1c$ & 6.39E-5 \\
$3c$ & 1.33E-4 \\
$5c$ & 1.27E-4 \\
$7c$ & 1.15E-4 \\
$9c$ & 1.22E-4 \\
$11c$ & 1.23E-4 \\
$13c$ & 1.22E-4 \\
$15c$ & 1.22E-4 \\
\hline
Direct approach using HORUS & 1.22E-4 \\
\hline
\end{tabular}
\label{table_naca_fwh_spl_convergence}
\end{table}

\subsection{Shape Optimization}

The shape of a NACA0012 airfoil is optimized to reduce the OASPL at a far-field observer located $10$ chord lengths below the trailing edge. The design parameters are maximum camber $c_{max}^{a}$ and its location $x_{c_{max}^{a}}$, maximum thickness $t_{max}^{a}$, and angle of attack $\alpha$, i.e. $\pmb{\mathcal{X}} = [c_{max}^{a}, x_{c_{max}^{a}}, t_{max}^{a}, \alpha]$. The maximum camber range is set to $c_{max}^{a} \in [-10, 10]$ as a percentage of the chord, with the distance from the airfoil leading edge in the range of $x_{c_{max}^{a}} \in [4, 9]$ as a tenth of the chord. The maximum thickness of the airfoil is within the range of $t_{max}^{a} \in [6,18]$ as a percentage of the chord. Finally, the angle of attack varies from $\alpha \in [0^\circ, 12^\circ]$. The objective function is defined as the overall sound pressure level at the observer with constraints on both the mean lift and mean drag coefficients. A quadratic penalty term is added to the objective function when the lift coefficient deviates from the baseline design, and an additional quadratic penalty term is added when the mean drag coefficient is above the baseline design. The objective function is defined as 
\begin{align}
&
\mathcal{F} = 
\begin{cases}
\text{OASPL} + \epsilon_1 \left( \overline{C_L} - \overline{C_{L,baseline}} \right)^2 + \epsilon_2 \left( \overline{C_D} - \overline{C_{D,baseline}} \right)^2 & \overline{C_D} > \overline{C_{D,baseline}} \\
\text{OASPL} + \epsilon_1 \left( \overline{C_L} - \overline{C_{L,baseline}} \right)^2 & \overline{C_D} \leq \overline{C_{D,baseline}}  \\
\end{cases}
&
\label{eq_obj_naca_3d}
\end{align}
where the constants $\epsilon_1$ and $\epsilon_2$ are set to $8,000$ and $400,000$, respectively, to compensate for the order of magnitude difference in OASPL, $\overline{C_L}$, and $\overline{C_D}$. The $\epsilon_1$ and $\epsilon_2$ constants ensure that aerodynamic penalties neither dominate nor are neglected during optimization. Notably, $\epsilon_2$ penalizes drag increase more aggressively than $\epsilon_1$ penalizes lift deviation, thus discouraging drag increase while allowing some lift variation. Thus, the defined objective function minimizes the overall sound pressure level while maintaining the mean lift coefficient, and ensures the optimized airfoil has a similar or lower mean drag coefficient. 

\begin{figure}
\centering
\begin{subfigure}{0.7\textwidth}
\includegraphics[width=\textwidth]{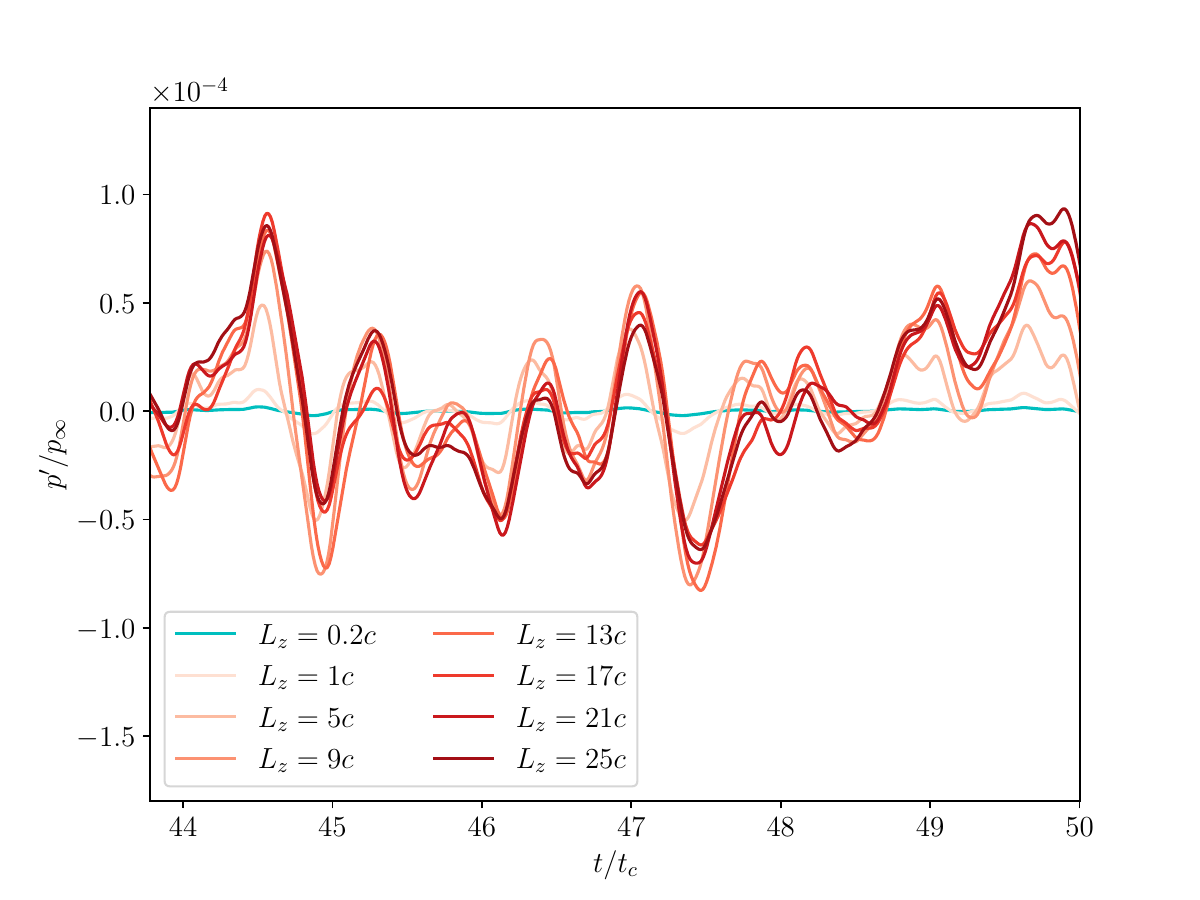}
\caption{The pressure perturbation time history.}
\label{fig_naca_fwh_baseline_pacs}
\end{subfigure}
\begin{subfigure}{0.7\textwidth}
\includegraphics[width=\textwidth]{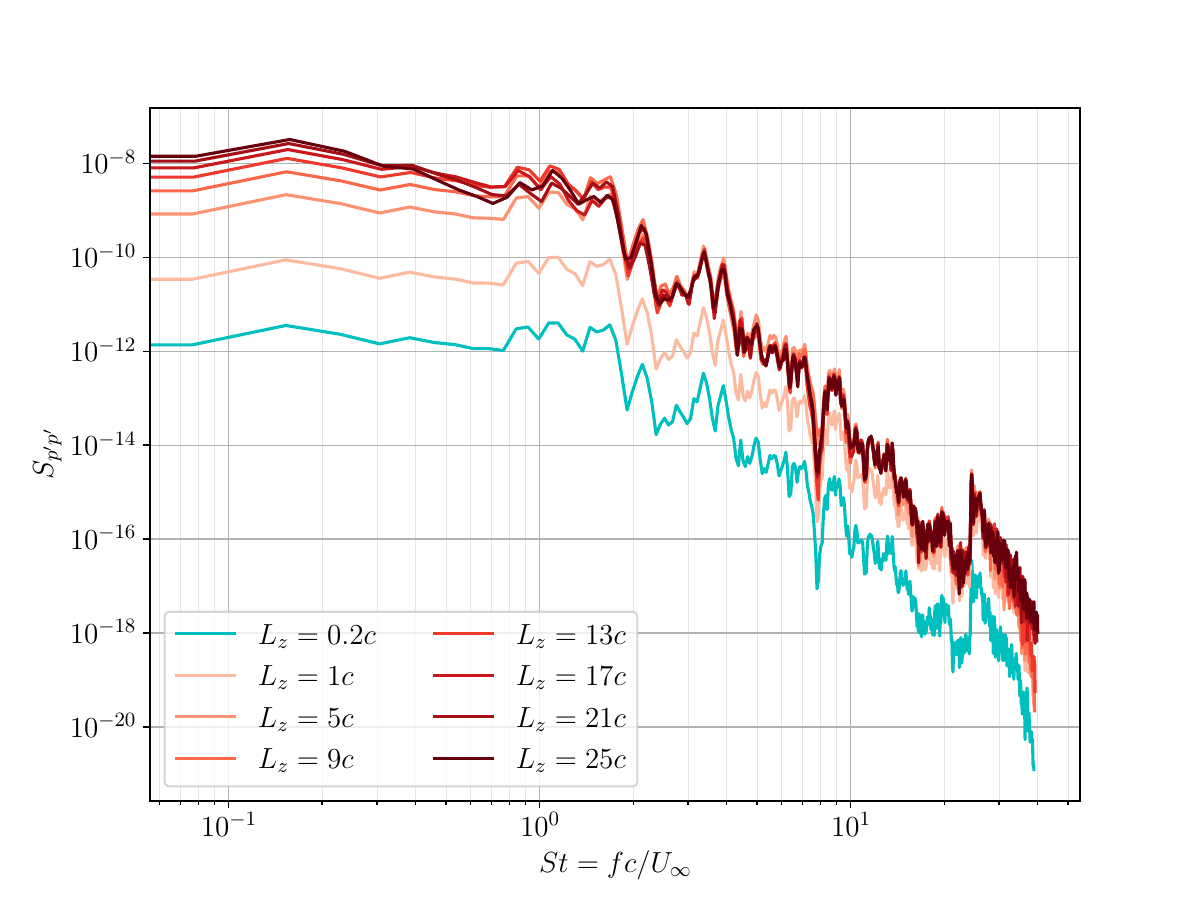}
\caption{The PSD of pressure perturbation.}
\label{fig_naca_fwh_baseline_psdspl}
\end{subfigure}
\caption{The convergence of the pressure perturbation time history and its corresponding PSD at the far-field observer using multiple sets of data surface duplications.}
\label{fig_naca_fwh_farfield_convergence_for_nduplications}
\end{figure}

In this study, the density, pressure, and velocity fields are gathered on the permeable data surface in HORUS and utilized as inputs for PyFWH solver. To ensure the accuracy of our acoustic analysis, we account for the potential influence of vortices crossing the data surface, which can introduce undesired noise artifacts. To mitigate this, the data surface is tilted to match the angle of attack, mirroring the orientation of the computational domain and effectively preventing vortices from crossing the data surface. Given that our observer is located in the far-field, we utilize various sets of data surface duplications to calculate the time history of pressure perturbations and its power spectral density as depicted in Figure \ref{fig_naca_fwh_farfield_convergence_for_nduplications}. 
Furthermore, Table \ref{table_naca_fwh_far_field_spl_convergence} provides a summary of $p^\prime_{rms}/p_\infty$ values obtained through different sets of data surface duplications.
These findings confirm that duplicating the data surface up to $L_z=25c$ adequately captures the far-field noise.

\begin{table}
\centering
\caption{The $p^\prime_{rms}/p_\infty$ of the far-field observer using different sets of data surface duplications.}
\begin{tabular}{cc}
\hline
Duplication Length $(L_z)$ & $p^\prime_{rms}/p_\infty$ \\
\hline
$0.2c$ & 9.89E-7 \\
$1c$ & 4.96E-6 \\
$5c$ & 2.43E-5 \\
$9c$ & 3.98E-5 \\
$13c$ & 4.72E-5 \\
$17c$ & 4.76E-5 \\
$21c$ & 4.87E-5 \\
$25c$ & 4.90E-5 \\
\hline
\end{tabular}
\label{table_naca_fwh_far_field_spl_convergence}
\end{table}

The aeroacoustic shape optimization for reducing far-field noise via PyFWH solver follows a sequential process. Initially, the flow field is resolved, and data on the data surface is collected using HORUS. Subsequently, the data surface is duplicated in the spanwise direction, extending over a distance of $L_z=25c$. This duplicated data surface is then utilized as inputs for the PyFWH solver. The subsequent steps involve computing pressure perturbations at the far-field observer point and evaluating the objective function. This function incorporates both the OASPL at the observer and the time-averaged lift and drag coefficients, as defined in Equation \ref{eq_obj_naca_3d}. The optimization results are presented in the following section.

\subsubsection{Results and Discussions}

CFD simulations are performed on the Narval supercomputing cluster with a runtime of approximately 12 hours on 4 NVIDIA A100-SXM4 GPUs. In a serial implementation, each optimization iteration would take 96 hours to complete, which is 8 CFD simulations of 12 hours runtime each. However, due to the multi-layer parallel implementation of the optimization framework, all 8 CFD simulations are performed concurrently, reducing the runtime of an optimization iteration to 12 hours, equivalent to that of a single CFD run. The optimization process converged after 26 iterations, consisting of a total of 208 objective function evaluations, resulting in a total wall-clock time of approximately 13 days. In contrast, the serial implementation would require over 3 months (104 days). This highlights the importance of the two-level parallelism in making high-fidelity gradient-free optimization feasible, provided enough computation resources are available. The computation cost of this optimization problem is approximately one GPU year and is the same in both serial and parallel implementations.

\begin{figure}
\centering
\begin{subfigure}{0.75\textwidth}
\centering
\includegraphics[width=\textwidth]{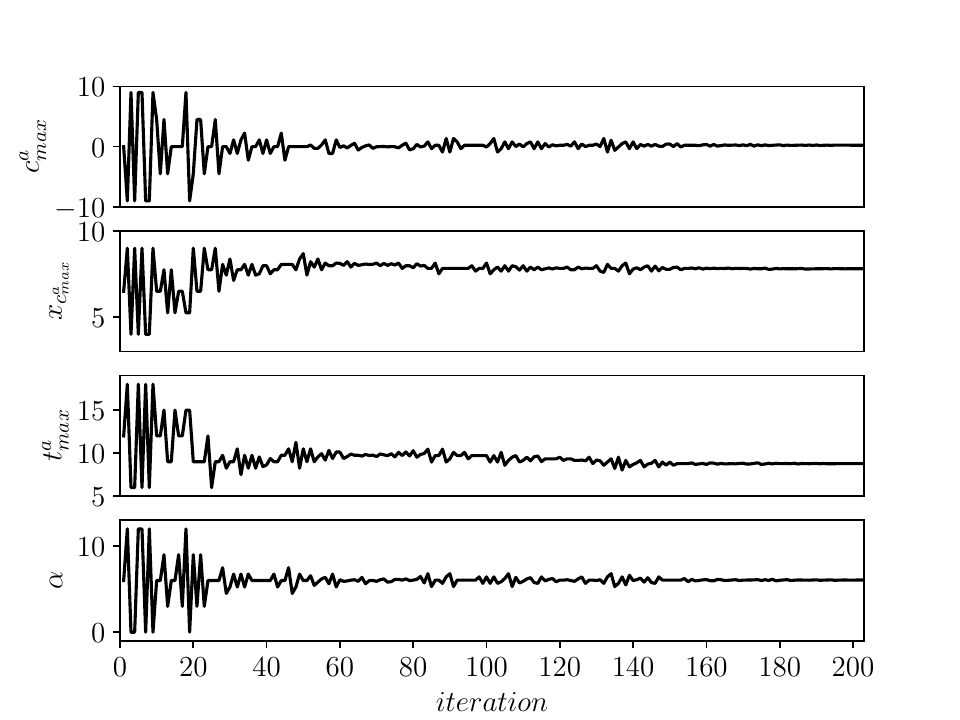}
\subcaption{The design space.}
\end{subfigure}
\begin{subfigure}{0.75\textwidth}
\centering
\includegraphics[width=\textwidth]{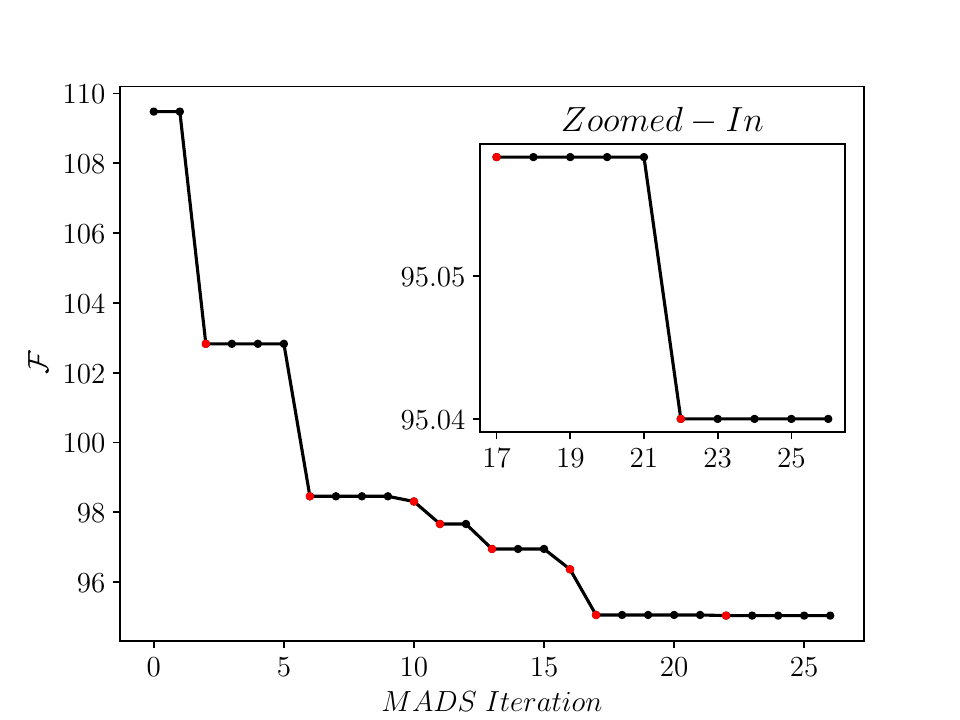}
\subcaption{The objective function convergence with the new incumbent design highlighted in red.}
\end{subfigure}
\caption{The design space and objective function convergence of the NACA 4-digit airfoil optimization.}
\label{fig_naca_fwh_optimization_convergence}
\end{figure}

The design space and the objective function convergence are depicted in Figure \ref{fig_naca_fwh_optimization_convergence}. The optimal airfoil design has a maximum camber of $c_{max}^{a} = 0.236206$ percent of the chord, at $7.8086$ tenths of the chord distance from the leading edge, with a thickness of $t_{max}^{a} = 8.783206$ percent of the chord, at an angle of attack of $\alpha = 6.054932^\circ$. The OASPL of the optimized airfoil is decreased by $5.9~dB$, the mean lift coefficient is slightly decreased by $0.69\%$ to $\overline{C_L} = 0.6489$, and finally, the mean drag coefficient is decreased by $14.07\%$ to $\overline{C_D} = 0.0475$. 

\begin{figure}
\centering
\includegraphics[width=0.5\textwidth]{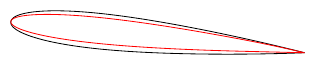}
\caption{The baseline, in black, and optimum, in red, designs of the NACA 4-digits airfoil.}
\label{fig_baseline_optimum_naca_fwh}
\end{figure}

The baseline and optimized airfoil shapes are shown in Figure \ref{fig_baseline_optimum_naca_fwh}, where the optimized airfoil features a slightly thinner profile and reduced camber compared to the baseline airfoil. The optimized airfoil features a more streamlined profile that likely reduces flow separation, resulting in lower drag and a less turbulent wake, improving the lift-to-drag ratio and reducing OASPL.

\begin{figure}
\centering
\begin{subfigure}{\textwidth}
\includegraphics[width=\textwidth]{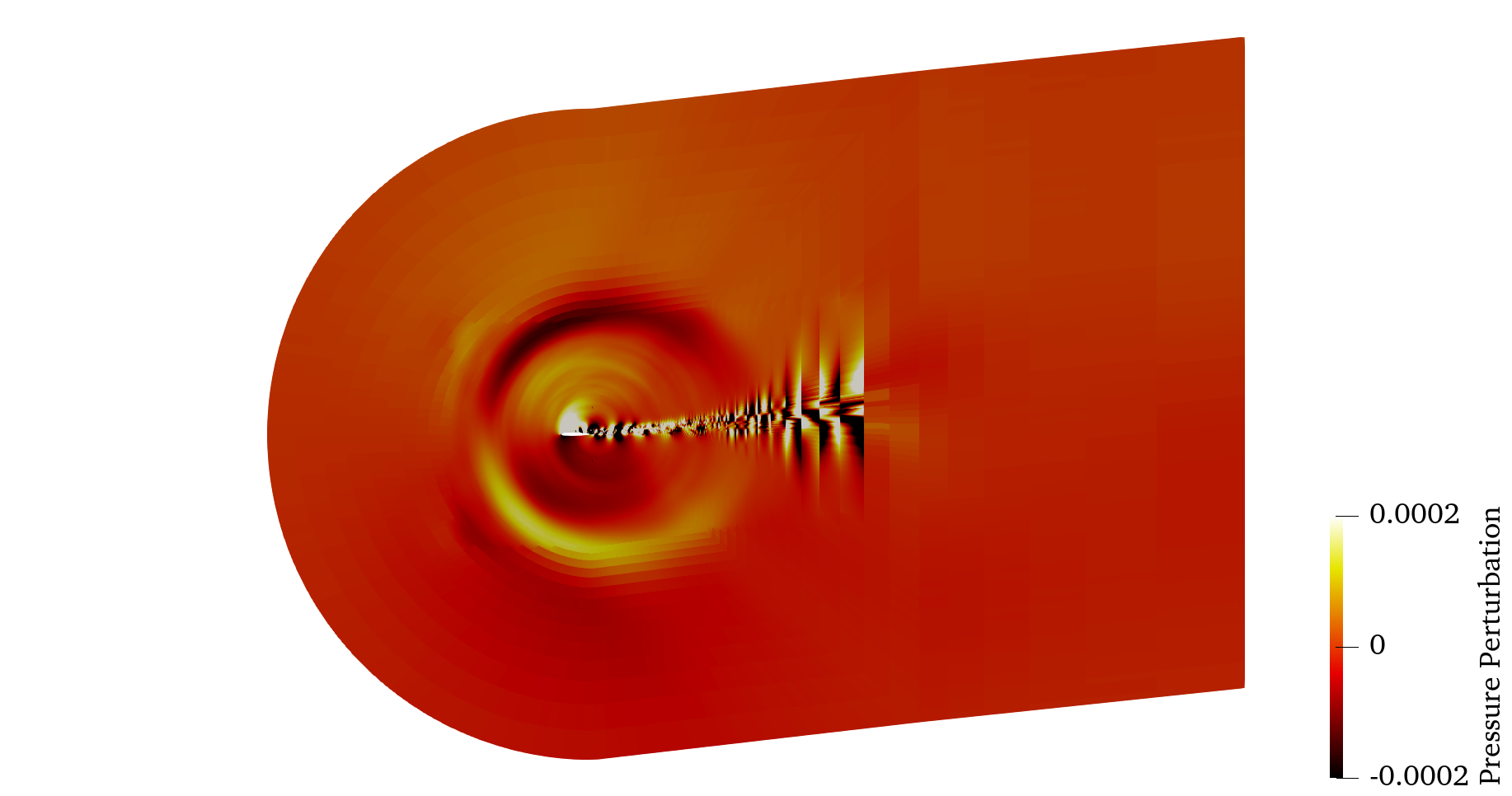}
\subcaption{Baseline design.}
\end{subfigure}
\begin{subfigure}{\textwidth}
\includegraphics[width=\textwidth]{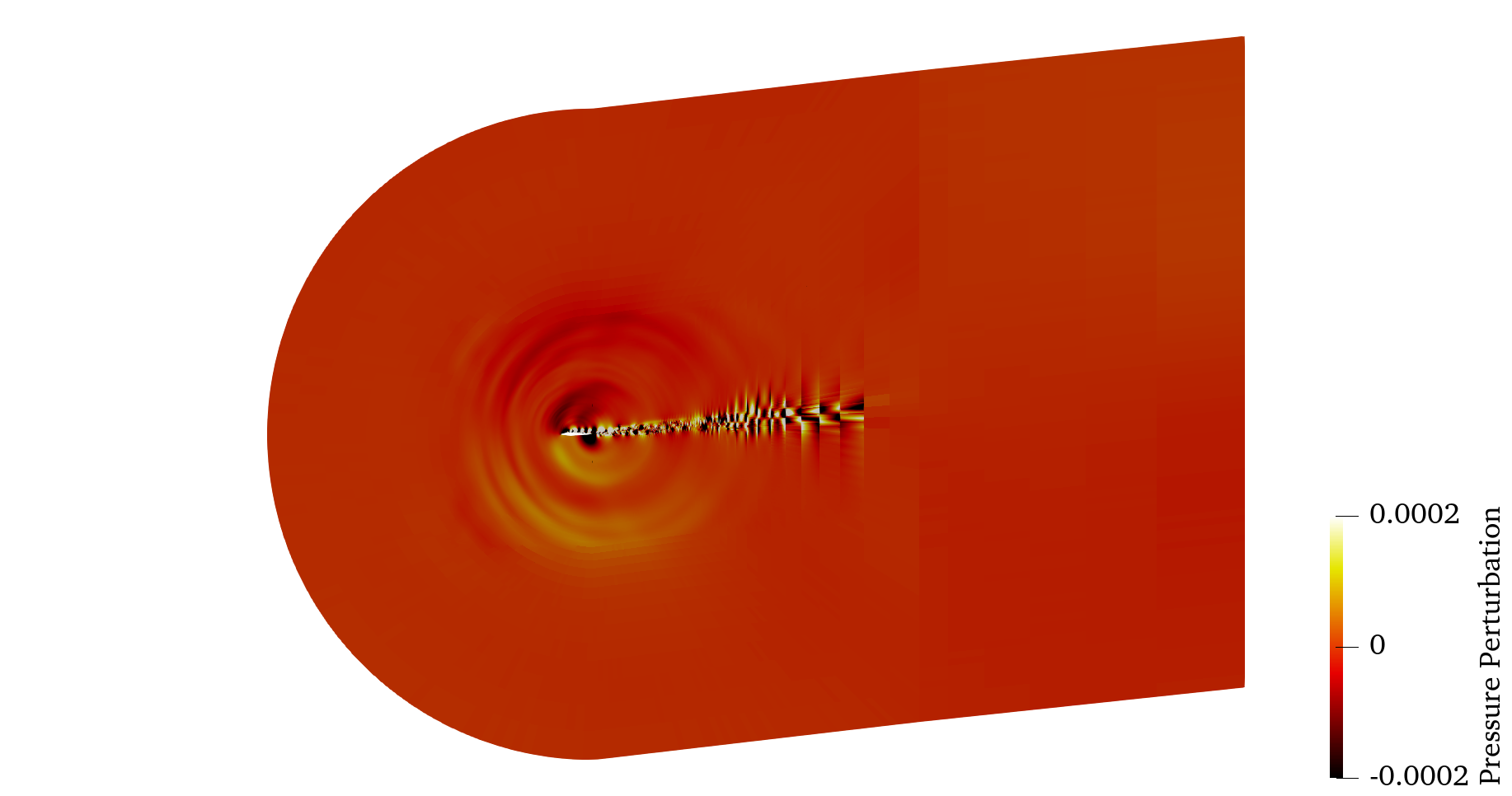}
\subcaption{Optimum design.}
\end{subfigure}
\caption{The acoustic pressure field at mid plane for baseline and optimum designs at $t_c=70$.}
\label{fig_naca_fwh_acoustic}
\end{figure}

\begin{figure}
\centering
\begin{subfigure}{0.7\textwidth}
\includegraphics[width=\textwidth]{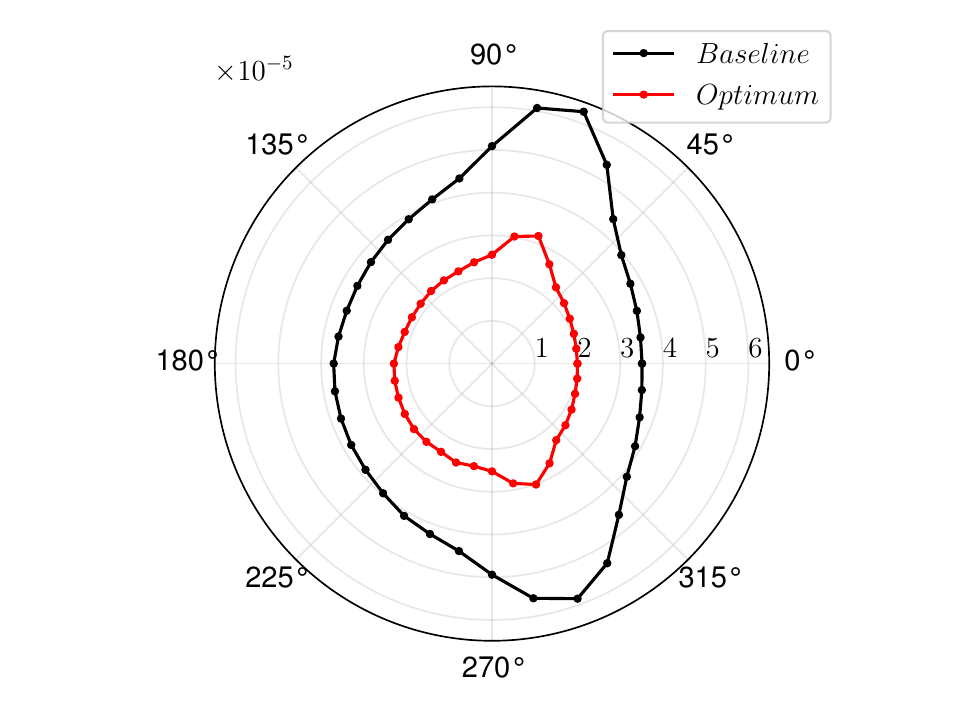}
\subcaption{The directivity of $p^\prime_{rms}/p_\infty$ at a radius of $r=10c$.}
\label{fig_fwh_dir_base_opt}
\end{subfigure}
\begin{subfigure}{0.7\textwidth}
\includegraphics[width=\textwidth]{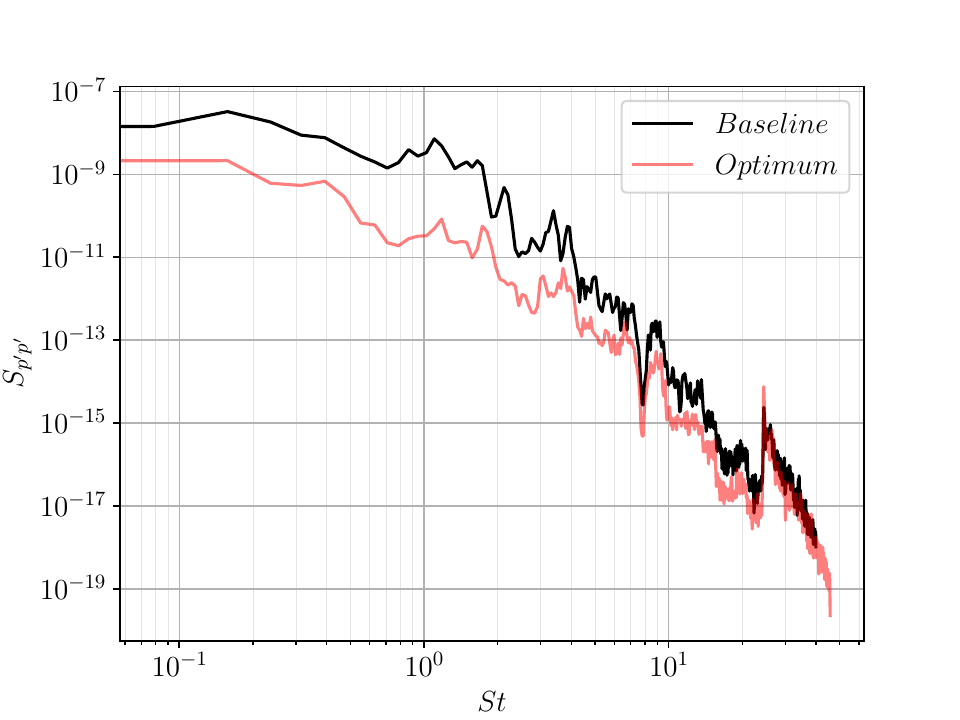}
\subcaption{The PSD of pressure perturbation at the far-field observer.}
\label{fig_fwh_spp_base_opt}
\end{subfigure}
\caption{The acoustic analysis of baseline and optimum designs.}
\label{fig_dir_and_fwhspp}
\end{figure}

The acoustic field is demonstrated in Figure \ref{fig_naca_fwh_acoustic}, where the pressure perturbation is shown in the computational domain. In the optimized design, the pressure perturbations are noticeably less significant compared to the baseline, highlighting the improvement in noise reduction. The absence of acoustic wave reflections off the non-physical boundaries confirms the effectiveness of the boundary treatments, ensuring the flow field is not contaminated. Furthermore, the $p^\prime_{rms}/p_\infty$ directivity results show that the optimized airfoil consistently reduces noise at all measurement angles compared to the baseline, without altering the main radiation pattern. Both designs have their maximum $p^\prime_{rms}/p_\infty$ directed downstream at $(45^\circ - 90^\circ)$ and $(270^\circ - 315^\circ)$, indicating that the optimization preserves the fundamental flow and acoustic characteristics. The power spectral density data confirms that the optimized airfoil achieves up to two orders of magnitude lower noise across low and mid frequencies $(St < 5)$, where noise is most critical for human perception and regulatory compliance. The similar high-frequency decay for both designs suggests that the core sound generation mechanisms are unchanged, but the optimal geometry weakens the source, likely through improved boundary layer properties (see Figure \ref{fig_dir_and_fwhspp}).

\begin{figure}
\centering
\begin{subfigure}{\textwidth}
\includegraphics[width=\textwidth]{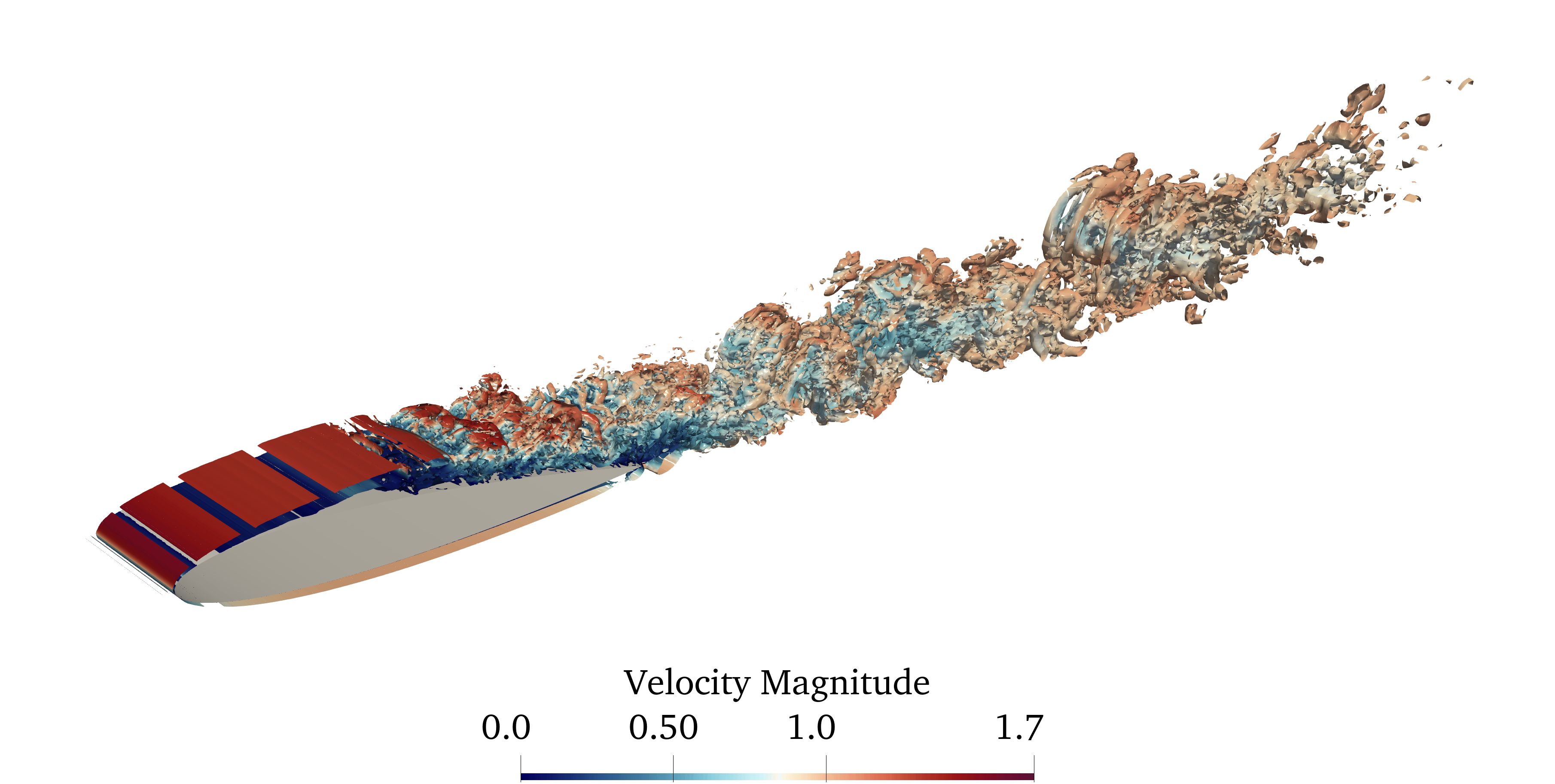}
\subcaption{Baseline design.}
\end{subfigure}
\begin{subfigure}{\textwidth}
\includegraphics[width=\textwidth]{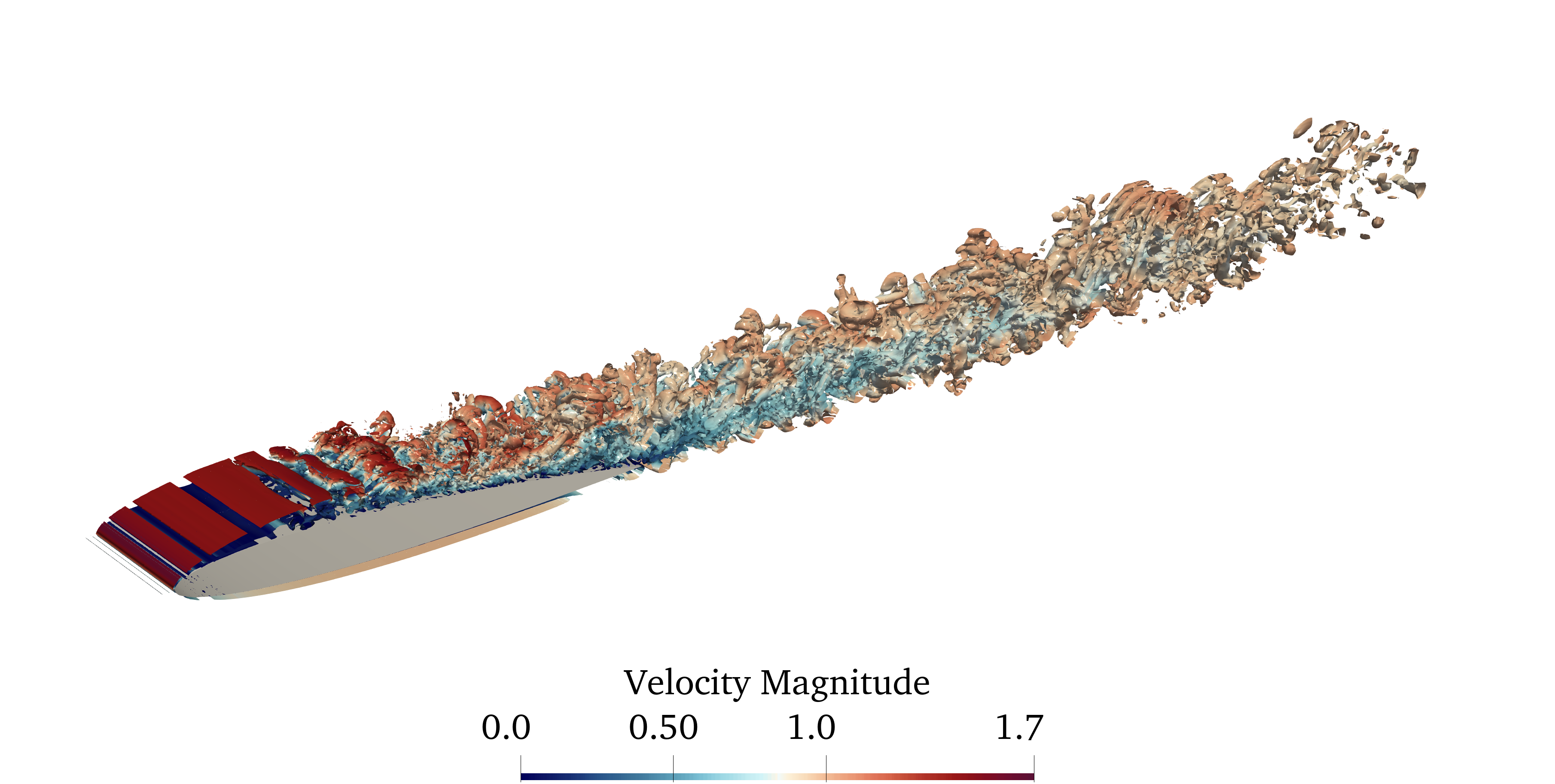}
\subcaption{Optimum design.}
\end{subfigure}
\caption{The Q-criterion coloured by velocity magnitude at mid plane for baseline and optimum designs at $t_c=70$.}
\label{fig_naca_fwh_qcriterion}
\end{figure}

Figure \ref{fig_naca_fwh_qcriterion} present the Q-criterion, colored by velocity magnitude, for both the baseline and optimized designs. In the baseline design, larger and more dominant vortical structures are visible in the wake region, indicating a higher level of turbulence. These vortices occupy a broader area in the wall-normal direction, reflecting a more chaotic and disturbed wake. Conversely, the optimized design exhibits smaller and more compact vortices. This reduction in turbulence and adverse pressure gradients leads to smoother flow separation. Thus, smaller vortices are generated, leading to a significant reduction in noise.

\begin{figure}
\centering
\begin{subfigure}{0.475\textwidth}
\includegraphics[width=\textwidth]{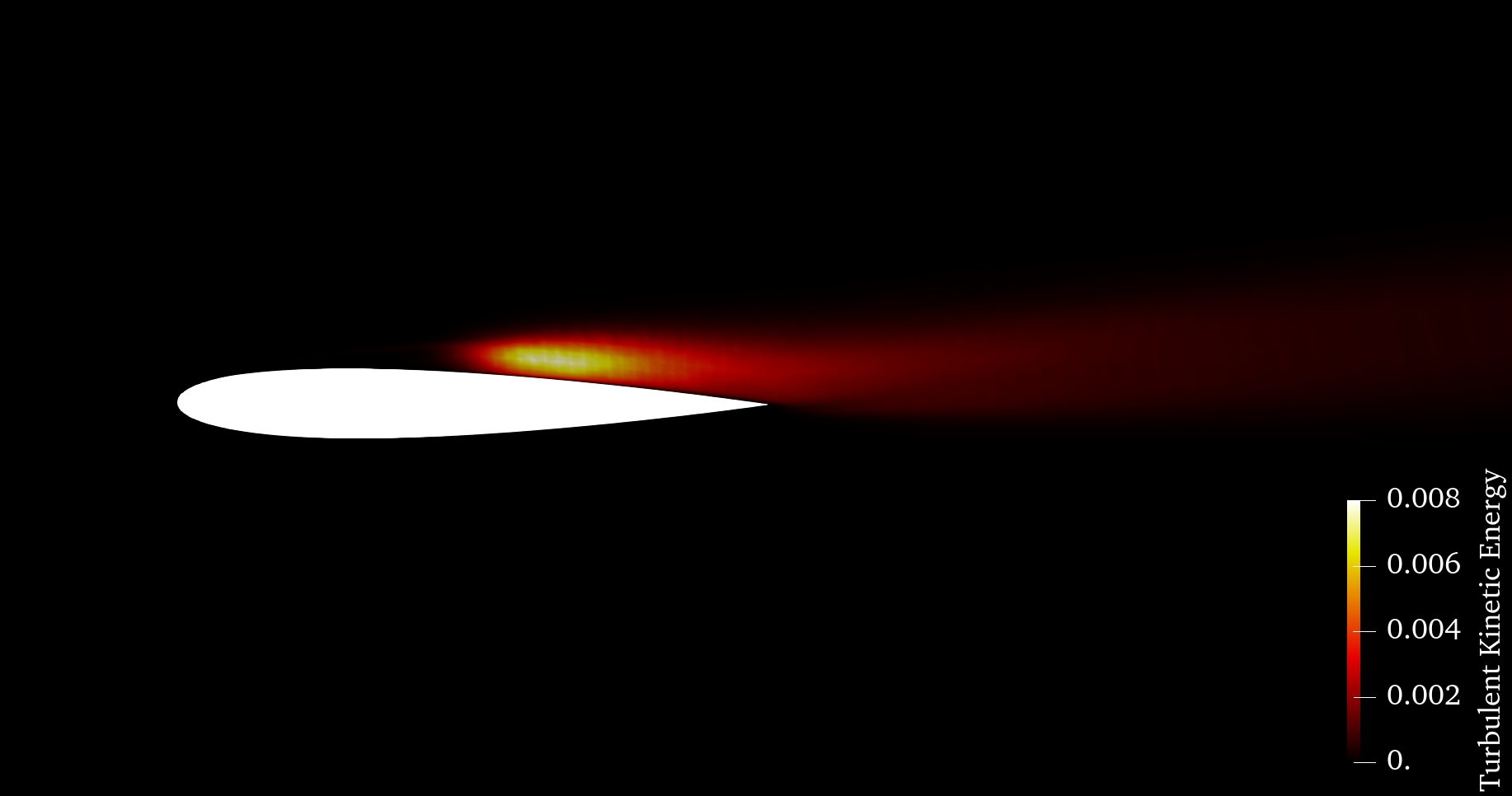}
\subcaption{Baseline design.}
\end{subfigure}
\begin{subfigure}{0.475\textwidth}
\includegraphics[width=\textwidth]{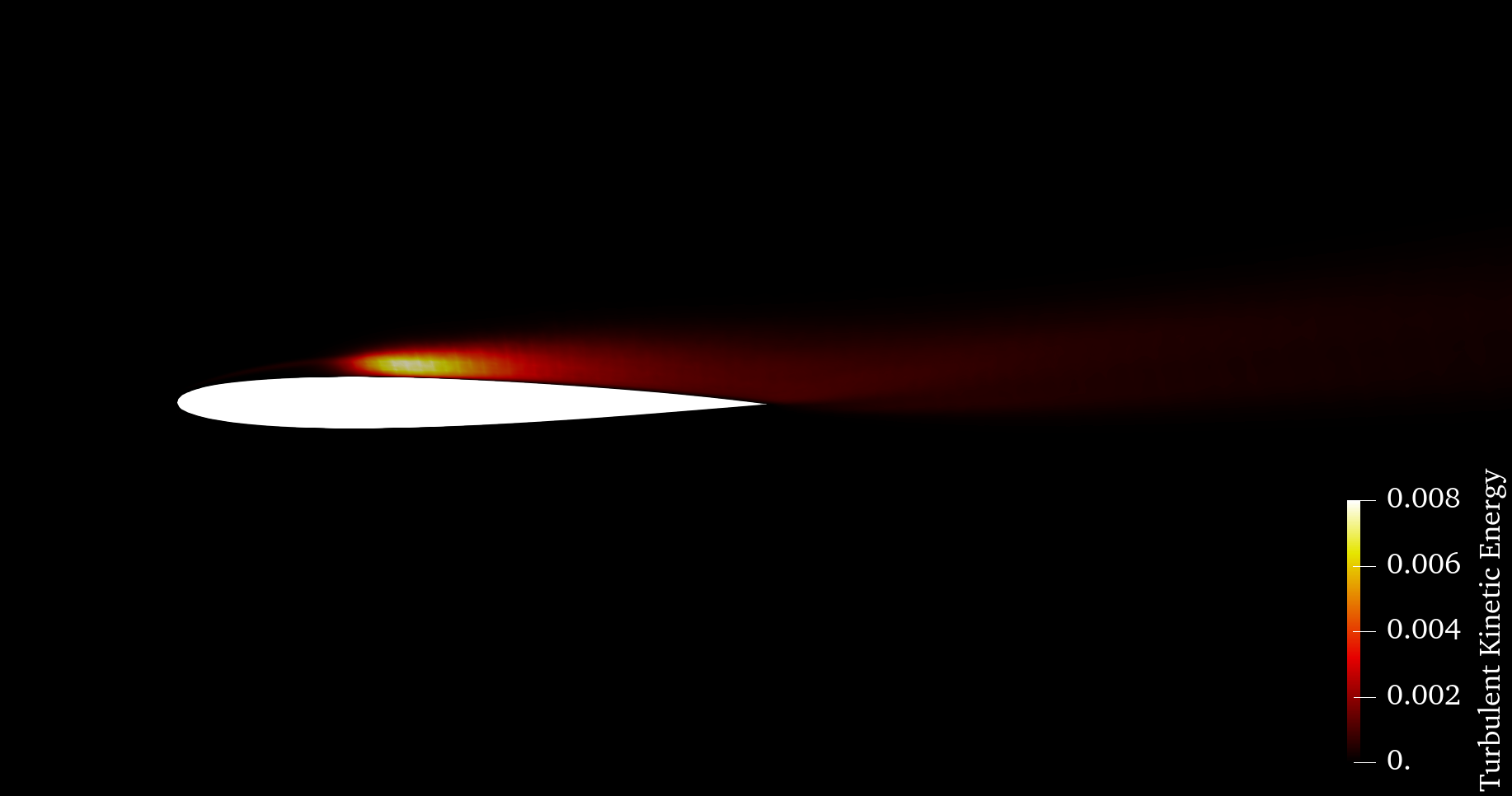}
\subcaption{Optimum design.}
\end{subfigure}
\caption{The turbulent kinetic energy for baseline and optimum designs at $t_c=70$.}
\label{fig_naca_fwh_tke}
\end{figure}

\begin{figure}
\centering

\begin{subfigure}{0.475\textwidth}
\includegraphics[width=\textwidth]{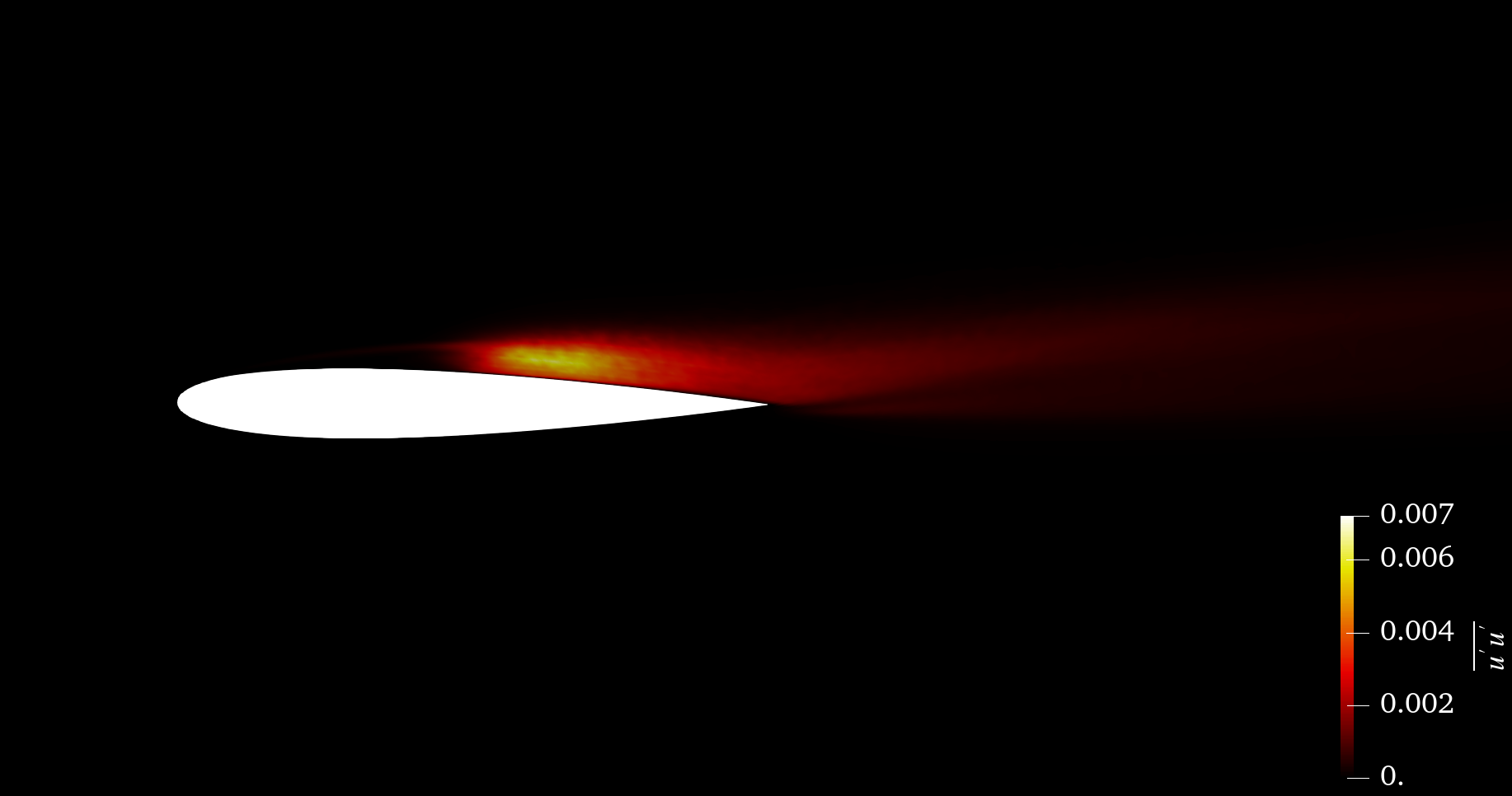}
\subcaption{$\overline{u^\prime u^\prime}$ for the baseline design.}
\end{subfigure}
\begin{subfigure}{0.475\textwidth}
\includegraphics[width=\textwidth]{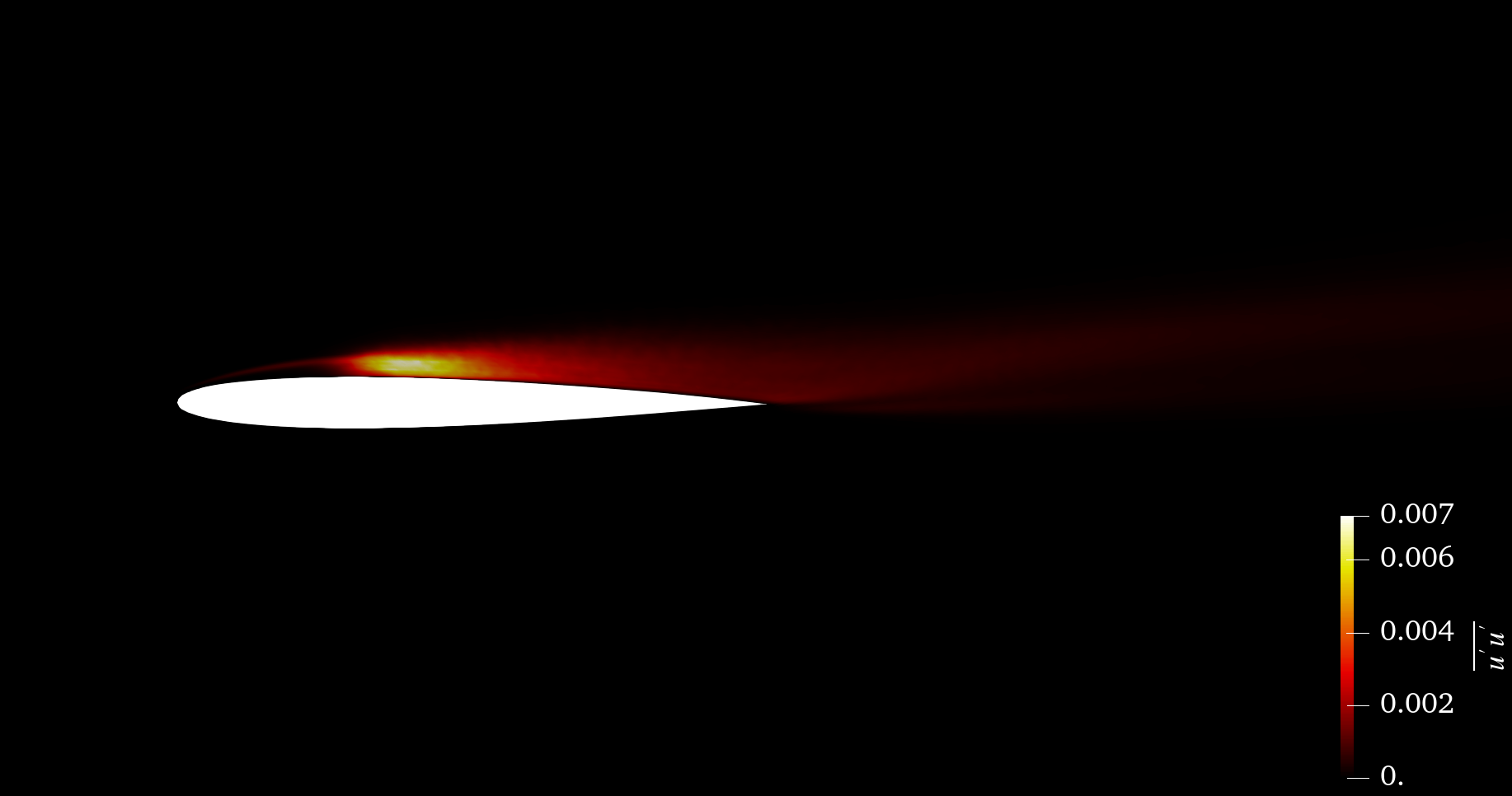}
\subcaption{$\overline{u^\prime u^\prime}$ for the optimum design.}
\end{subfigure}

\begin{subfigure}{0.475\textwidth}
\includegraphics[width=\textwidth]{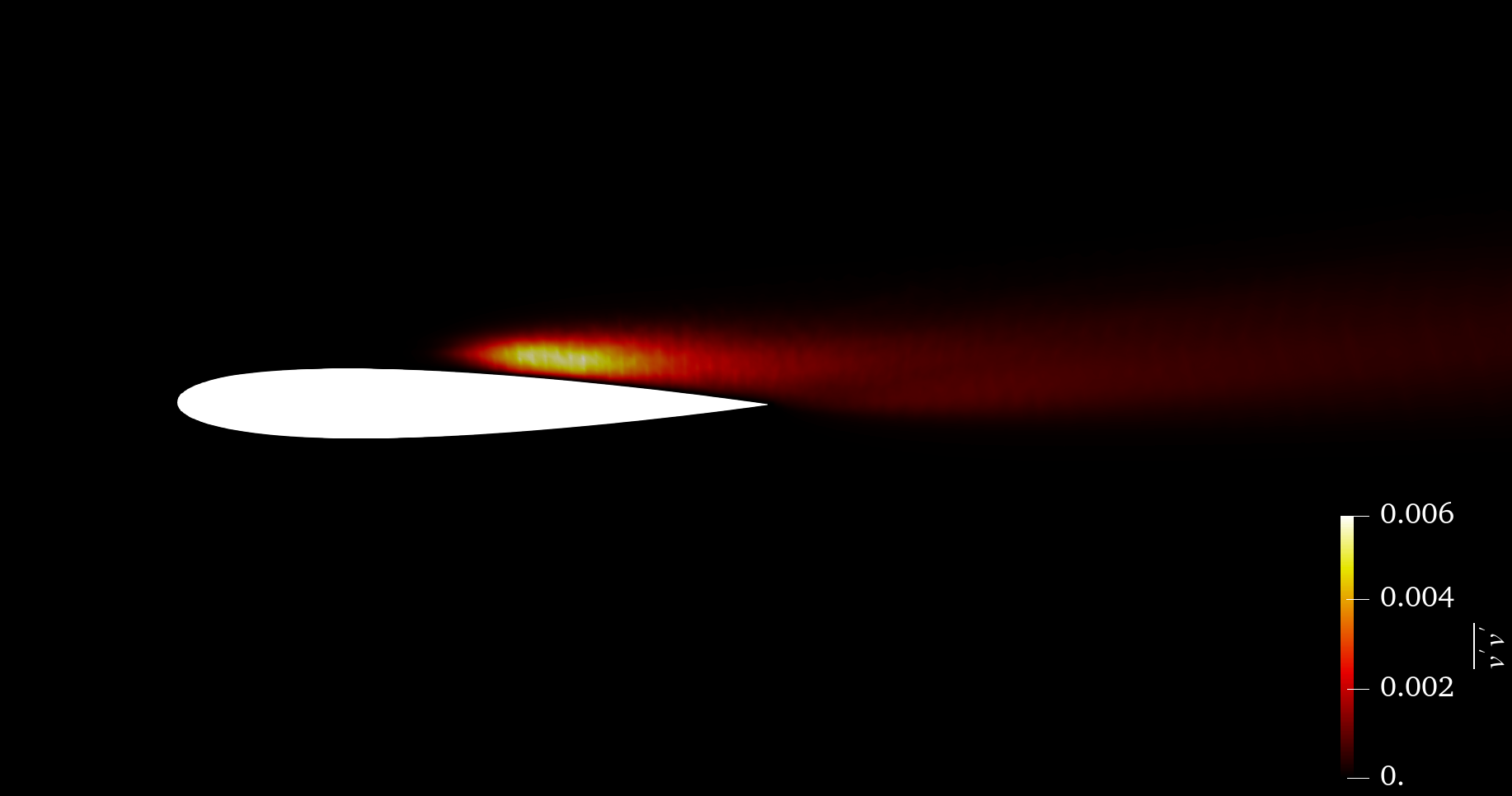}
\subcaption{$\overline{v^\prime v^\prime}$ for the baseline design.}
\end{subfigure}
\begin{subfigure}{0.475\textwidth}
\includegraphics[width=\textwidth]{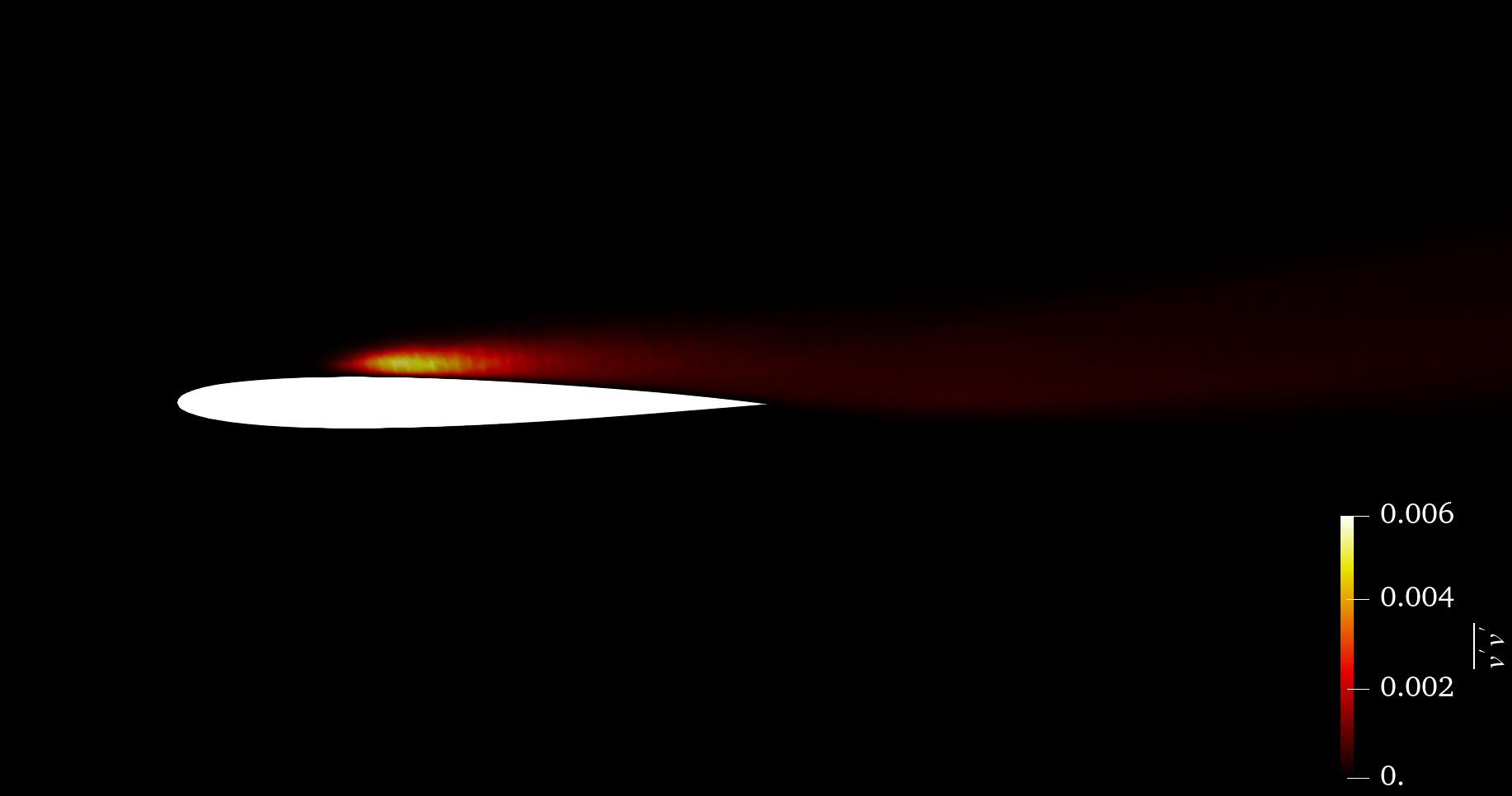}
\subcaption{$\overline{v^\prime v^\prime}$ for the optimum design.}
\end{subfigure}

\begin{subfigure}{0.475\textwidth}
\includegraphics[width=\textwidth]{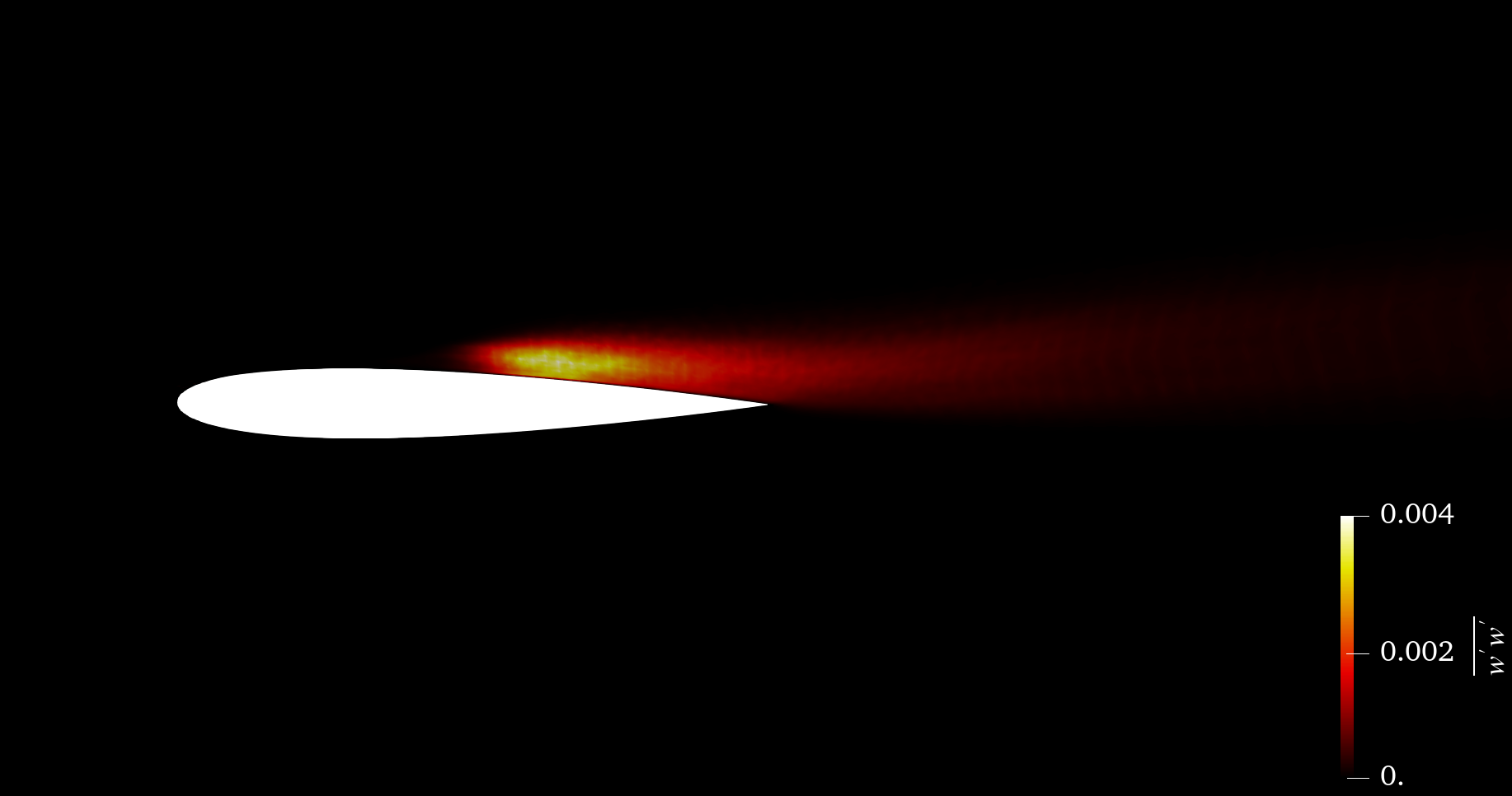}
\subcaption{$\overline{w^\prime w^\prime}$ for the baseline design.}
\end{subfigure}
\begin{subfigure}{0.475\textwidth}
\includegraphics[width=\textwidth]{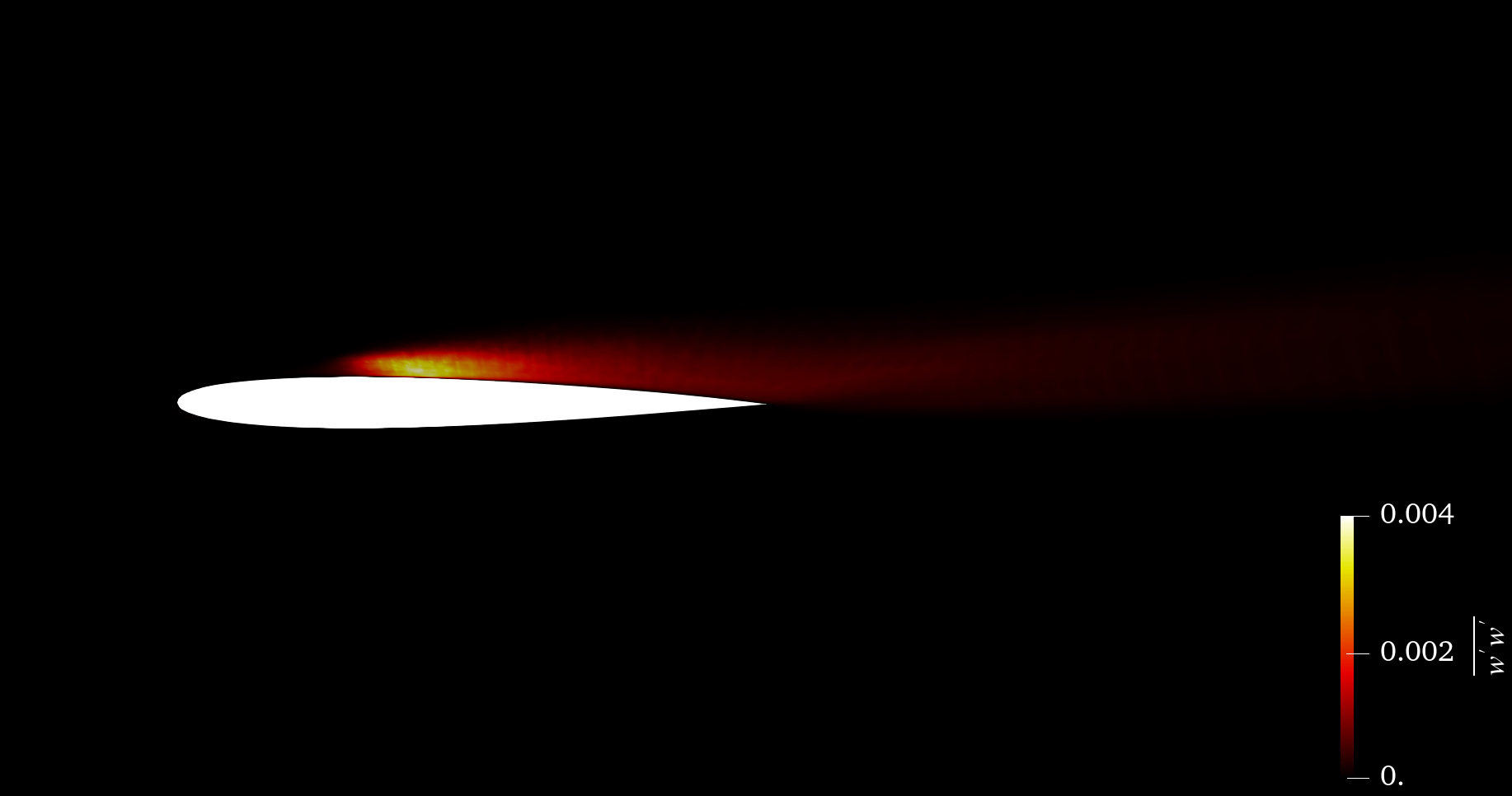}
\subcaption{$\overline{w^\prime w^\prime}$ for the optimum design.}
\end{subfigure}

\caption{The normal components of the Reynolds stresses for baseline and optimum designs at $t_c=70$.}
\label{fig_naca_fwh_normalRe}

\end{figure}

\begin{figure}
\centering

\begin{subfigure}{0.475\textwidth}
\includegraphics[width=\textwidth]{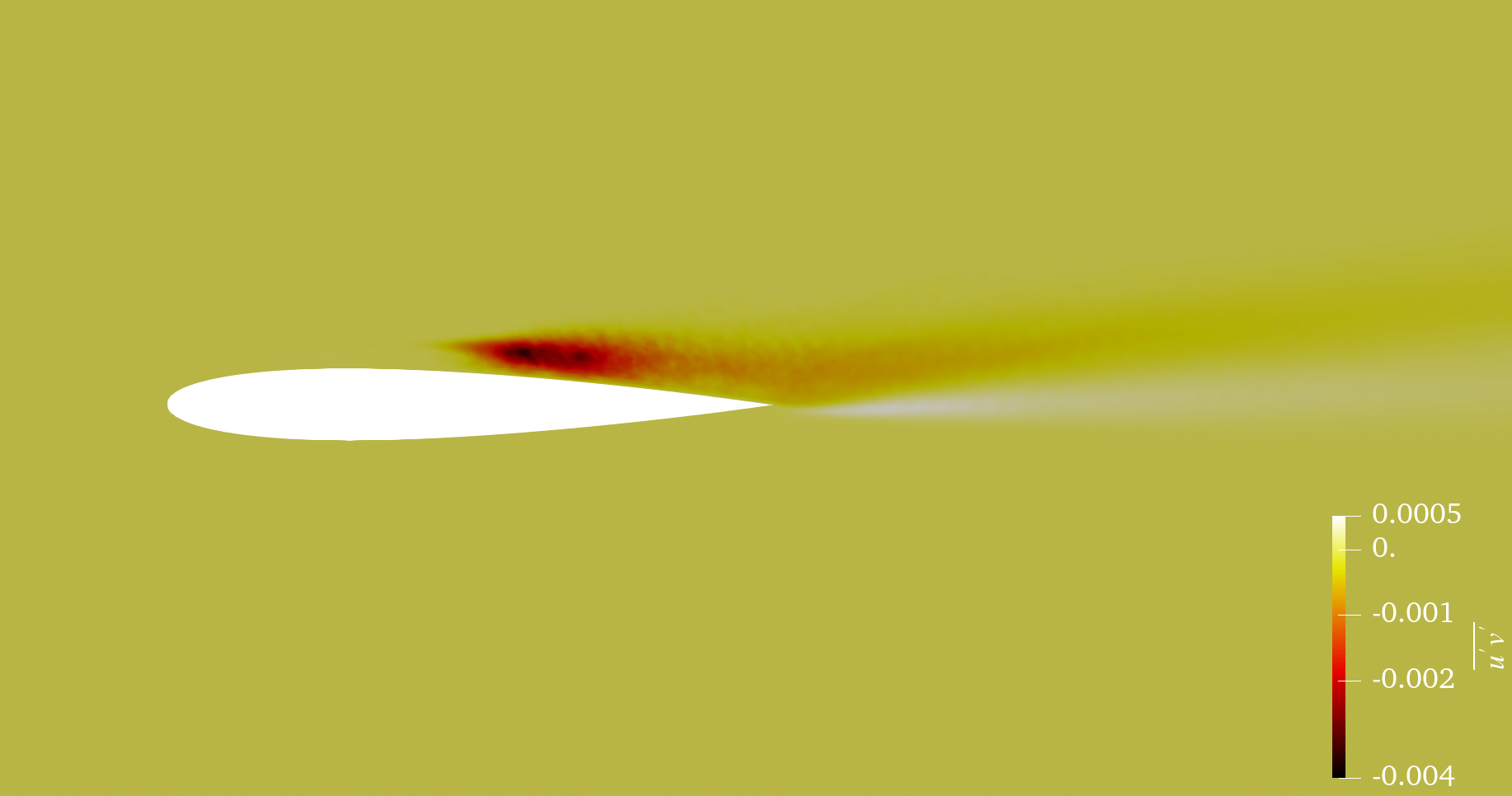}
\subcaption{$\overline{u^\prime v^\prime}$ for the baseline design.}
\end{subfigure}
\begin{subfigure}{0.475\textwidth}
\includegraphics[width=\textwidth]{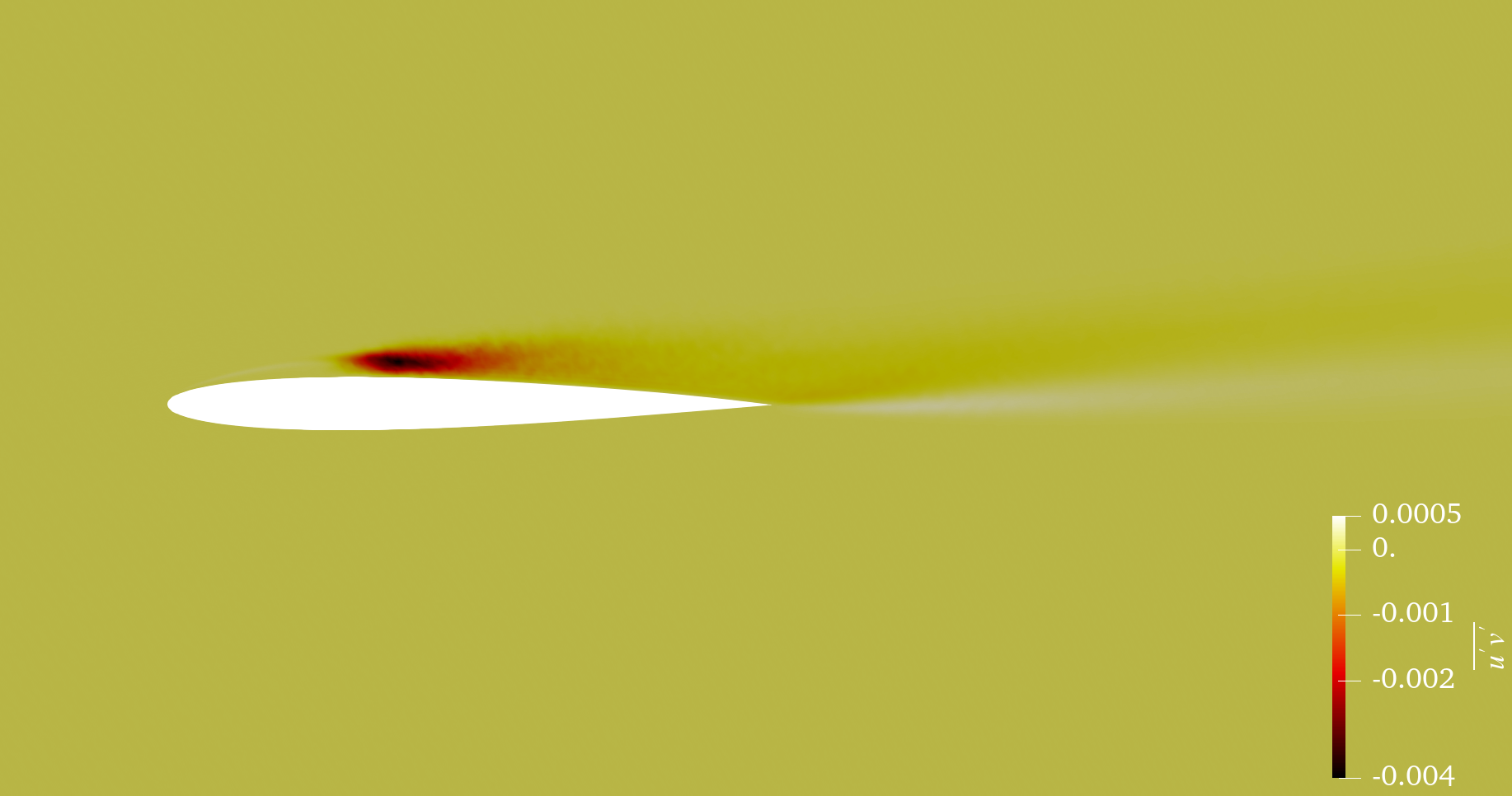}
\subcaption{$\overline{u^\prime v^\prime}$ for the optimum design.}
\end{subfigure}

\caption{The cross terms of the Reynolds stresses for baseline and optimum designs at $t_c=70$.}
\label{fig_naca_fwh_crossRe}

\end{figure}

The Turbulent Kinetic Energy (TKE) is shown in Figure \ref{fig_naca_fwh_tke}, and the normal components and cross term of the Reynolds stresses are shown in Figures \ref{fig_naca_fwh_normalRe} and \ref{fig_naca_fwh_crossRe}, respectively. From these contours, it is evident that in the optimum design, the peak of TKE and Reynolds stresses have moved closer to the leading edge of the airfoil compared to the baseline design. This shift indicates that the boundary layer separates earlier, leading to less energy being available for turbulent fluctuations, which weakens the turbulence in the wake. In the baseline design, the peak of Reynolds stress occurs further downstream, suggesting that the turbulent boundary layer persists longer and creates stronger turbulence in the wake. The earlier separation in the optimum design results in lower turbulence levels behind the airfoil, which has a direct impact on both drag and noise reduction. With reduced turbulence in the wake, there is less flow resistance acting on the airfoil, leading to a decrease in the drag coefficient. Additionally, the turbulent fluctuations in the streamwise, vertical, and spanwise directions reveal significantly lower turbulence fluctuations in the optimum design. This reduction in turbulence results in less pronounced unsteady pressure forces acting on the airfoil surface, leading to a smoother pressure field and reduced acoustic radiation. The spanwise direction has the lowest energy, while the streamwise direction exhibits the highest. Furthermore, the cross term in the Reynolds stresses show high values near the trailing edge and separation point. These values indicate weak correlations between velocity fluctuations in different directions, contributing to the formation of vortical structures. The weaker wake turbulence in the optimum design also contributes to lower noise levels, as aeroacoustic noise primarily originates from unsteady wake interactions and vortex shedding. The reduction in turbulence intensity in the wake minimizes these noise sources, resulting in a lower OASPL. Thus, the changes in the distribution of TKE and Reynolds stresses in the optimum design lead to improved aerodynamic performance through drag reduction and quieter operation by reducing noise.

\begin{figure}
\centering
\includegraphics[width=0.75\textwidth]{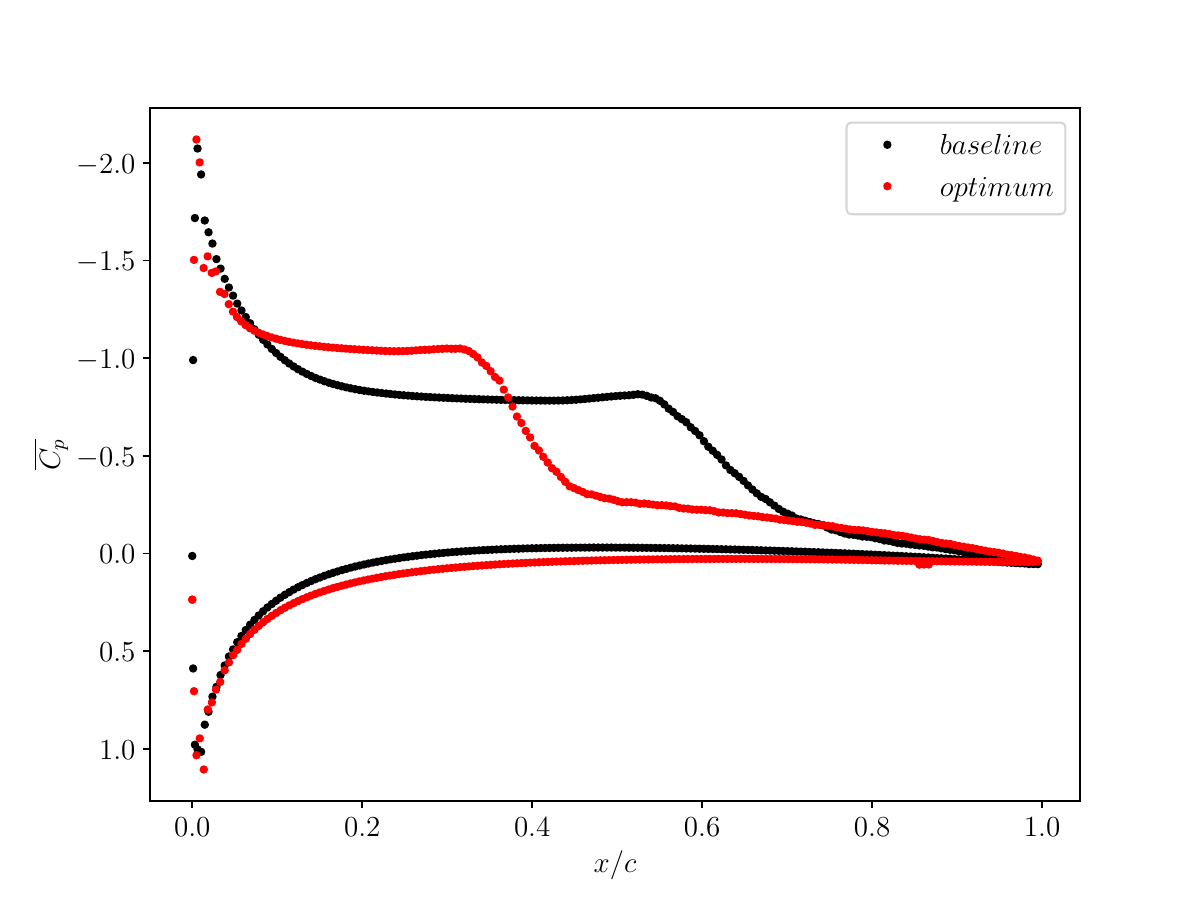}
\caption{The time-averaged pressure coefficient for both baseline and optimum designs at $t_c=70$.}
\label{fig_naca_fwh_cp}
\end{figure}

The time-averaged pressure coefficient distribution is illustrated in Figure \ref{fig_naca_fwh_cp}, showing key differences in the aerodynamic and aeroacoustic behavior of the two designs. In the baseline design, the pressure drop along the upper surface is more gradual, indicating weaker suction and a slower acceleration of flow, which contributes to higher drag. The pressure recovery towards the trailing edge is also more gradual, suggesting increased turbulence in the wake. These features not only increase drag but also contribute to higher noise levels, as turbulence and vortex shedding in the wake are primary sources of aeroacoustic noise. In contrast, the optimum airfoil demonstrates a much stronger suction on the upper surface, with a sharper pressure gradient near the leading edge. Additionally, the sharper pressure recovery near the trailing edge points to a more stable flow pattern, leading to weaker wake turbulence and lower drag. The more consistent positive $\overline{C_p}$ on the lower surface of the optimum design helps maintain a favorable pressure difference, further enhancing the aerodynamic performance. From an aeroacoustic perspective, the smoother and sharper pressure recovery in the optimum airfoil reduces the unsteady pressure forces that drive noise generation. By minimizing wake turbulence and vortex shedding, the optimum design is likely to produce significantly lower sound pressure levels compared to the baseline. Overall, the differences in $\overline{C_p}$ distribution between the two designs explain the improved aerodynamic efficiency and reduced OASPL in the optimum airfoil.

\begin{figure}
\centering
\includegraphics[width=0.75\textwidth]{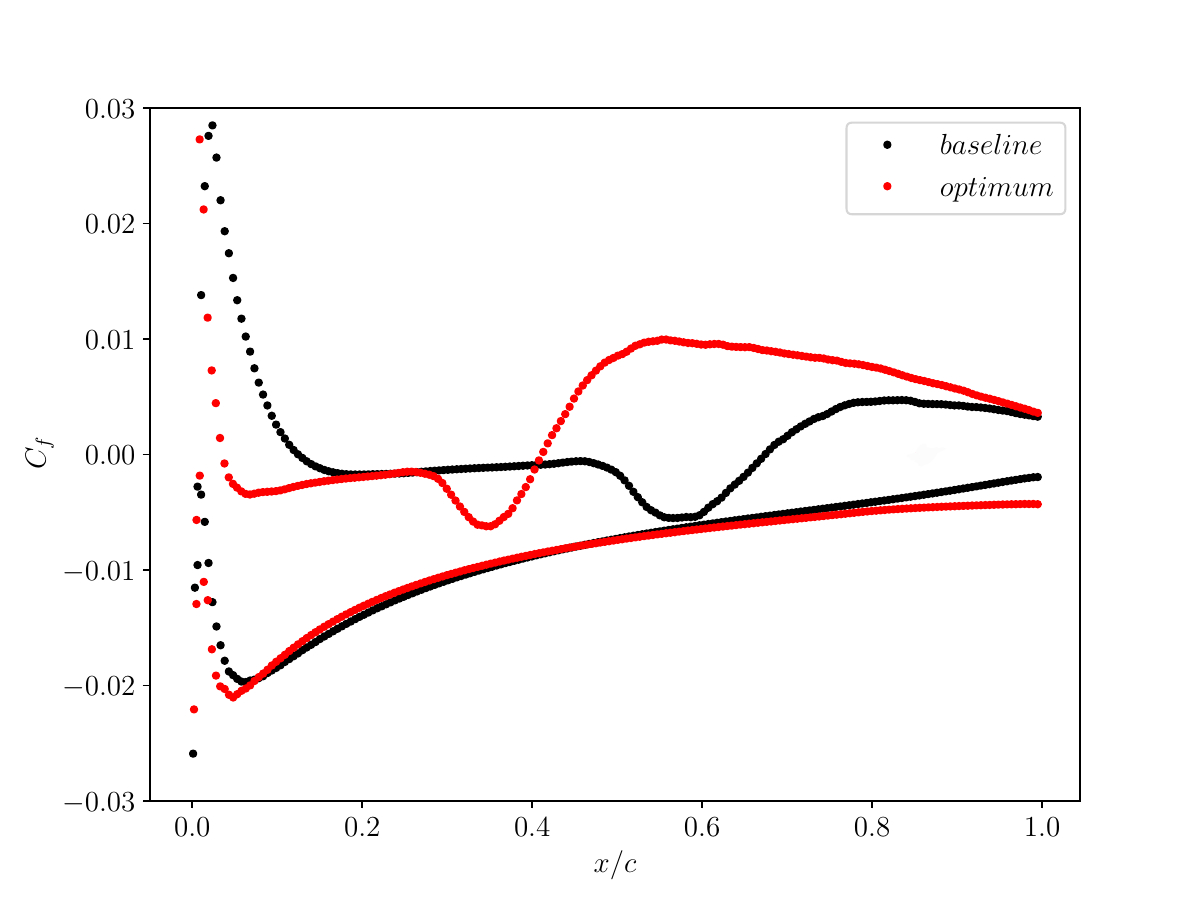}
\caption{The skin friction coefficient for both baseline and optimum designs at $t_c=70$.}
\label{fig_naca_fwh_cf}
\end{figure}

The skin friction coefficient $C_f$ distribution, illustrated in Figure \ref{fig_naca_fwh_cf}, shows key differences between the baseline and optimum airfoils, with significant implications for aerodynamic performance. In the baseline design, higher $C_f$ values close to the leading edge indicate stronger surface shear forces and higher skin friction drag, suggesting that the boundary layer remains attached longer before separating. In contrast, the optimum design exhibits lower $C_f$ values, particularly near the leading edge, indicating reduced surface shear stress and earlier boundary layer separation, which leads to lower skin friction drag. Notably, the optimum design shows a smaller region of negative $C_f$ on the suction side, which indicates a less extended flow separation region. The less pronounced negative $C_f$ values near the trailing edge in the optimum design suggest more controlled separation, further reducing form drag. Overall, the lower $C_f$ in the optimum design contributes to reduced drag and smoother boundary layer behavior, which also helps minimize unsteady flow structures that could generate noise, thereby improving both aerodynamic efficiency and reducing aeroacoustic noise.

\section{Conclusions}
\label{sec:Conclusions}

In conclusion, we implemented a far-field aeroacoustic prediction solver using the FW-H formulation for moving medium problems in the time domain. This solver undergoes verification with analytical test cases and validation through a high-order flow solver for both inviscid and viscous flows. Serving as a post-processing tool for three-dimensional problems, it is coupled with the high-order flow solver, HORUS, employing ILES for turbulence modeling. These solvers are further integrated into a parallelized gradient-free optimization framework, effectively reducing OASPL at a far-field observer for NACA 4-digit airfoils. Notably, our proposed framework eliminates runtime dependency on the number of design parameters in gradient-free optimization algorithms. Through parallel implementation, a consistent runtime is maintained for each optimization iteration, akin to a single CFD simulation, provided adequate computational resources are available. Numerical results for the optimized NACA 4-digit airfoil highlight significant improvements across key performance metrics, including reduced noise levels and drag coefficient, alongside maintaining lift coefficient, indicating a comprehensive enhancement of both aerodynamic and acoustic efficiency.

The feasibility of the proposed aeroacoustic shape optimization framework can be assessed through testing at higher Reynolds numbers and addressing more industry-relevant problems. This research suggests potential improvements in aeroacoustic shape optimization methods, with significant implications for the development of quieter and more efficient aerodynamic designs.

\section*{Data Statement}

Data relating to the results in this manuscript can be downloaded from the publication’s website under a CC-BY-NC-ND 4.0 license.

\section*{CRediT authorship contribution statement}

\textbf{Mohsen Hamedi:} Conceptualization; Data curation; Formal analysis; Investigation; Methodology; Software; Validation; Visualization; Writing - original draft. 
\textbf{Brian Vermeire:} Conceptualization; Funding acquisition; Investigation; Methodology; Project administration; Resources; Software; Supervision; Writing - review and editing.

\section*{Declaration of competing interest}

The authors declare that they have no known competing financial interests or personal relationships that could have appeared to influence the work reported in this paper.

\section*{Acknowledgements}

The authors acknowledge support from the Natural Sciences and Engineering Research Council of Canada (NSERC) [RGPIN-2017-06773] and the Fonds de recherche du Québec (FRQNT) via the nouveaux chercheurs program. This research was enabled in part by support provided by Calcul Qu\'ebec (www.calculquebec.ca) and the Digital Research Alliance of Canada (www.alliancecan.ca) via a Resources for Research Groups allocation. M.H acknowledges Fonds de Recherche du Qu\'ebec - Nature et Technologie (FRQNT) via a B2X scholarship.

\begin{appendices}
\label{sec:appendix}

\section{Ffowcs Williams and Hawkings Formulation}
\label{sec:AcousticFormulation}

\renewcommand{\theequation}{A.\arabic{equation}}
\renewcommand{\thefigure}{A.\arabic{figure}}
\renewcommand{\thetable}{A.\arabic{table}}
\setcounter{equation}{0}
\setcounter{figure}{0}
\setcounter{table}{0}

The FW-H equation, an exact rearrangement of continuity and Navier-Stokes equations, yields an inhomogeneous wave equation with surface source terms, including monopole and dipole, and a volume source term, namely the quadrupole. Although the computational costs for volume integration of the quadrupole is notably higher, its impact can be neglected in many subsonic applications under certain conditions \cite{brentner1998analytical}. There are different solutions to the FW-H equation depending on the problem under investigation. The well-known Formulations 1 and 1A by Farassat \cite{brentner1986prediction, farassat1980review} assume sound wave propagation in a stationary medium, while Najafi-Yazdi et al. \cite{najafi2011acoustic} and Ghorbaniasl et al. \cite{ghorbaniasl2012moving} introduced formulations more suitable for CFD simulations, considering a moving medium. In this paper, the time-domain moving medium formulation is implemented, following the formulation proposed by Ghorbaniasl \cite{ghorbaniasl2012moving}. 

In the FW-H acoustic analogy, we define a data surface on the solid boundaries of the body, referred to as a solid data surface, or within the flow, encompassing the body, known as a permeable data surface. While computationally attractive, placing the permeable data surface too close to the body may lead to predictions suffering from solid data surface disadvantages \cite{ribeiro2023lessons}. Conversely, enclosing an expansive volume increases the need for fine spatial and temporal resolutions, further elevating computational costs. In general, this data surface is defined in space by a function $f(\pmb{x}, t)$, as 
\begin{equation}
f(\pmb{x}, t) 
\begin{cases}
< 0 \quad \mbox{inside the boundary}, \\
= 0 \quad \mbox{on the boundary}, \\
> 0 \quad \mbox{outside the boundary},
\end{cases}
\end{equation}
and it is assumed that 
\begin{equation}
| \pmb{\nabla} f | = 1,
\end{equation}
and $f$ is smooth, without discontinuities, so that 
\begin{equation}
\frac{\partial f}{\partial x_i} = \hat{\mathbf{n}}_i
\end{equation}
is the local outer normal of the data surface. 

The initial step in the derivation of the FW-H equation involves multiplying the Heaviside function by the conservation of mass and momentum equations. This operation confines the application of these equations exclusively to regions outside the data surface. Subsequently, employing the principles of generalized function theory, these equations are transformed into non-homogeneous wave equations, as detailed in \cite{farassat1994introduction}. Thus, the conservation of mass will be
\begin{equation}
\frac{D}{Dt} \left[ \left( \rho - \rho_0 \right) H (f) \right] + \frac{\partial}{\partial x_i} \left[ \rho u_i H (f) \right] = Q  \delta (f) ,
\label{eq:continuity}
\end{equation}
with
\begin{equation}
Q = \rho \left( u_n + U_{\infty n} - v_n \right) + \rho_0 \left( v_n - U_{\infty n} \right) ,
\end{equation}
where $\rho$ is the density of the fluid, $\rho_0$ denotes the fluid density at rest, $H(f)$ is the Heaviside function, $u_i$ are the velocity components, $Q$ is the source term for the continuity equation known as the thickness term and accounts for the flux of mass across the surface, and $\delta (f)$ is the Dirac's delta function of $f(\pmb{x}, t)$. Finally, the subscript $n$ denotes the local normal term of the data surface. Thus, $u_n = u_i \hat{n}_i$,  $U_{\infty n} = U_{\infty i} \hat{n}_i$, and $v_n = v_i \hat{n}_i$. $U_{\infty i}$ being the $i^{th}$ component of the mean flow velocity and $v_i$ being the $i^{th}$ component of the data surface velocity which is zero throughout this study. Note that Equation \ref{eq:continuity} returns zero inside the data surface. 

Applying the same methodology, the non-linear momentum equation yields the following
\begin{equation}
\frac{D}{Dt} \left[ \rho u_i H (f) \right] + \frac{\partial}{\partial x_j} \left[ \rho u_i u_j H (f) \right] + \frac{\partial}{\partial x_j} \left[ \left( p \delta_{ij} - \sigma_{ij} \right) H(f) \right] = L_i \delta (f) ,
\label{eq:momentum}
\end{equation}
with
\begin{equation}
L_i = P_{ij} \hat{n}_j + \rho u_i \left( u_n + U_{\infty n} - v_n \right) , 
\end{equation}
and
\begin{equation}
P_{ij} = \left( p - p_0 \right) \delta_{ij} - \sigma_{ij} ,
\end{equation}
where $p$ is the static pressure, $\sigma_{ij}$ is the viscous stress tensor, $L_i$ is the source term for the non-linear momentum equation known as the loading term and accounts for the flux of momentum across the surface, and $P_{ij}$ is the compressive stress tensor.

The equation for propagation of noise is obtained via taking the time derivative of Equation \ref{eq:continuity} and subtracting the divergence of Equation \ref{eq:momentum}, and is
\begin{equation}
\left( \frac{1}{c_{0}^{2}} \frac{D^2}{Dt^2} - \nabla^2 \right) \left( p^{\prime} \left( \pmb{x}, t \right) H(f) \right) = \frac{D}{Dt} \left( Q \delta (f) \right) - \frac{\partial}{\partial x_i} \left( L_i \delta (f) \right) + \frac{\partial^2}{\partial x_i \partial x_j} \left( T_{ij} H(f) \right) ,
\label{eq:NSnoise}
\end{equation}
where $T_{ij}$ is the Lighthill's stress tensor and defined as
\begin{equation}
T_{ij} = \rho u_i u_j + \left[ \left( p - p_0 \right) - c_0^2 \left( \rho - \rho_0 \right) \right] \delta_{ij} - \sigma_{ij} .
\end{equation}
On the right-hand side of Equation \ref{eq:NSnoise}, the first two terms represent the monopole (thickness) and dipole (loading) sources, respectively, acting on the surface $f=0$, presented with the Dirac delta function, $\delta(f)$. The third term corresponds to the quadrupole source acting on the volume outside of the data surface, as indicated by the Heaviside function, $H(f)$. This convective wave equation, Equation \ref{eq:NSnoise}, can be solved either on a solid data surface \cite{brentner1988prediction, brentner2003modeling, farassat2007derivation} with the drawback of involving costly volume integrals, or on a permeable data surface \cite{lyrintzis2003surface, di1997new}, in either the time domain \cite{farassat2007derivation, farassat1975theory} or frequency domain \cite{lockard2002comparison, lockard2000efficient, ghorbaniasl2015analytical}. Additionally, it can be addressed for stationary medium problems using the well-established Farassat's Formulations 1 and 1A \cite{farassat1975theory, farassat1980review, brentner1997numerical}. Alternatively, it can account for the presence of mean flow using formulations such as Najafi-Yazdi et al.'s \cite{najafi2011acoustic} or Ghorbaniasl et al.'s \cite{ghorbaniasl2012moving}. 

\section{Solution to the FW-H Equations}
\label{sec:AcousticSolver}

\renewcommand{\theequation}{B.\arabic{equation}}
\renewcommand{\thefigure}{B.\arabic{figure}}
\renewcommand{\thetable}{B.\arabic{table}}
\setcounter{equation}{0}
\setcounter{figure}{0}
\setcounter{table}{0}

Given the resemblance of CFD simulations to wind tunnels with a mean flow, we adopt a formulation similar to Najafi-Yazdi et al. \cite{najafi2011acoustic} and Ghorbaniasl et al. \cite{ghorbaniasl2012moving}. This approach addresses the presence of mean flow in wind tunnel problems with a moving medium by solving a convective wave equation, initially derived by Wells and Han \cite{wells1995acoustics}. In this paper, we utilize a time-domain formulation with a moving medium and a stationary permeable data surface approach, following the Ghorbaniasl's formulation \cite{ghorbaniasl2012moving}.

The numerical computation of the flow field is performed using our in-house high-order flow solver, HORUS. After predicting density, pressure, and velocity fields, and collecting data on a predefined data surface, this information is input into the FW-H formulation. Subsequently, the pressure perturbation propagates to the observer location, and acoustic pressure is computed as post-processing tools to HORUS via the FW-H formulation.

The acoustic pressure consists of three sources, namely, thickness, loading, and quadrupole sources \cite{ghorbaniasl2012moving}, 
\begin{equation}
p^\prime (\pmb{x}, t, \pmb{M}_\infty) = p^\prime_T (\pmb{x}, t, \pmb{M}_\infty) + p^\prime_L (\pmb{x}, t, \pmb{M}_\infty) + p^\prime_Q (\pmb{x}, t, \pmb{M}_\infty) ,
\end{equation}
where $p^\prime_T$ and $p^\prime_L$ are the thickness and loading pressures, respectively, computed via surface integration with low computational cost. The quadrupole pressure, $p^\prime_Q$, involves computationally expensive volume integration. Using a permeable data surface, which encloses a limited volume adjacent to the body and covers all non-linear flow field and noise sources, allows the neglect of quadrupole terms. This enhances efficiency and reduces computational costs in the acoustic analogy. Therefore, many derivations assume all noise sources are within the permeable data surface, leading to the omission of volume integration, specifically the quadrupole noise source \cite{najafi2011acoustic}.

The thickness and loading pressures are expressed as \cite{ghorbaniasl2012moving},
\begin{equation}
4 \pi p^{\prime}_T (\pmb{x}, t, \pmb{M}_\infty) = \int_S \left[ \frac{\left( 1 - M_{\infty R} \right) \dot{Q}}{R^{\star}} \right]_e dS - \int_S \left[ Q \frac{c_0 M_{\infty R^\star}}{R^{\star 2}} \right]_e dS ,
\label{eq:pt}
\end{equation}
and
\begin{equation}
4 \pi p^{\prime}_L (\pmb{x}, t, \pmb{M}_\infty) = \frac{1}{c_0} \int_S \left[ \frac{\dot{L}_R}{R^{\star}} \right]_e dS + \int_S \left[ \frac{L_{R^\star}}{R^{\star 2}} \right]_e dS ,
\label{eq:pl}
\end{equation}
where the dot over quantities denotes the temporal derivative with respect to the source time $\tau$, and $c_0$ is the speed of sound. The integrands in Equations \ref{eq:pt} and \ref{eq:pl} are defined as
\begin{equation}
M_{\infty R} = M_{\infty i} \tilde{R}_i ,
\end{equation}
\begin{equation}
M_{\infty R^\star} = M_{\infty i} \tilde{R}_i^\star ,
\end{equation}
\begin{equation}
\dot{L}_R = \dot{L}_i \tilde{R}_i ,
\end{equation}
\begin{equation}
L_{R^\star} = L_i \tilde{R}_i^\star ,
\end{equation}
\begin{equation}
R^\star = \frac{1}{\gamma} \sqrt{|\pmb{x}-\pmb{y}|^2 + \gamma^2 \left( \pmb{M}_\infty \cdot \left( \pmb{x} - \pmb{y} \right) \right)^2} = \frac{1}{\gamma} \sqrt{r^2 + \gamma^2 \left( \pmb{M}_\infty \cdot \pmb{r} \right)^2} ,
\label{eq:Rstar}
\end{equation}
\begin{equation}
R = \gamma^2 \left( R^\star - \pmb{M}_\infty \cdot \pmb{r} \right) ,
\label{eq:R}
\end{equation}
\begin{equation}
\gamma^2 = \frac{1}{1 - | \pmb{M}_\infty |^2} , 
\end{equation}
\begin{equation}
\tilde{R}_i^\star = \frac{\partial R^\star}{\partial	x_i} = \frac{r_i + \gamma^2 \left( M_{\infty j} r_j \right) M_{\infty i}}{\gamma^2 R^\star} ,
\end{equation}
\begin{equation}
\tilde{R}_i = \frac{\partial R}{\partial x_i} = \gamma^2 \left( \tilde{R}^\star_i - M_{\infty i} \right) ,
\end{equation}
where $R^\star$ and $R$ are called the amplitude and phase radii, respectively, $\pmb{r}=\pmb{x}-\pmb{y}$ is the distance between the observer position, $\pmb{x}$, and the source position, $\pmb{y}$, and, finally, $\tau=t-R/c_0$ is the source time with $t$ being the observer time. 

In Equations \ref{eq:pt} and \ref{eq:pl}, the subscripts $e$ denote integration at the source time, $\tau$, where all quantities are computed via HORUS. The right-hand side is in the source time frame, and the left-hand side is in the observer time frame. Two main numerical approaches exist for solving Equations \ref{eq:pt} and \ref{eq:pl}, namely, the retarded-time approach and the advanced-time approach \cite{brentner2003modeling}. This study employs the advanced-time approach, also known as the source-time-dominant approach.

In the advanced-time approach, the source time corresponds to the time history obtained from CFD simulations. On the permeable data surface, each panel, with a single point at its center, emits noise to the observer from a unique source time. Considering a single snapshot of the flow field, the contribution of each point on the data surface does not reach the observer simultaneously due to varying distances between these points and the observer. Thus, the noise contribution from each point, at a single snapshot, reaches the observer at different times. Consequently, for each point on the permeable data surface, a distinct and unique time history is obtained. The observer time, which is unique for every individual point on the permeable data surface, is computed via
\begin{equation}
t = \tau + \frac{R}{c_0}.
\end{equation}

For each point on the data surface, a distinct observer time history is computed. To unify these individual time histories into a single observer time history, we determine an observer time history that ensures the first entry aligns with the moment when contributions from all other points reach the observer. Similarly, the last entry of the unified observer time history aligns with the moment when the contribution from closest point to the observer ends. Once a unified array of observer times is obtained, the next step involves interpolating each integrand in Equations \ref{eq:pt} and \ref{eq:pl}, i.e.
\begin{equation}
4 \pi p^\prime (\pmb{x}, t^\star, \pmb{M}_\infty ) \approx \sum_{i=1}^{n_p}  \mathcal{I} \left( I_i(t), t^\star \right) ,
\end{equation}
where $p^\prime$ is either $p^\prime_T$ or $p^\prime_L$, $t^\star$ is the desired observer time, $n_p$ is number of points on the permeable data surface, $\mathcal{I}$ is an interpolation operator, and $I_i(t)$ is the right-hand side of either Equation \ref{eq:pt} or \ref{eq:pl}. Brentner et al. \cite{bres2004maneuvering} showed that the advanced-time approach requires significantly less operation than the retarded-time approach and, thus, is more computationally efficient. Following the interpolation of integrands, surface integrations are performed to calculate the $p^\prime_T$ and $p^\prime_L$. Subsequently, the time history of acoustic pressure at the observer is obtained by summing the thickness and loading pressures.

The aeroacoustic solver, employing the FW-H formulation, undergoes verification through analytical test cases. Subsequently, validation takes place by solving both the Euler and Navier-Stokes equations. The details of this verification and validation processes are explained in the following sections.

\section{Verification}
\label{sec:AcousticVerification}

\renewcommand{\theequation}{C.\arabic{equation}}
\renewcommand{\thefigure}{C.\arabic{figure}}
\renewcommand{\thetable}{C.\arabic{table}}
\setcounter{equation}{0}
\setcounter{figure}{0}
\setcounter{table}{0}

Verification of the acoustic solver is conducted by examining two analytical test cases, specifically wind tunnel scenarios featuring stationary sources — a monopole source and a dipole source.

\subsection{Stationary Monopole}

\begin{figure}
\centering
\begin{subfigure}{0.45\textwidth}
\includegraphics[width=\textwidth]{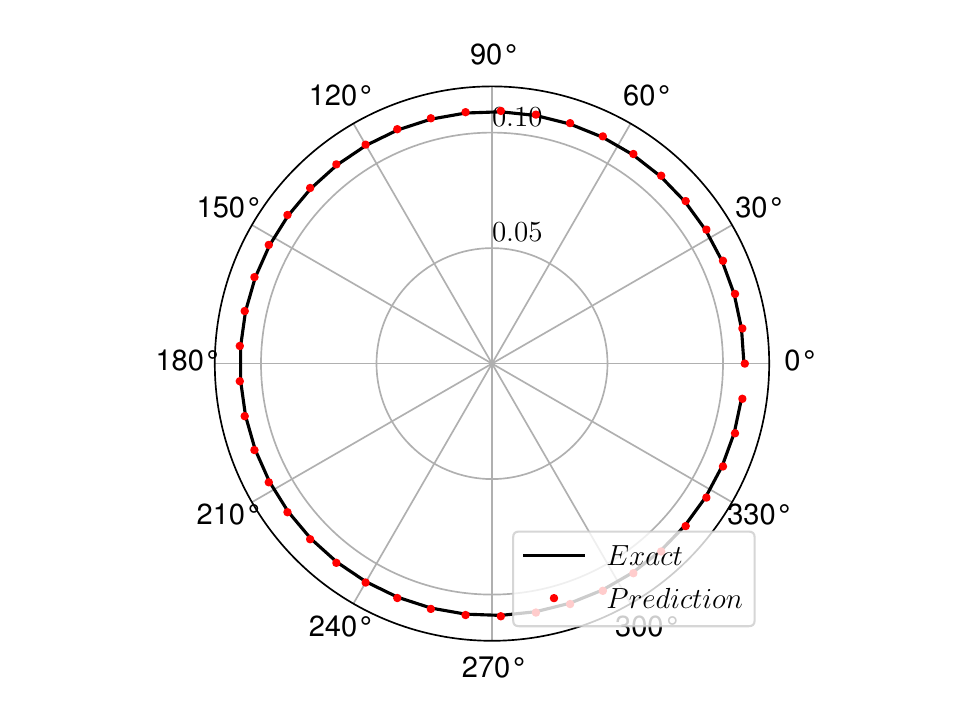}
\caption{$\pmb{M}_{\infty}=[0.0,0.0,0.0]$}
\end{subfigure}
\begin{subfigure}{0.45\textwidth}
\includegraphics[width=\textwidth]{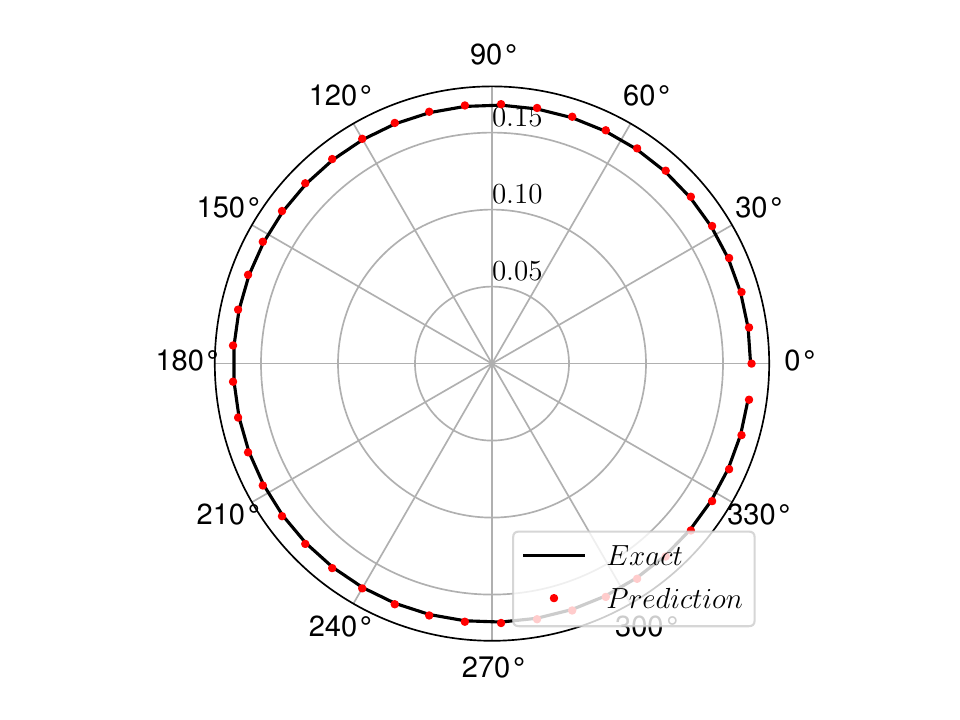}
\caption{$\pmb{M}_{\infty}=[0.0,0.0,0.5]$}
\end{subfigure}
\begin{subfigure}{0.45\textwidth}
\includegraphics[width=\textwidth]{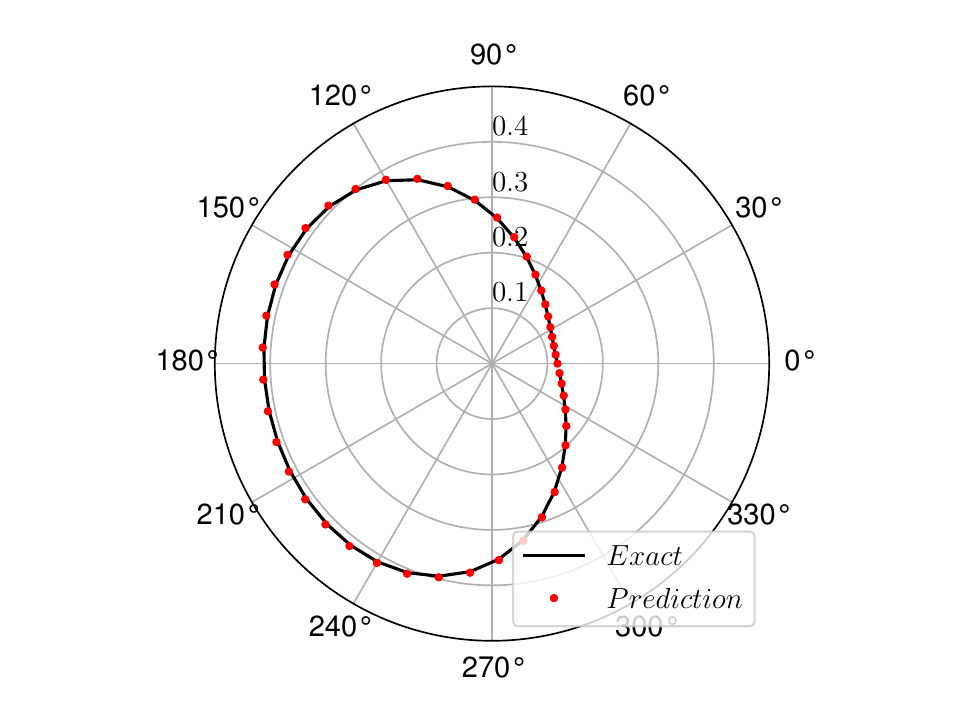}
\caption{$\pmb{M}_{\infty}=[0.5,0.1,0.5]$}
\end{subfigure}
\begin{subfigure}{0.45\textwidth}
\includegraphics[width=\textwidth]{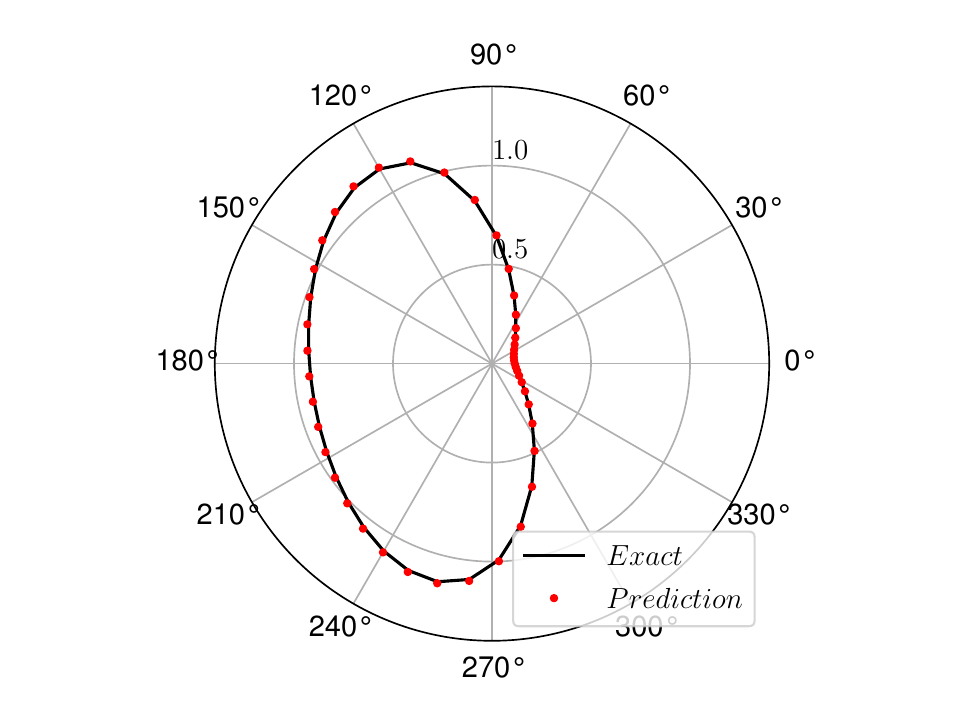}
\caption{$\pmb{M}_{\infty}=[0.7,0.1,0.5]$}
\end{subfigure}
\caption{Comparison of the root-mean-squared of the predicted acoustic pressure with the exact solution for different Mach number flows.}
\label{fig:DirectivityMonopole}
\end{figure}

\begin{figure}[htbp]
\centering
\begin{subfigure}{0.45\textwidth}
\includegraphics[width=\textwidth]{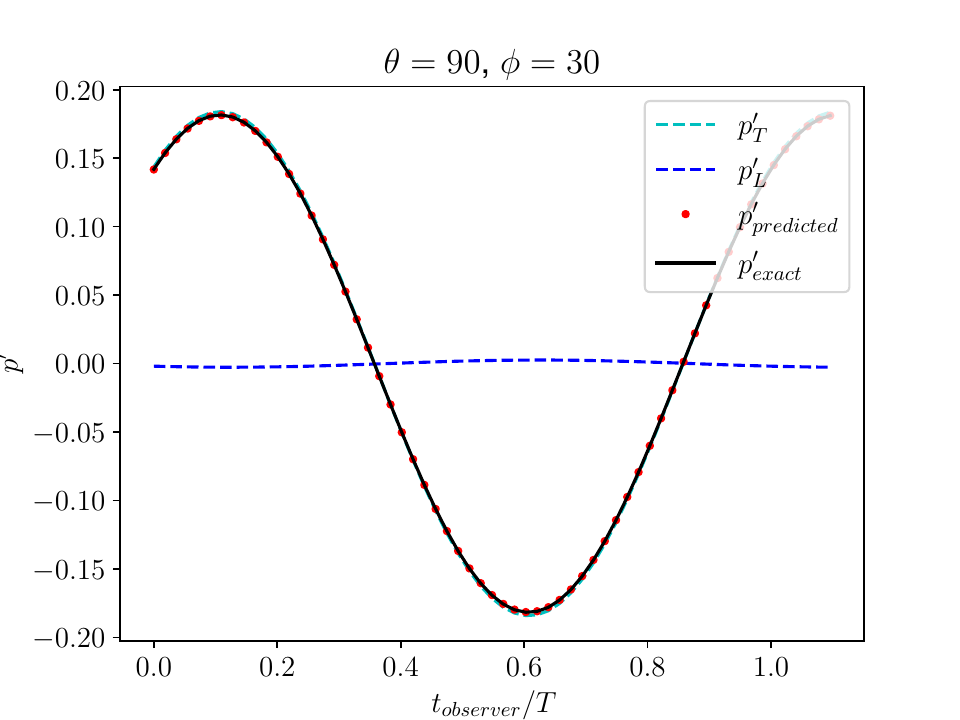}
\caption{ }
\end{subfigure}
\begin{subfigure}{0.45\textwidth}
\includegraphics[width=\textwidth]{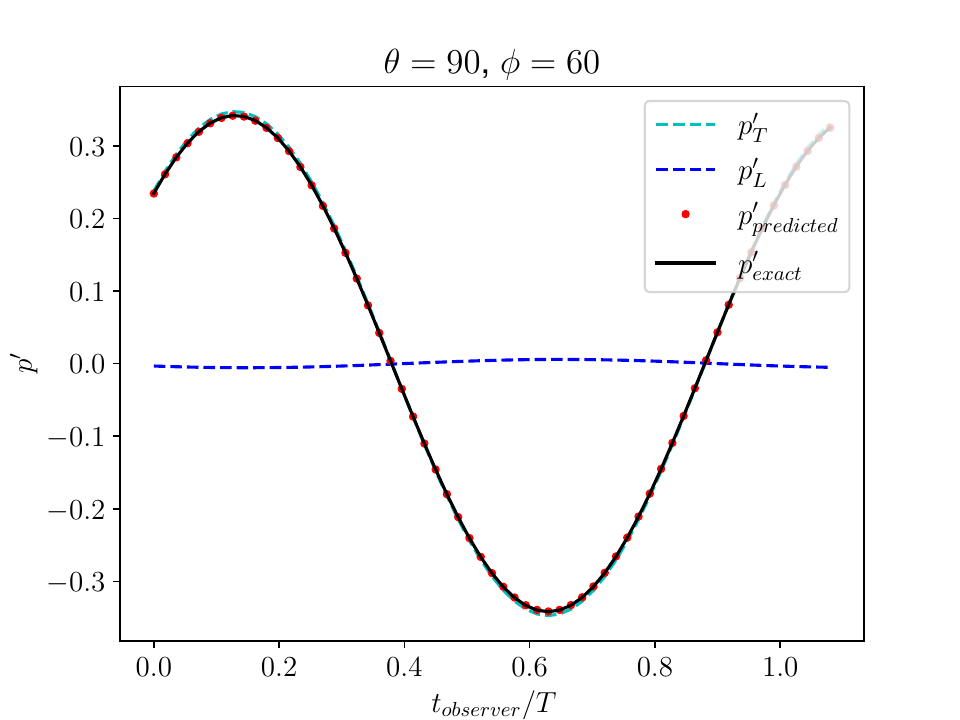}
\caption{ }
\end{subfigure}
\begin{subfigure}{0.45\textwidth}
\includegraphics[width=\textwidth]{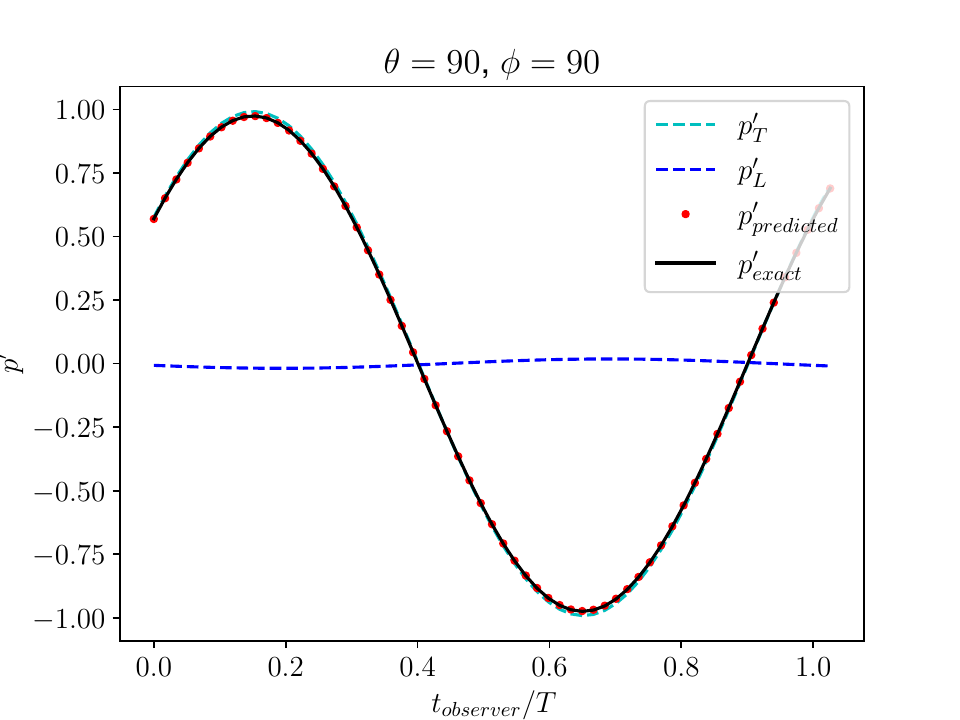}
\caption{ }
\end{subfigure}
\begin{subfigure}{0.45\textwidth}
\includegraphics[width=\textwidth]{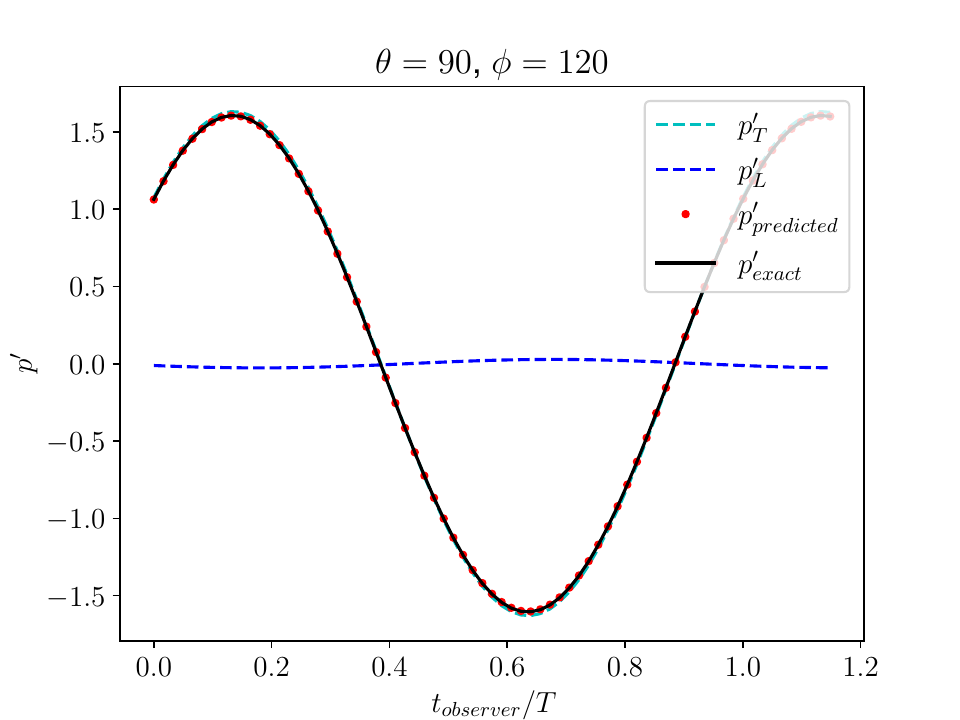}
\caption{ }
\end{subfigure}
\caption{Comparison of the predicted and exact acoustic pressure time histories for inflow Mach number of $\pmb{M}_\infty = [0.7,0.1,0.5]$.}
\label{fig:TimeHistoryMonopole}
\end{figure}

A stationary single-frequency monopole source is positioned at the origin of a medium moving at a constant velocity. The complex velocity potential, denoted as $\varphi_m$, initially derived for the monopole in a uniform flow along the $x_1$-direction \cite{dowling1984sound}, is extended to arbitrary orientations as \cite{ghorbaniasl2012moving},
\begin{equation}
\varphi_m \left( \pmb{x}, t \right) = A \frac{1}{4 \pi R^\star} \exp \left[ i \omega \left( t - \frac{R}{c_0} \right) \right] ,
\end{equation}
where $R^\star$ and $R$ are computed via Equations \ref{eq:Rstar} and \ref{eq:R}, respectively. Then, the acoustic particle velocity and the acoustic pressure are obtained via 
\begin{equation}
u_i^\prime \left( \pmb{x}, t \right) = \frac{\partial \varphi_m \left( \pmb{x} , t \right)}{\partial x_i} ,
\end{equation}
and
\begin{equation}
p^\prime \left( \pmb{x}, t \right) = - \rho_0 \left( \frac{\partial \varphi_m \left( \pmb{x} , t \right)}{\partial t} + c_0 M_{\infty i} \frac{\partial \varphi_m \left( \pmb{x} , t \right)}{\partial x_i} \right) = - \rho_0 \left( i \omega + c_0 M_{\infty i} \frac{\partial}{\partial x_i} \right) \varphi_m \left( \pmb{x} , t \right) ,
\end{equation}
respectively. And finally, the induced density is 
\begin{equation}
\rho^\prime \left( \pmb{x} , t \right) = \frac{p^\prime \left( \pmb{x} , t \right)}{c_0^2} .
\end{equation}
Here, the velocity potential amplitude is $A=1 m^2/s$, the angular frequency of the source is $\omega = 10 \pi ~ rad/s$, the ambient speed of sound is $c_0=340.75 m/s$, the free-stream flow density is $\rho_0 = 1.234 kg/m^3$, and the specific heat ratio of air is $\gamma = 1.4$. Thus, the free-stream pressure is obtained via the ideal gas law as
\begin{equation}
p_0 = \rho_0 R_g T_0 \xrightarrow{c_0 = \sqrt{\gamma R_g T_0}} p_0 = \frac{\rho_0 c_0^2}{\gamma} ,
\end{equation} 
where $R_g$ is a gas constant. A permeable data surface in the form of a sphere with a radius $r=1$ is utilized. The sphere is discretized into $30$ polar sections, ensuring a constant spacing of $2 \pi / 45$ between data points along each section. This uniform distribution guarantees equal area for each data panel. For adequate temporal resolution, a value of $\Delta t/T = 0.02$ is chosen, with $T$ representing the period of the source signal. At a distance of $20~m$ from the source, the radiated sound pressure is recorded for various mean flow orientations. The root-mean-squared value of the monopole acoustic pressure is computed over a duration of $10$ periods. Figure \ref{fig:DirectivityMonopole} illustrates these values for different mean flow orientations. Additionally, Figure \ref{fig:TimeHistoryMonopole} compares the calculated monopole acoustic pressure time history with the exact solution, showing an exact match between the predicted pressure perturbation, determined using the acoustic solver, and the analytical values. Both figures affirm the accuracy of the acoustic solver for monopole-like sources.

\subsection{Stationary Dipole}

\begin{figure}
\centering
\begin{subfigure}{0.45\textwidth}
\includegraphics[width=\textwidth]{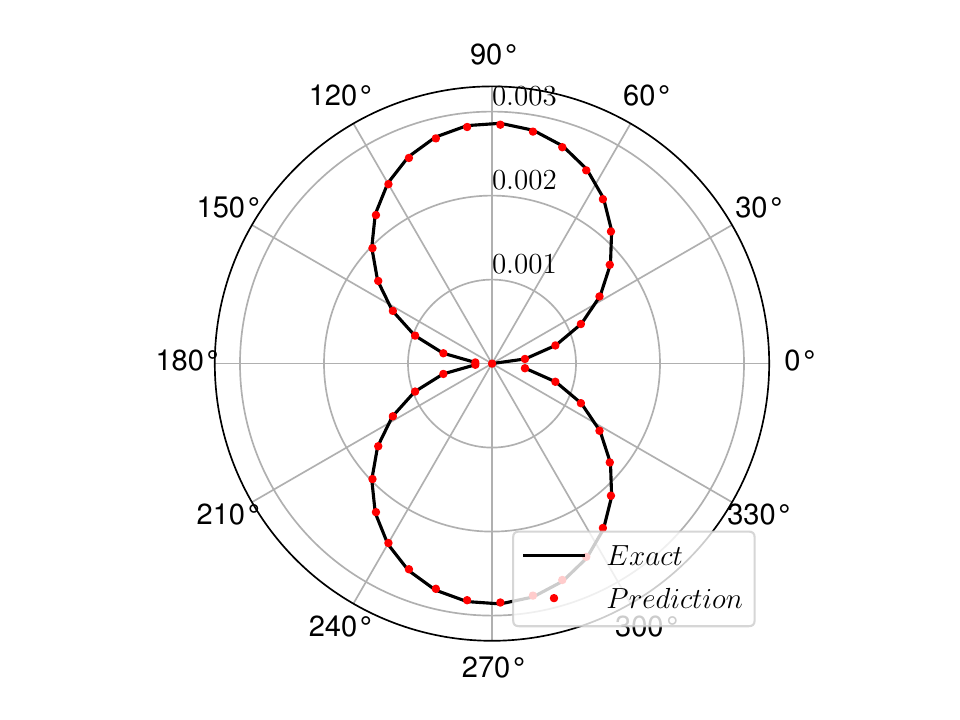}
\caption{$\pmb{M}_{\infty}=[0.0,0.0,0.4]$}
\end{subfigure}
\begin{subfigure}{0.45\textwidth}
\includegraphics[width=\textwidth]{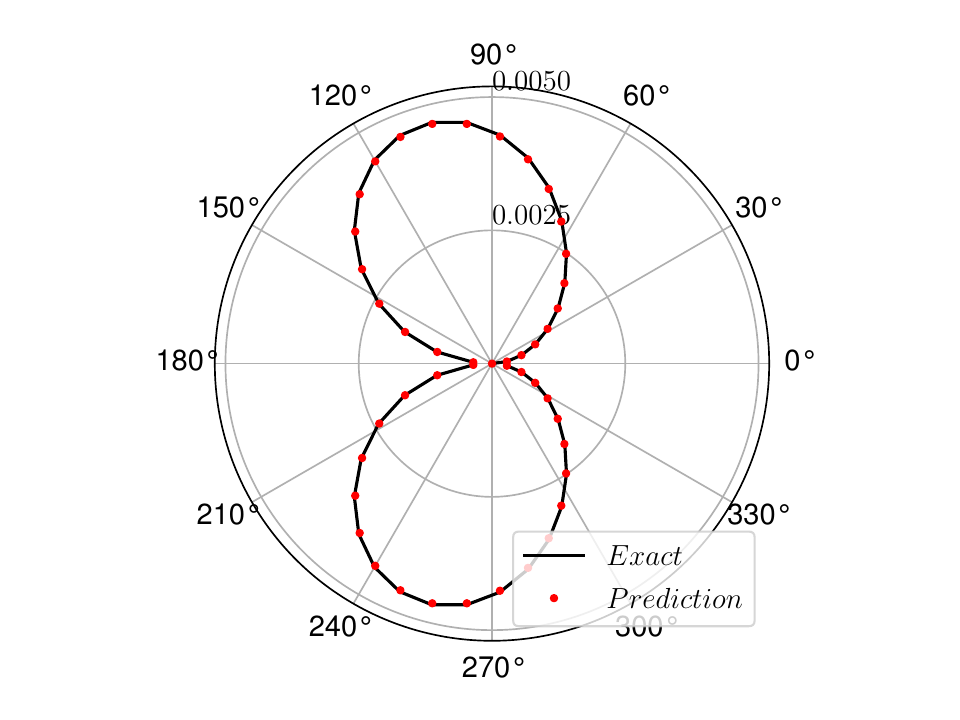}
\caption{$\pmb{M}_{\infty}=[0.4,0.0,0.4]$}
\end{subfigure}
\begin{subfigure}{0.45\textwidth}
\includegraphics[width=\textwidth]{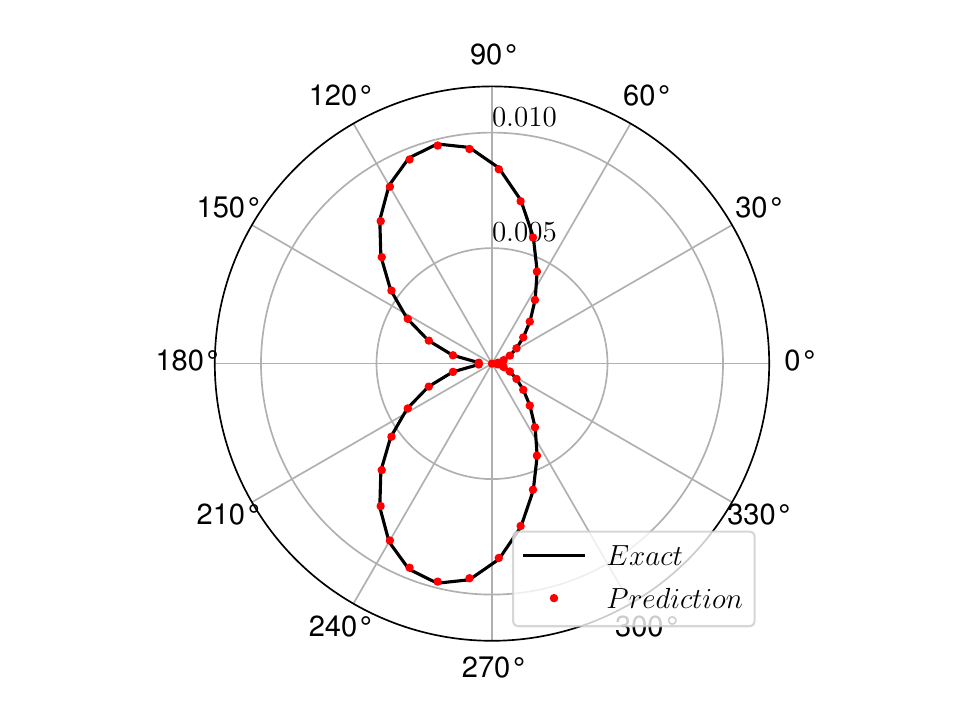}
\caption{$\pmb{M}_{\infty}=[0.6,0.0,0.4]$}
\end{subfigure}
\begin{subfigure}{0.45\textwidth}
\includegraphics[width=\textwidth]{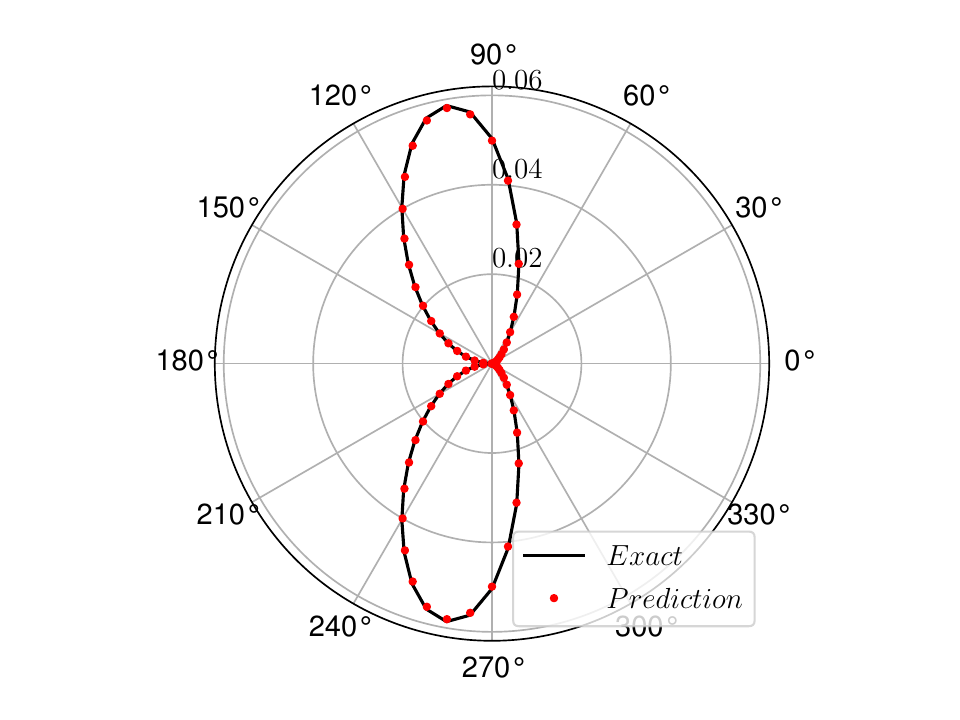}
\caption{$\pmb{M}_{\infty}=[0.8,0.0,0.4]$}
\end{subfigure}
\caption{Comparison of the root-mean-squared of the predicted acoustic pressure with the exact solution for different Mach number flows.}
\label{fig:DirectivityDipole}
\end{figure}

\begin{figure}[htbp]
\centering
\begin{subfigure}{0.45\textwidth}
\includegraphics[width=\textwidth]{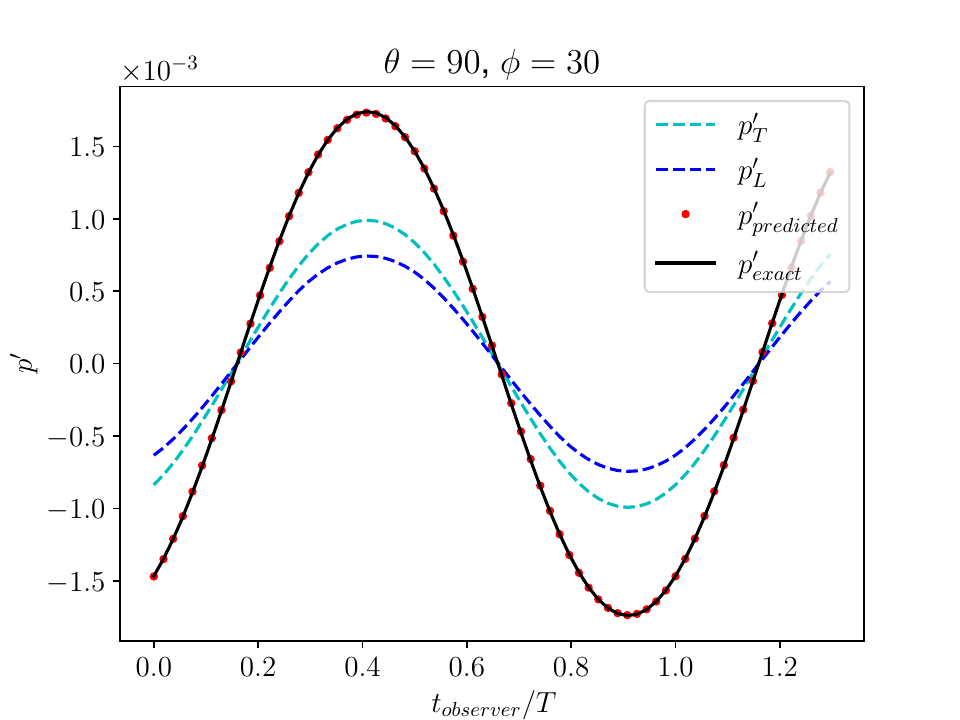}
\caption{ }
\end{subfigure}
\begin{subfigure}{0.45\textwidth}
\includegraphics[width=\textwidth]{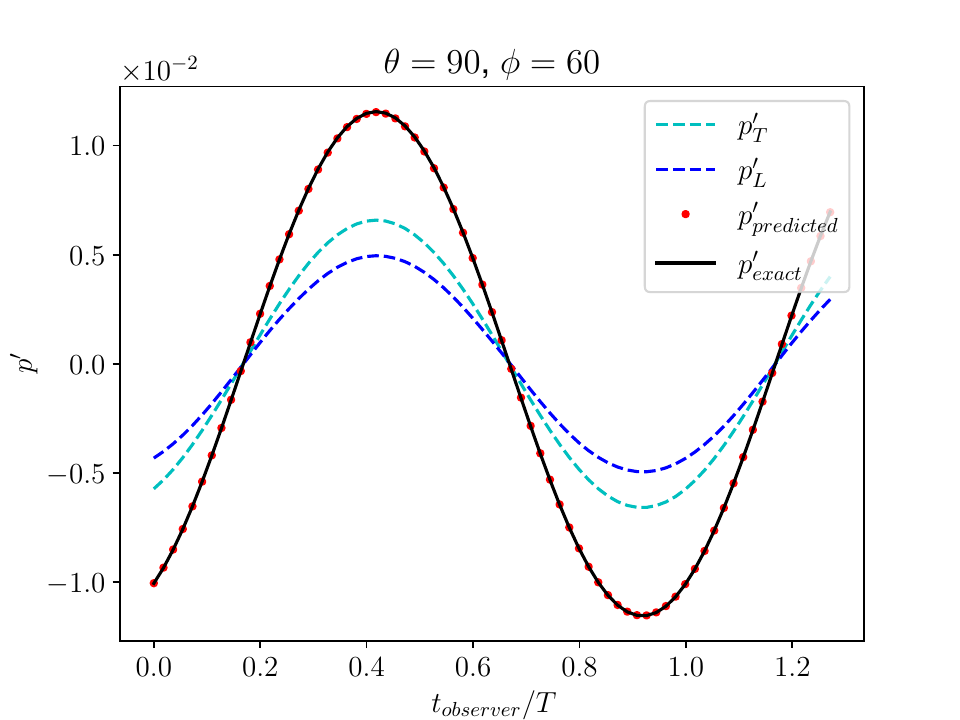}
\caption{ }
\end{subfigure}
\begin{subfigure}{0.45\textwidth}
\includegraphics[width=\textwidth]{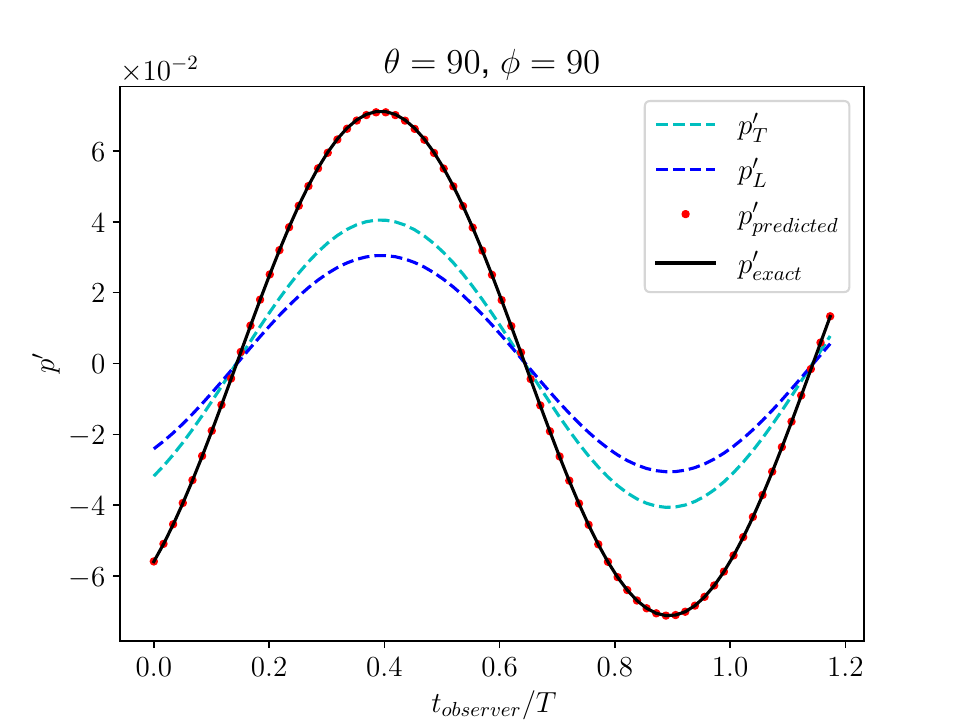}
\caption{ }
\end{subfigure}
\begin{subfigure}{0.45\textwidth}
\includegraphics[width=\textwidth]{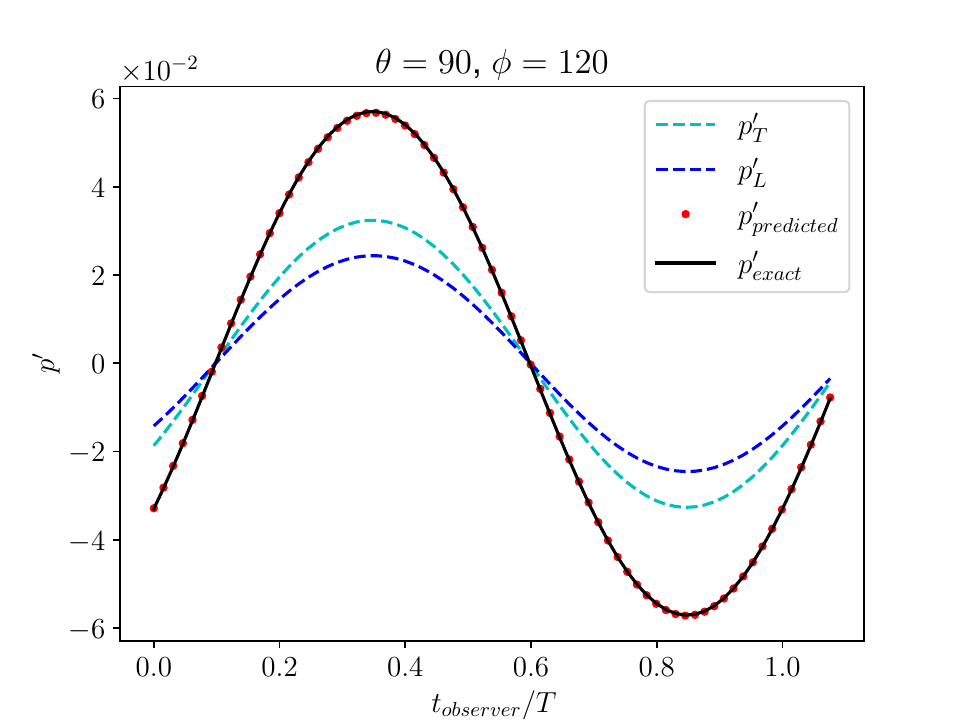}
\caption{ }
\end{subfigure}
\caption{Comparison of the predicted and exact acoustic pressure time histories for inflow Mach number of $\pmb{M}_\infty = [0.8,0.0,0.4]$.}
\label{fig:TimeHistoryDipole}
\end{figure}

The second verification test for the acoustic solver involves a stationary dipole positioned at the origin of a medium moving at a constant velocity with an arbitrary orientation. We assume the dipole's axis aligns with the $x_2$-axis. In this scenario, the complex velocity potential for the dipole can be expressed as the derivative of the monopole's complex velocity potential with respect to $x_2$, 
\begin{equation}
\varphi_d \left( \pmb{x}, t \right) = \frac{\partial}{\partial x_2} \varphi_{m} \left( \pmb{x}, t \right) .
\end{equation}
The calculation of acoustic particle velocity, pressure, and induced density follows a similar procedure to the monopole case. Utilizing a spherical data surface with a radius of $r=1$, mirroring the monopole approach, this surface is discretized into $30$ sections in the polar direction, and the azimuthal direction employs a grid size of $2 \pi / 45$. Temporal calculations maintain a resolution of $\Delta t/T = 0.02$. The radiated sound pressure is recorded $100~m$ from the dipole source. Subsequently, the root-mean-squared value of the acoustic pressure is computed over a span of $10$ periods. These computations, conducted for various mean flow orientations, are illustrated in Figure \ref{fig:DirectivityDipole}. Additionally, Figure \ref{fig:TimeHistoryDipole} displays the time history of the acoustic pressure. Both figures exhibit an exact match between the FW-H prediction and the analytical data, affirming the accuracy of the acoustic solver for dipole-like sources.

\section{Validation}
\label{sec:AcousticValidation}

\renewcommand{\theequation}{D.\arabic{equation}}
\renewcommand{\thefigure}{D.\arabic{figure}}
\renewcommand{\thetable}{D.\arabic{table}}
\setcounter{equation}{0}
\setcounter{figure}{0}
\setcounter{table}{0}

Having successfully verified our acoustic solver against analytical test cases, the next step involves its validation against the direct acoustic approach, where acoustic pressure is computed directly from the flow solver. In this section, our validation process focuses on comparing the acoustic pressure obtained through our acoustic solver with that computed directly via HORUS. Initially, we validate the acoustic solver in an inviscid flow scenario, devoid of vortices, where the flow solver solves the Euler equations. Two test cases are employed: the first involves a single monopole positioned at the center of a cubic box, and the second introduces multiple monopoles placed near the center of the cubic box. The inflow Mach number is set to zero, ensuring a quiescent flow, and a source term is incorporated into the energy equation to emulate a monopole.

\subsection{Single Monopole in Quiescent Flow}

The source term for the single monopole is defined as
\begin{equation}
s(\pmb{x},t) = A \exp \left( {-k \left[ \left(x-x_s \right)^2 + \left(y-y_s \right)^2 + \left(z-z_s \right)^2 \right] } \right) \sin ( 2 \pi \omega t ),
\label{eq:EulerSource}
\end{equation}
where $A=0.05$ is the amplitude, $k=100~1/m^2$ is the range factor, $[x_s,y_s,z_s]=[0,0,0]$ is the location of the source or monopole, and $\omega=0.5~1/s$ is the frequency. 

In this problem, the source term exhibits characteristics similar to a Gaussian bump and undergoes oscillations within the domain, creating a fluctuating pressure field around the source point. The absence of vortices, attributed to a zero inflow Mach number and inviscid flow conditions, eliminates challenges associated with boundary treatments. Consequently, this configuration provides a robust validation for the acoustic solver.

\begin{figure}
\centering
\includegraphics[width=0.3\textwidth]{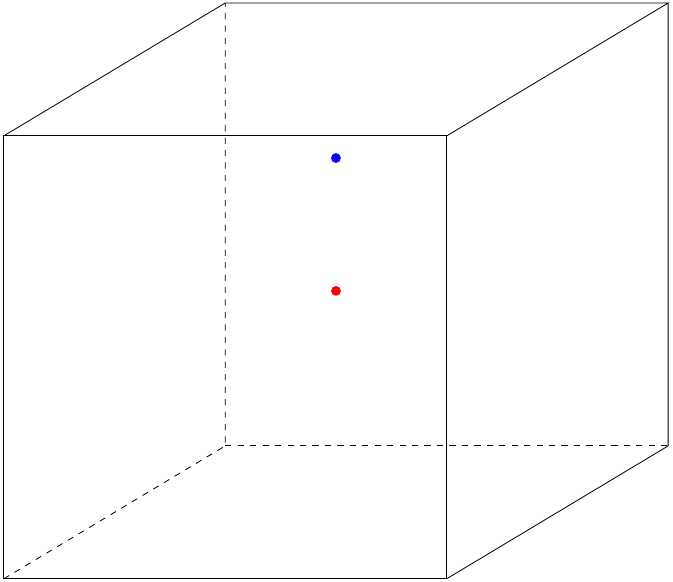}
\caption{The computational domain with the monopole in red and the observer in blue.}
\label{figure:eulermonopoledomain}
\end{figure}

A $[10 \times 10 \times 10]$ cube is discretized into $125,000$ structured hexahedral elements with applied Riemann invariant boundary conditions. An observer is positioned at $[x_{obs},y_{obs},z_{obs}]=[0,3,0]$, located above the monopole. The Euler equations are solved using $\mathcal{P}3$ simulation, and the flow data is collected on a spherical data surface of radius $r=1.5$. Figure \ref{figure:eulermonopoledomain} visualizes the computational domain, monopole, and observer position. Acoustic pressure at the observer point is determined through two approaches. First, directly computed from the flow solver using a $\mathcal{P}3$ simulation, and second, obtained by collecting flow data on the data surface through $\mathcal{P}1,~\mathcal{P}2,$ and $\mathcal{P}3$ simulations, which is then input into the acoustic solver. The resulting acoustic pressure fields from these approaches are compared for analysis.

\begin{figure}
\centering
\begin{subfigure}{0.45\textwidth}
\includegraphics[width=\textwidth]{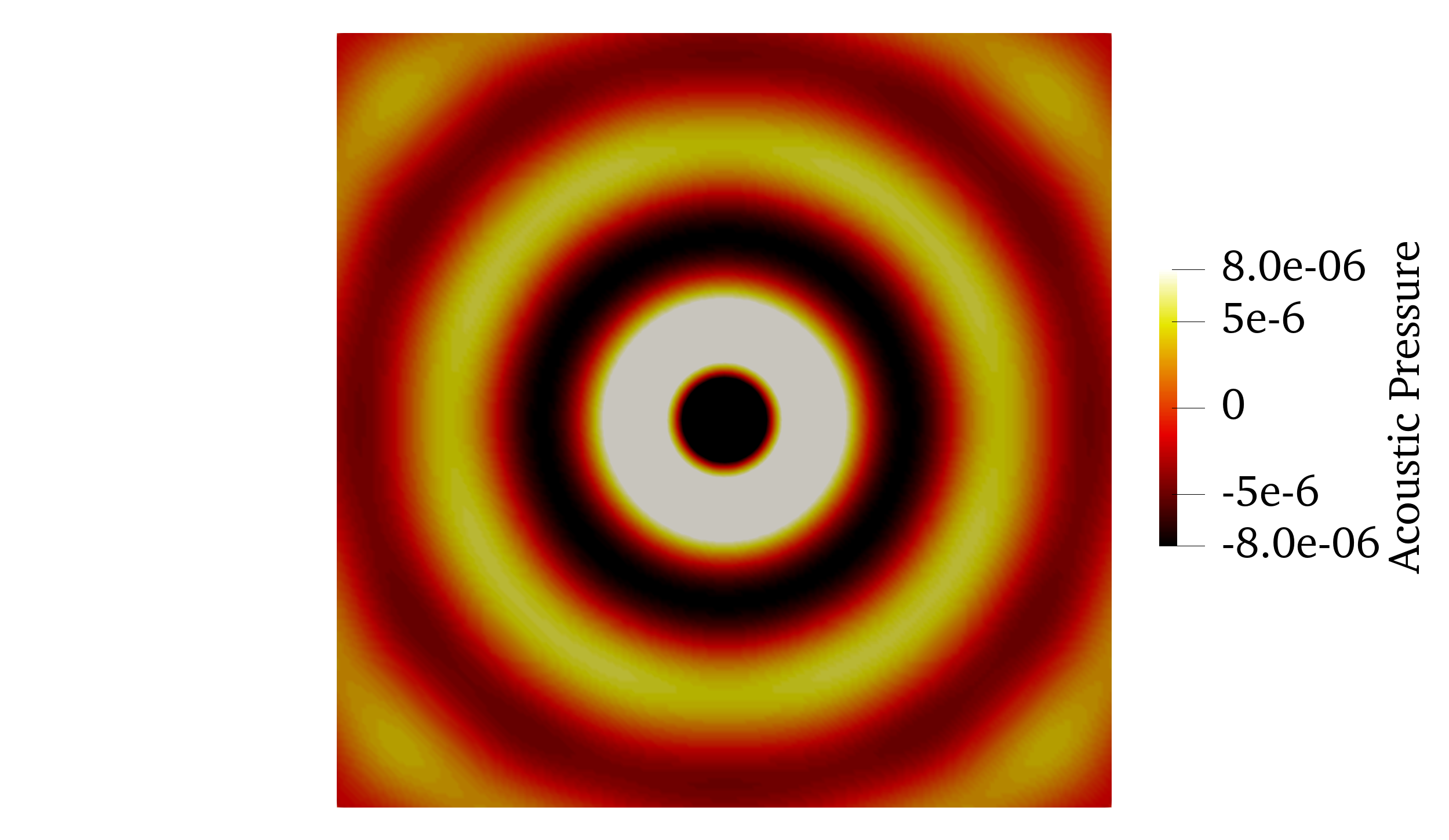}
\subcaption{HORUS.}
\end{subfigure}
\begin{subfigure}{0.45\textwidth}
\includegraphics[width=\textwidth]{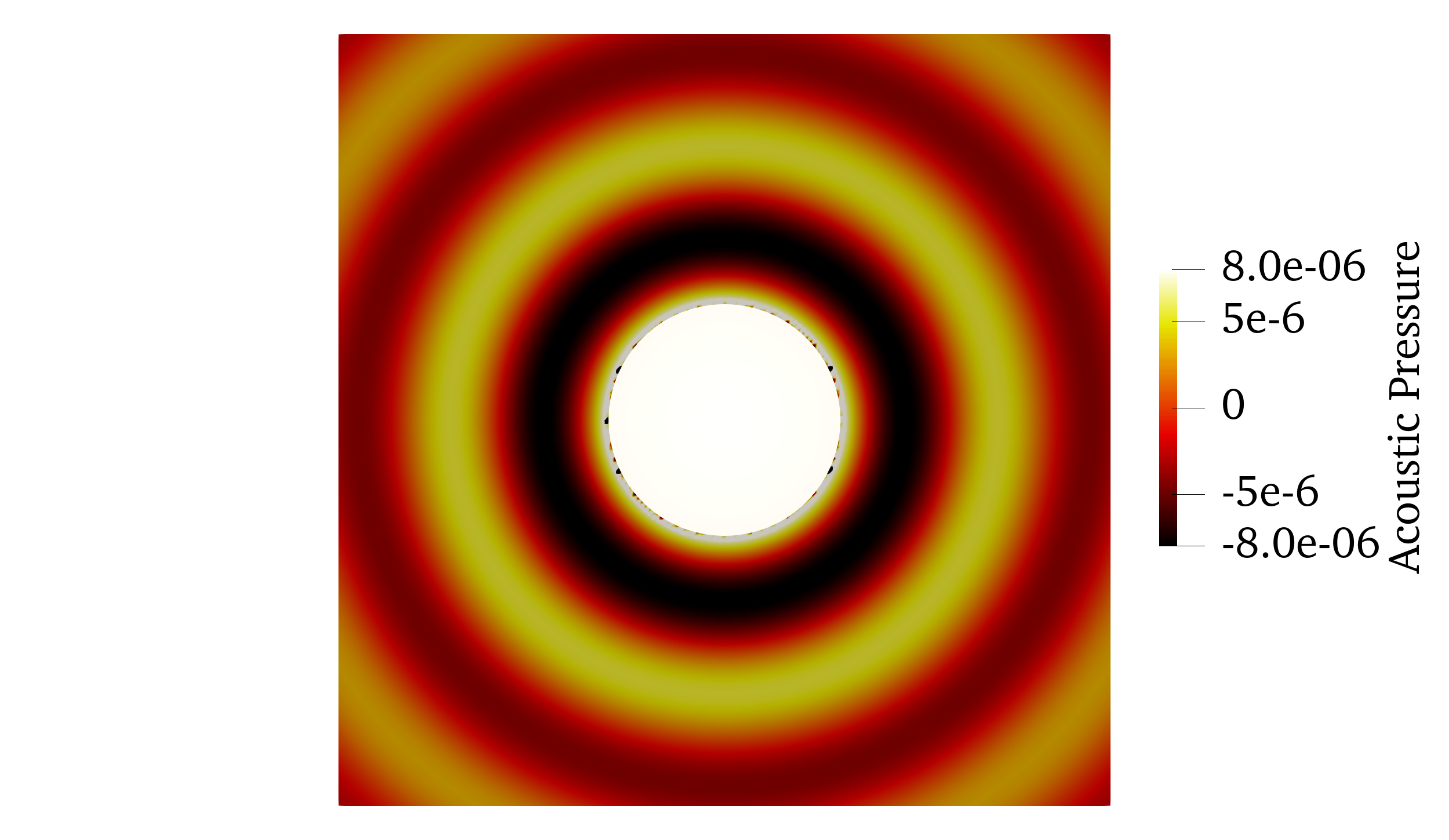}
\subcaption{PyFWH.}
\end{subfigure}
\begin{subfigure}{0.3\textwidth}
\includegraphics[width=\textwidth]{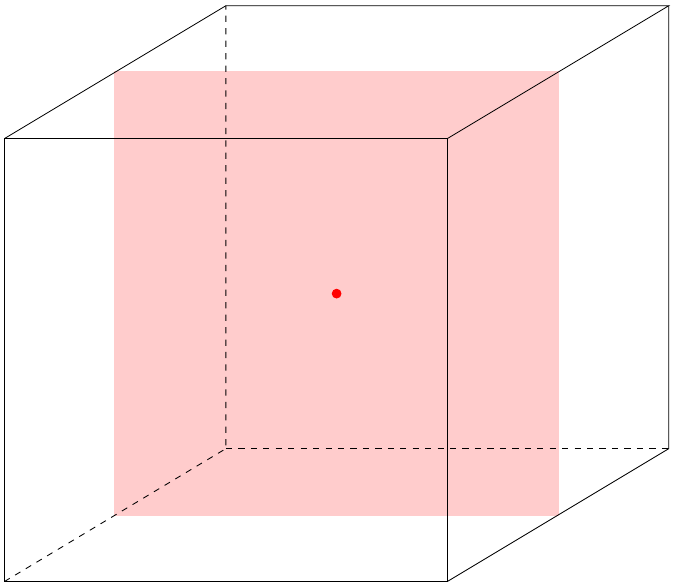}
\subcaption{The slice through the domain.}
\end{subfigure}
\caption{The acoustic pressure field obtained via direct and hybrid approaches using $\mathcal{P}3$ simulations.}
\label{figure:acousticfieldss}
\end{figure}

\begin{figure}
\centering
\includegraphics[width=0.7\textwidth]{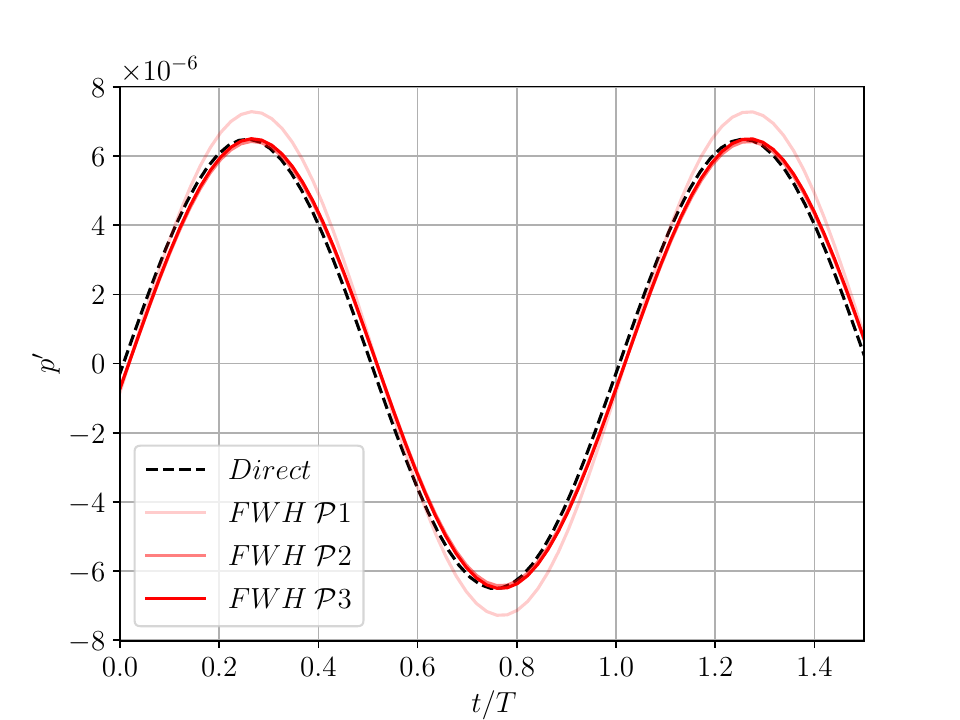}
\caption{Comparison of the time history of acoustic pressure obtained through the $\mathcal{P}3$ direct approach with that derived from hybrid approaches employing $\mathcal{P}1$, $\mathcal{P}2$, and $\mathcal{P}3$ CFD simulations as input for the acoustic solver for a single monopole.}
\label{figure:eulertimehistorySS}
\end{figure}

Figure \ref{figure:acousticfieldss} illustrates the acoustic pressure field obtained from HORUS through $\mathcal{P}3$ simulation, alongside the output from the acoustic solver driven by $\mathcal{P}3$ inputs. This is presented on a slice through the domain.
Figure \ref{figure:eulertimehistorySS} illustrates the acoustic pressure time history at the observer location for both approaches. Notably, the $\mathcal{P}1$ hybrid calculation exhibits an over-prediction of the acoustic pressure. However, a more favorable agreement with the $\mathcal{P}3$ direct approach is observed with the $\mathcal{P}2$ and $\mathcal{P}3$ hybrid approaches, where more accurate inputs are supplied for the acoustic solver. This underscores the substantial impact of flow solver accuracy on acoustic prediction, emphasizing the critical need for precise data in the acoustic solver.

\subsection{Multiple Monopoles in Quiescent Flow}

\begin{figure}
\centering
\begin{subfigure}{0.49\textwidth}
\includegraphics[width=\textwidth]{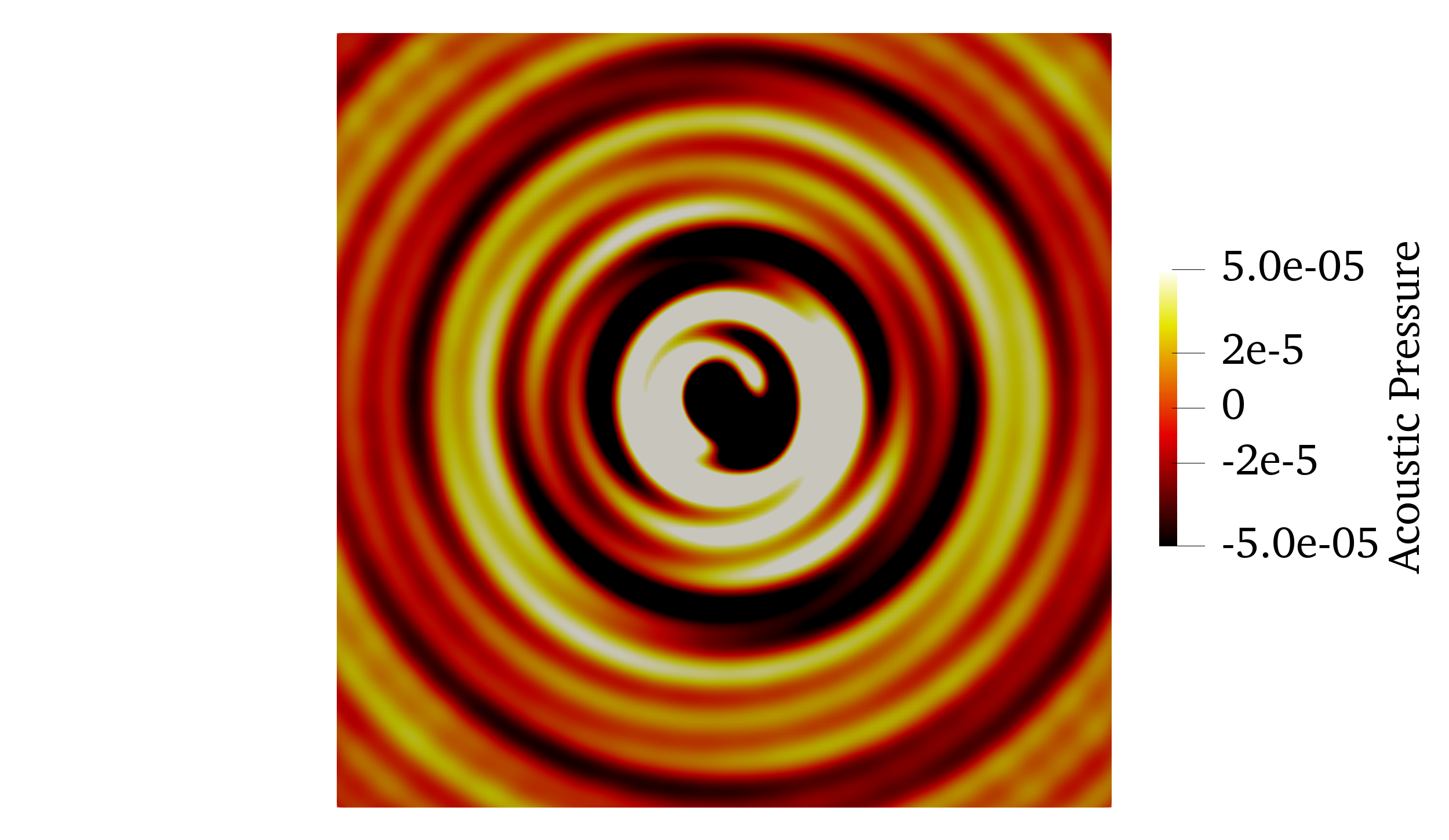}
\subcaption{HORUS.}
\end{subfigure}
\begin{subfigure}{0.49\textwidth}
\includegraphics[width=\textwidth]{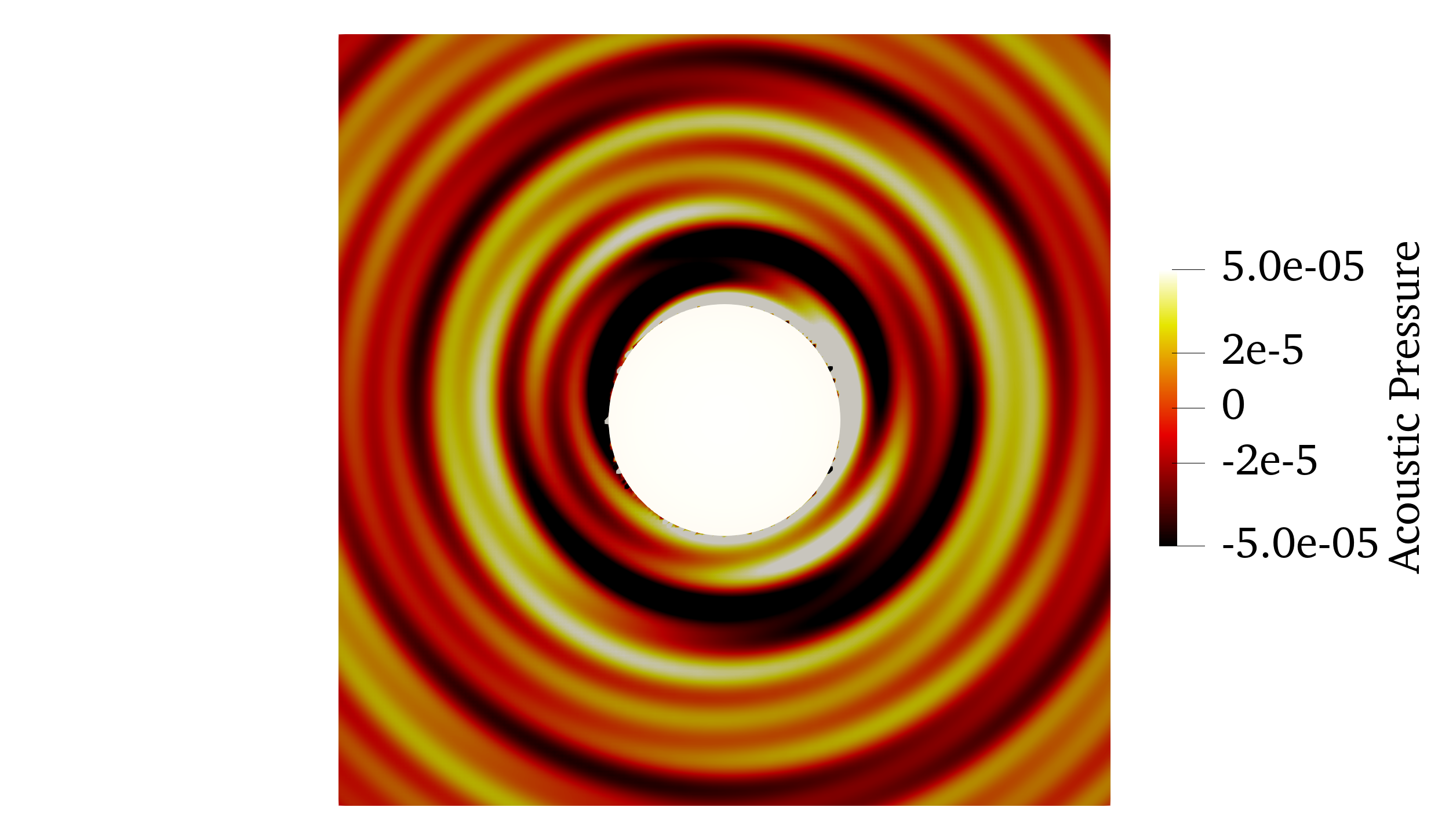}
\subcaption{PyFWH.}
\end{subfigure}
\begin{subfigure}{0.3\textwidth}
\includegraphics[width=\textwidth]{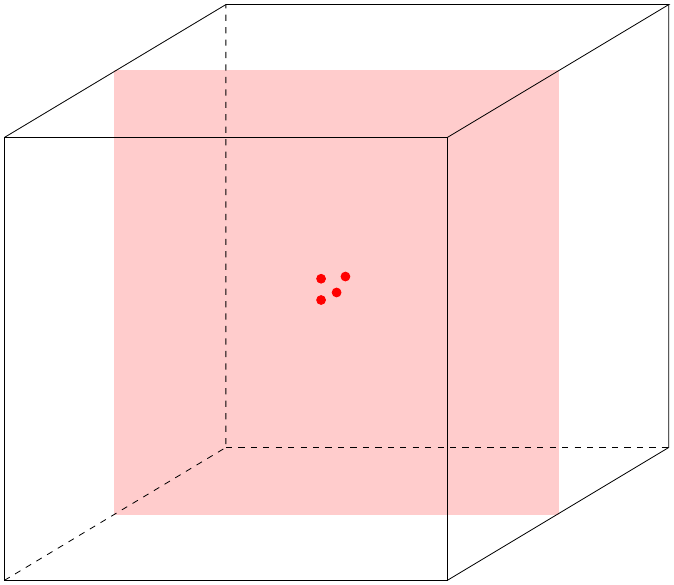}
\subcaption{The slice through the domain.}
\end{subfigure}
\caption{The acoustic pressure field obtained via direct and hybrid approaches using $\mathcal{P}3$ simulations.}
\label{figure:acousticfieldMS}
\end{figure}

\begin{figure}[htbp]
\centering
\includegraphics[width=0.7\textwidth]{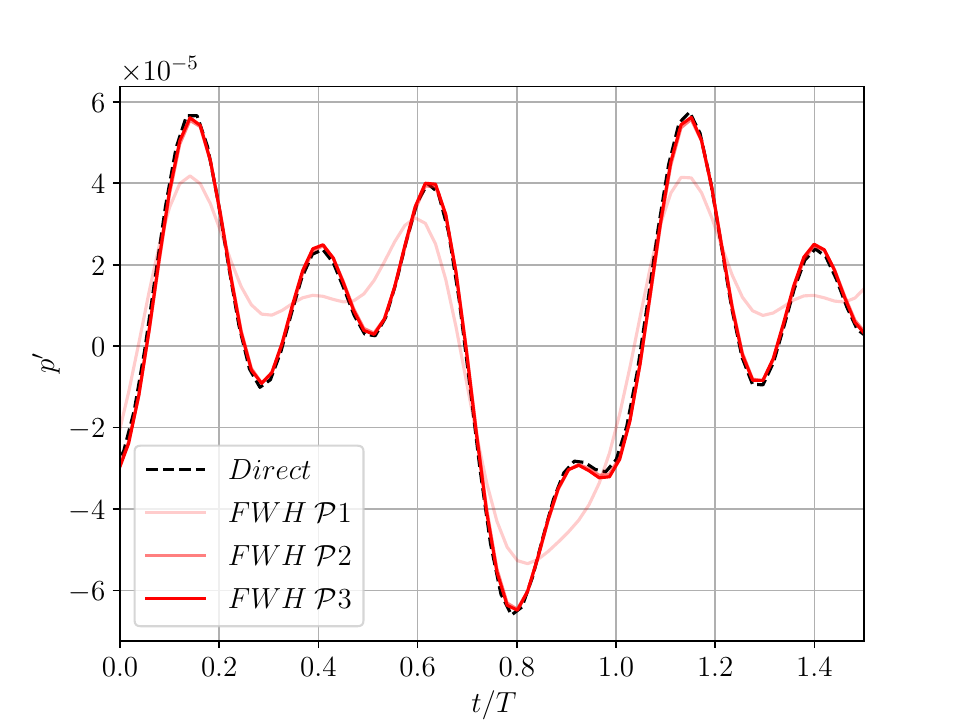}
\caption{Comparison of the time history of acoustic pressure obtained through the $\mathcal{P}3$ direct approach with that derived from hybrid approaches employing $\mathcal{P}1$, $\mathcal{P}2$, and $\mathcal{P}3$ CFD simulations as input for the acoustic solver for multiple monopoles.}
\label{figure:eulertimehistoryMS}
\end{figure}

To add complexity to the acoustic field, the preceding problem is replicated using four monopoles, each characterized by distinct amplitudes and frequencies, situated in close proximity to the origin. The source term incorporated into the energy equation is defined in the same manner as Equation \ref{eq:EulerSource},
\begin{dmath}
s(\pmb{x},t) = A \exp \left( {-k \left[ \left(x-x_{s_1} \right)^2 + \left(y-y_{s_1} \right)^2 + \left(z-z_{s_1} \right)^2 \right] } \right) \sin ( 2 \pi \omega t ) + A \exp \left( {-k \left[ \left(x-x_{s_2} \right)^2 + \left(y-y_{s_2} \right)^2 + \left(z-z_{s_2} \right)^2 \right] } \right) \sin ( 8 \pi \omega t ) + 2 A \exp \left({-k \left[ \left(x-x_{s_3} \right)^2 + \left(y-y_{s_3} \right)^2 + \left(z-z_{s_3} \right)^2 \right] } \right) \sin ( 4 \pi \omega t ) + 4 A \exp \left( {-k \left[ \left(x-x_{s_1} \right)^2 + \left(y-y_{s_1} \right)^2 + \left(z-z_{s_1} \right)^2 \right] } \right) \sin ( 2 \pi \omega t ) ,
\end{dmath}
where the monopoles are located at $[x_{s_1}, y_{s_1}, z_{s_1}] = [0, 0, 0]$, $[x_{s_2}, y_{s_2}, z_{s_2}] = [0.1, 0.3, 0.2]$, $[x_{s_3}, y_{s_3}, z_{s_3}] = [-0.2, 0.4, -0.3]$, and $[x_{s_4}, y_{s_4}, z_{s_4}] = [-0.4, -0.2, 0.1]$. The acoustic pressure field snapshots, depicted in Figure \ref{figure:acousticfieldMS}, demonstrate a qualitative agreement between results obtained from the flow solver and the acoustic solver. Furthermore, Figure \ref{figure:eulertimehistoryMS} presents the temporal evolution of the acoustic pressure, reflecting behavior akin to that of a single monopole source. Significantly, increasing the polynomial degree in the CFD simulation enhances the accuracy of the acoustic solver outcomes.

\section{OrthoMADS Optimization Algorithm}
\label{sec:OrthoMads}

The MADS optimization technique falls between the Generalized Pattern Search (GPS) \cite{torczon1997convergence} and the Coope and Price frame-based methods \cite{coope2001convergence}. Unlike GPS, MADS allows for a more flexible exploration of the design space during the optimization process, which makes it a more effective solution for both unconstrained and linearly constrained optimization \cite{audet2006mesh}. A major advantage of MADS over GPS is the flexible local exploration, known as poll directions, rather than a fixed set of directions. Two parameters are defined in the context of the MADS optimization: the mesh size parameter, $\Delta^m$, and the poll size parameter, $\Delta^p$.  The mesh size parameter determines the resolution of the design space mesh. A higher resolution leads to a more precise search while a lower resolution allows for a wider search and a higher chance of finding the global optimal solution.  The poll size parameter determines the neighborhood size around the incumbent point for selecting new trial points. The number of trial points per design cycle can be either $n+1$, known as minimal positive basis, or $2n$, known as maximal positive basis \cite{audet2006mesh}, where $n$ is the number of design variables. In this study, the maximal positive basis construction is used. 

The MADS algorithm consists of two sequential steps in each iteration: the search step and the poll step. Initially, the optimization procedure starts with the search step at the initial design point, $\pmb{\mathcal{X}}_0 = [\mathcal{X}^1_0, \mathcal{X}^2_0, ..., \mathcal{X}^n_0]$, where the subscript is the optimization iteration and the superscript denotes each design parameter. Pseudo-random trial points are generated, and infeasible ones, which are points within the design space not meeting the optimization problem's constraints, are discarded. The trial points are generated based on the current mesh and the direction vectors, $d_j \in \mathcal{D}$ (for $j = 1,2,...,n$), where $\mathcal{D}$ is the design space. $\mathcal{D}$ must be a positive spanning set \cite{davis1954theory}, and each direction, $d_j$, must be the product of some fixed non-singular generating matrix by an integer vector \cite{audet2006mesh}. The mesh at iteration $k$ is defined as \cite{audet2006mesh}
\begin{equation}
\mathcal{M}_k = \bigcup_{\mathcal{X} \in \mathcal{S}_k} \left\lbrace \mathcal{X}  + \Delta^m_k \mathcal{D} z : z \in \mathbb{N}^{n_{\mathcal{D}}} \right\rbrace ,
\end{equation}
where $\mathcal{S}_k$ is the set of trial points that the objective function is evaluated at, in iteration $k$. The mesh $\mathcal{M}_k$ is constructed from a finite set of $n_{\mathcal{D}}$ directions, $\mathcal{D} \subset \mathbb{R}^n$, scaled by a mesh size parameter $\Delta^m_k \in \mathbb{R}_{+}$. The objective function is evaluated at these trial points. The iteration terminates either after evaluating the objective function at all trial points or upon finding a lower objective function, where the former is employed in this study. Then, the next iteration starts with a new incumbent solution $\pmb{\mathcal{X}}_{k+1} \in \Omega$ with objective function of $\mathcal{F}(\pmb{\mathcal{X}}_{k+1}) < \mathcal{F}(\pmb{\mathcal{X}}_k)$, and a mesh size parameter $\Delta^m_{k+1} \geq \Delta_k^m$. The maximum value of the mesh size parameter, at any iteration, is set to one, $\Delta^m_{max}=1$. Note that the design space of each design variable is scaled to one, and a mesh size parameter of one can cover the entire design space.

On the other hand, if the search step fails to find a new optimum, the poll step is invoked before terminating the current optimization iteration. In the poll step, the mesh size parameter is reduced to define a new set of trial points closer to the incumbent design. The key difference between GPS and MADS is the new poll size parameter, $\Delta^p_k \in \mathbb{R}_{+}$, that controls the magnitude of the distance between trial points generated by the poll step to the incumbent point. This new set of trial points defined in the poll step is called a frame. The MADS frame at iteration $k$ is defined to be \cite{audet2006mesh}
\begin{equation}
P_k = \left\lbrace \mathcal{X}_k + \Delta^m_k d: d \in \mathcal{D}_k \right\rbrace \subset \mathcal{M}_k ,
\end{equation}
where $\mathcal{D}_k$ is a positive spanning set. In each MADS iteration, the mesh and poll size parameters are defined. The mesh size parameter of the new iteration is defined as \cite{audet2006mesh}
\begin{equation}
\Delta_{k+1}^m = 
\begin{cases}
\frac{1}{4} \Delta_k^m & \text{if the search step fails to find an improved design point,} \\
4 \Delta_k^m & \text{if an improved design point is found, and if } \Delta_k^m \leq \frac{1}{4}, \\
\Delta_k^m & \text{otherwise.} 
\end{cases}
\end{equation}
These rules ensure $\Delta_k^m$ is always a power of $4$ and never exceeds $1$. The poll size parameter is also defined as \cite{audet2006mesh}
\begin{equation}
\Delta_{k+1}^p = 
\begin{cases}
n \sqrt{\Delta_k^m} & \text{if the minimal positive basis construction is used,} \\
\sqrt{\Delta_k^m} & \text{if the maximal positive basis construction is used.} 
\end{cases}
\label{eq:poll_and_mesh_parameters}
\end{equation}

OrthoMADS is a deterministic variant of MADS that replaces the randomly generated polling directions with structured, orthogonal directions. It uses a combination of Halton sequences and Householder transformations to generate a maximal positive basis of polling directions that are both orthogonal and mesh-compatible. This results in well-distributed trial points, and enhances directional coverage compared to stochastic versions. OrthoMADS retains the same underlying mesh framework and adaptive strategy as MADS but improves efficiency and convergence consistency by eliminating randomness in direction generation.

\end{appendices}

\clearpage

\bibliography{arXiv}

\end{document}